\documentclass[twocolumn]{aastex62}

\usepackage{graphics}
\usepackage{float}
\usepackage{comment}
\usepackage{aas_macros}

\newcommand{\MZAMS}{\ensuremath{M_{\rm ZAMS}}}

\newcommand{\LOVC}{\ensuremath{\rm L_{A\,}}}
\newcommand{\MOVC}{\ensuremath{\rm M_{A\,}}}

\usepackage{bm}
\renewcommand{\vec}[1]{\ensuremath{\bm{#1}}}
\renewcommand{\tensor}[1]{\ensuremath{\bm{#1}}}

\newcommand{\KT}{\textcolor{black}}

\graphicspath{{./}{figures/}}

\received{January 1, 2018}
\revised{\today}
\accepted{}
\submitjournal{ApJ}

\shorttitle{Multi-D simulations of Oxygen shell burning}
\shortauthors{Yoshida et al.}

\begin{document}

\title{One-, Two-, and Three-dimensional Simulations of Oxygen Shell Burning Just Before the Core-Collapse of Massive Stars}

\correspondingauthor{Takashi Yoshida}
\email{tyoshida@astron.s.u-tokyo.ac.jp}

\author[0000-0002-8967-7063]{Takashi Yoshida}
\affil{Department of Astronomy, Graduate School of Science, University of Tokyo, 7-3-1 Hongo, Bunkyo-ku, Tokyo 113-0033, Japan}

\author[0000-0003-0304-9283]{Tomoya Takiwaki}
\affiliation{Division of Theoretical Astronomy, National Astronomical Observatory of Japan, 2-21-1 Osawa, Mitaka, Tokyo 181-8588, Japan}

\author[0000-0003-2456-6183]{Kei Kotake}
\affiliation{Department of Applied Physics \& Research Institute of Stellar Explosive Phenomena, Fukuoka University, Fukuoka 814-0180, Japan}

\author[0000-0002-6705-6303]{Koh Takahashi}
\affiliation{Argelander-Institute f\"ur Astronomie, Universit\"ate Bonn, D-53121 Bonn, Germany}
\affiliation{Max Planck Institute for Gravitational Physics, D-14476 Potsdam, Germany}

\author[0000-0002-8734-2147]{Ko Nakamura}
\affiliation{Department of Applied Physics, Fukuoka University, Fukuoka 814-0180, Japan}

\author{Hideyuki Umeda}
\affiliation{Department of Astronomy, Graduate School of Science, University of Tokyo, 7-3-1 Hongo, Bunkyo-ku, Tokyo 113-0033, Japan}

\begin{abstract}
We perform two- (2D) and three-dimensional (3D) hydrodynamics simulations of convective oxygen shell-burning that takes place deep inside a massive progenitor star of a core-collapse supernova.
Using one dimensional (1D) stellar evolution code,  we first calculate the evolution of massive stars with an initial mass of 9--40 $M_\odot$.
Four different overshoot parameters are applied, and CO core mass trend similar to previous works is obtained in the 1D models. Selecting eleven 1D models that have a silicon and oxygen coexisting layer, we perform 2D hydrodynamics simulations of the evolution for $\sim$100 s until the onset of core-collapse.
We find that convection with large-scale eddies and the turbulent Mach number $\sim$0.1 is obtained in the models having a Si/O layer with a scale of 10$^8$ cm, whereas most models that have an extended O/Si layer up to a few $\times 10^9$ cm exhibit lower turbulent velocity. Our results indicate that the supernova progenitors that possess a thick Si/O layer could provide a preferable condition for perturbation-aided explosions. We perform 3D simulation of a 25 $M_\odot$ model, which exhibits large-scale convection in the 2D models. The 3D model develops large-scale ($\ell = 2$) convection similar to the 2D model, however, the turbulent velocity is lower.
By estimating the neutrino emission properties of the 3D model, we point out that a time modulation of the event rates, if observed in KamLAND and Hyper-Kamiokande, would provide an important information about structural changes in the presupernova convective layer. 

\end{abstract}


\section{Introduction}\label{sec:intro}

From theory and observations, it is almost certain that the explosions of massive stars as core-collapse supernovae (CCSNe) are generically multi-dimensional (multi-D) phenomena (see \citet{foglizzo_15,janka16_rev,patat_book} for reviews).
To facilitate the neutrino-driven mechanism of CCSNe \citep{bethe85}, multi-D hydrodynamics instabilities such as neutrino-driven 
convection and the standing accretion shock instability \citep{Blondin03} play a pivotal role in enhancing the neutrino heating efficiency to trigger the onset of the explosion.
In fact, a growing number of self-consistent 
models in two or three spatial dimensions 
(2D, 3D) now report revival of the stalled 
bounce shock into explosion for a wide mass range of progenitors (see, e.g., \citet{vartanyan19,oconnor18b,bernhard17,Roberts16,Nakamura16,melson15a,Summa16,Lentz15,Takiwaki14,Hanke13} for collective references therein).

These successes, however, provide further motivation for exploring missing ingredients in the neutrino mechanism, partly because the estimated explosion energies obtained in the multi-D models generally do not reach the typically observed value 
 \citep[e.g., $\sim$10$^{51}$erg,][]{Tanaka09}. 
 Various possible candidates to obtain more robust explosions have recently been proposed, including multi-D effects 
during the final stage of the presupernova evolution
(see \citet{couch_book} for a review), general relativity (GR, e.g., \citet{BMuller12a,Ott13,KurodaT12,KurodaT16}),
rapid rotation (e.g., \citet{Marek09,Suwa10,Takiwaki16,Summa18,harada18}) and/or magnetic
fields (e.g., \citet{martin06,moesta14,jerome15,masada15,martin17}), and sophistication in the neutrino opacities \citep{melson15b,bollig17,Burrows18,kotake18} and in the neutrino transport schemes (e.g., \citet{sumi12,richers17,nagakura18,just18}). In this work, we focus on the first item listed in the above list. 

\citet{couch_ott13} were the first to demonstrate that the inhomogeneities seeded by convective shell burning fosters the onset of a neutrino-driven explosion (see also \citet{rodorigo14,couch_ott15,bernhard15,Burrows18}). This is because the infalling perturbation that could be amplified in the supersonic accretion
\citep{takahashi14,nagakura13,nagakura19}  enhances turbulence behind the postshock material  leading to the reduction of the critical neutrino luminosity for shock revival (e.g., \citet{bernhard15,ernazar16}). In these studies, the non-spherical structures in the burning shells, although physically motivated, were treated in a parametric manner due to the paucity of the multi-D stellar evolution models covering the lifespan of massive stars up to the iron core-collapse. Currently one-dimensional (1D) stellar evolution calculations are the only way to accomplish this \citep{woosley02,Woosley07,Sukhbold18}, where the errors introduced from the omission of multi-D effects are absorbed into the free parameters of MLT, namely the mixing length theory, (e.g., \citet{kippenhahn12}). 
 
The {\it truly} multi-D hydrodynamics stellar evolution calculations have been  done over several turnover timescales of convection (limited by the affordable computational resources) in selected burning shells (e.g., \citet{meakin07}; \citet{viallet13,simon15,cristini17}; \citet{cristini19} for different burning shells, and see \citet{arnett16} for a review).
Pushed by the observation of SN1987A,  2D and 3D stellar evolution simulations focusing on the late burning stages have been extensively carried out since the 1990s \citep{arnett94,bazan94,bazan98,asida00,kuhlen03,meakin06,meakin07,arnett11,chatz14,chatz16,jones17}. 
  
More recently, ground-breaking attempts to evolve convective shells in 3D prior to the onset of collapse have been first reported by \citet{Couch15} for silicon shell burning in a 15 $M_{\odot}$ star and by \citet{bernhard16_prog} for oxygen shell burning in an 18 $M_{\odot}$ star (and also in 11.8, 12, and 12.5 $M_{\odot}$ stars by \citet{bernhard18_prog}). \citet{Couch15} obtained earlier onset of a neutrino-driven explosion for the 3D progenitor model of the $15 M_{\odot}$ star compared to that in the corresponding 1D progenitor model. 
By performing 3D GR simulations with more advanced neutrino transport scheme, \citet{bernhard17} obtained a neutrino-driven explosion with the seed perturbations.
In comparison, this shock was not revived in the corresponding 1D progenitor model. These studies clearly show that convective seed perturbations could potentially have a favorable impact on the neutrino-driven explosions. 
In order to clarify the criteria for precollapse seed perturbation growth, \citet{Collins18} recently reported a detailed analysis on the convective oxygen and silicon burning shells by performing a broad range of 1D presupernova calculations. Using the prescription of the MLT theory in 1D, they pointed out that the extended oxygen burning shells between $\sim$16 and 26 $M_{\odot}$ are most likely to exhibit large-scale convective overturn with high convective Mach numbers, leading to the most favorable condition for perturbation-aided explosions. In fact the 3D progenitor model of the 18 $M_{\odot}$ star \citep{bernhard16_prog} is in the predicted mass range.

Joining in these efforts, we investigate in this study how the asphericities could grow, particularly  driven by
the convective oxygen shell burning in the O and Si-rich layer. 
First we perform a series of 1D stellar evolution calculation with zero-age main sequence (ZAMS) masses between 9 and 40 $M_{\odot}$ with the {\bf HOSHI} code developed by \citet{Takahashi16,Takahashi18}. Based on the 1D results, 
we select $\sim$ ten 1D progenitors that have extended and enriched O and Si
layers, presumably leading to vigorous convection.
At a time of $\sim 100$ s before the onset of collapse, the 1D evolution models are mapped to multi-D hydrodynamics code (a branch of {\bf 3DnSNe}, e.g., \citet{Takiwaki16,Nakamura16,kotake18}). 
We perform axisymmetric (2D) simulations for the selected progenitors having an extended O and Si-rich layer and investigate the features of their convective motion,  especially the convective-eddy scale and the turbulent Mach number. 
We then move on to perform a 3D simulation by choosing one of the progenitors that exhibits strong convective activity in 2D.  We make an analysis to investigate how the convective features between 3D and 2D differs and discuss its possible implication to the explosion dynamics.

The paper is organized as follows. Section \ref{sec2} starts with a brief description of the numerical methods employed in our 1D stellar evolution calculation as well as 2D and 3D hydrodynamics simulations.
In Section \ref{sec3}, we present the results of the 1D stellar evolution models in Section \ref{1d}, which is followed in order by
 2D (Section \ref{2d}) and 3D (Section \ref{3d}) results, respectively. 
 In Section \ref{sec4}, we 
 summarize with a discussion of the 
 possible implications.
 Appendices address the comparison of our 1D stellar evolution code with other reference codes and the sensitivity of our results with respect to the different  parameters.  

\section{SETUP and NUMERICAL METHODs} \label{sec2}

In this section, we briefly summarize the numerical setups of our stellar evolution calculations in 1D (Section \ref{1d}), 
2D (Section \ref{2d}), and 3D (Section \ref{3d}).

\subsection{1D Stellar Evolution} \label{subsec:1d-method}

\begin{table}[t]
\centering
\caption{Stellar evolution model sets and the corresponding overshoot parameters.
$f_{\rm ov}$ is the overshoot parameter during the H and He burning phases. 
$f_{\rm ov,A}$ is the one during the more advanced stages, namely after the He burning phase.
For the meanings of the notations ``L" and ``M", see the text.
Convective overshoot during the advanced stages is considered for model sets with the subscript ``$_{\rm A}$".
} \label{table1}
\begin{tabular}{ccc}
\tablewidth{0pt}
\hline
\hline
Model & $f_{\rm ov}$ & $f_{\rm ov,A}$ \\
\hline
L & 0.03 & 0 \\
L$_{\rm A}$ & 0.03 & 0.002 \\
M & 0.01 & 0 \\
M$_{\rm A}$ & 0.01 & 0.002 \\
\hline
\end{tabular}
\end{table}

We calculate the 1D evolution of solar-metallicity massive stars with the ZAMS masses between 9 and 40 $M_{\odot}$ up to the onset of collapse of the iron core. 
The calculations are performed 
using an up-to-date version of the 1D stellar evolution code, 
{\bf HOSHI} ({\bf HO}ngo {\bf S}tellar {\bf H}ydrodynamics {\bf I}nvestigator) code\footnote{"HOSHI" is a noun in Japanese meaning a star.} \citep[e.g.][]{Takahashi16,Takahashi18}.
In the code, a 300-species nuclear reaction network\footnote{The employed nuclear species are as follows:
$^1$n, $^{1,2}$H, $^{3,4}$He, $^{6,7}$Li, $^{7,9}$Be, $^{8,10,11}$B, $^{11-16}$C, $^{13-18}$N, $^{14-20}$O, $^{17-22}$F, $^{18-24}$Ne, $^{21-26}$Na, $^{22-28}$Mg, $^{25-30}$Al, $^{26-32}$Si, $^{27-34}$P, $^{30-37}$S, $^{32-38}$Cl, $^{34-43}$Ar, $^{36-45}$K, $^{38-48}$Ca, $^{40-49}$Sc, $^{42-51}$Ti, $^{44-53}$V, $^{46-55}$Cr, $^{48-57}$Mn, $^{50-61}$Fe, $^{51-62}$Co, $^{54-66}$Ni, $^{56-68}$Cu, $^{59-71}$Zn, $^{61-73}$Ga, $^{63-75}$Ge, $^{65-76}$As, $^{67-78}$Se, $^{69-79}$Br.
The isomeric state of $^{26}$Al is also included.} is included, the rates of which are 
 taken from JINA REACLIB v1 \citep{Cyburt10} except for $^{12}$C($\alpha,\gamma)^{16}$O 
 \citep[see][for details]{Takahashi16}.

The mass of the helium (He), carbon-oxygen (CO), and iron (Fe) cores as well as the advanced stage evolution depend on the treatment of convection.
We use the Ledoux criterion for convective instability.
Inside the convective region, we treat the chemical mixing by means of the MLT using the diffusion coefficients as described in \citet{Takahashi18}. 

In order to take into account the chemical mixing by convective overshoot,
 an exponentially decaying coefficient,
\begin{equation}
    D_{\rm cv}^{\rm ov} = D_{\rm cv,0} \exp \left( -2 \frac{\Delta r}{f_{\rm ov} H_{P0}}\right),
\label{eq:overshoot}
\end{equation}
 is included, where $D_{\rm cv,0}$, $\Delta r$, $f_{\rm ov}$, $H_{P0}$ are the diffusion coefficient at the convective boundary, the distance from the convective boundary, the overshoot parameter, and the pressure scale-height at the convective boundary, respectively
 \citep[e.g.,][]{Takahashi16}. 
We consider the following four models 
to see the impacts of the different overshoot parameter ($f_{\rm ov}$) on the 1D stellar evolution. First we consider the two cases during the H and He core burning phases ($f_{\rm ov}$ = 0.03 or 0.01, see Table 1). The former and the latter values are determined based 
on the calibrations to early B-type stars in the Large Magellanic Cloud (LMC) \citep{Brott11} and the main-sequence width observed for AB stars in open clusters of the Milky-Way Galaxy \citep{Maeder89,Ekstroem12}, respectively. We name the former model as model ``L" after the LMC and the latter model as model "M" after the Milky-Way Galaxy.
Similarly, in order to investigate the impact in more advanced stages, we test two different overshooting parameters of $f_{\rm ov,A}$ = 0 or 0.002 (see Table 1) for the advanced stages including core carbon burning phase.
The convective overshoot during advanced stages is considered for both core and shell convective regions.
The subscript ``A" is added to the models with $f_{\rm ov,A} = 0.002$.
When a star model has a ZAMS mass of x $M_\odot$ and belongs to model L($_{\rm A}$) or M($_{\rm A}$), we set the model name to be xL($_{\rm A}$) or xM($_{\rm A}$).
We also name the set of models contained within models xL($_{\rm A}$) or xM($_{\rm A}$) after Set L($_{\rm A}$) or Set M($_{\rm A}$).

We note that the stellar evolution and the final structure also depend on the metallicity and rotation.
However, the main purpose of this study is to investigate
precollapse inhomogeneities for canonical CCSNe with solar-metallicity.
We leave the investigation of the metallicity and rotation dependence for the future study.

The mass loss rate at different evolution stages is important for determining the final mass and the He and CO core masses for high mass stars.
We adopt \citet{Vink01} as the mass loss rate of a main sequence star where the effective temperature is higher than $\log T_{\rm eff} = 4.05$ and the surface hydrogen mass fraction $X_{\rm H}$ is higher than or equal to 0.3.
The mass loss rate of \citet{Nugis00} is adopted for Wolf-Rayet (WR) stars where the effective temperature is higher than $\log T_{\rm eff} = 4.05$ and the surface hydrogen mass fraction $X_{\rm H}$ is lower than 0.3.
When the surface temperature is lower than $\log T_{\rm eff}$ = 3.90, we adopt \citet{deJager88} as the red supergiant mass loss rate.

\subsection{Multi-D Stellar Hydrodynamics Simulations}
\label{subsec:md-method}

We compute 2D and 3D 
models with our hydrodynamics code {\bf 3DnSEV}
({\bf 3} {\bf D}imensional {\bf n}uclear hydrodynamic simulation code for {\bf S}tellar {\bf EV}olution),
that is a branch of the {\it 3DnSNe} code 
(see \citet{Takiwaki16,Nakamura16,Sasaki17,kotake18} 
for the recent code development).
Similar to the base code,
{\it 3DnSEV} solves Newtonian hydrodynamics equations using spherical polar coordinates as follows:
\begin{eqnarray}
    \partial_t \rho + \nabla\cdot \left(\rho \vec{v}\right) &=& 0,\\
    \partial_t\left(\rho \vec{v}\right)  + \nabla\cdot\left(\rho \vec{v}\vec{v}+p\tensor{\delta} \right) &=& -\rho \nabla \Phi,\\
    \partial_t e_{\rm tot} + \nabla\cdot \left[(e_{\rm tot}+p) \vec{v}\right] &=& -\rho \vec{v}\cdot\nabla\Phi\nonumber\\
    & & +\rho \epsilon_{\rm burn}+C,\\
    \partial_t \rho X_i + \nabla\cdot \left(\rho X_i \vec{v}\right) &=& \gamma_{\rm burn},
\end{eqnarray}
where $\rho,\vec{v},p,e_{\rm tot},\Phi$ are density, velocity, pressure, total energy density (sum of internal energy and kinetic energy) and gravitational potential, respectively.
$X_i$ denotes mass fraction of $i$-th isotopes and  $\epsilon_{\rm burn}$ is the energy generation due to the change of composition, $\gamma_{\rm burn}$, by the nuclear burning. $C$ is the energy loss by neutrino emission.
The sub-grid scale physics is handled by implicit numerical
diffusion instead of solving
filtered hydrodynamic equations and creating a sub-grid
model for the dissipation of kinetic energy as Large-Eddy Simulation.
A piecewise linear method with the geometrical correction of the spherical coordinates is used to reconstruct variables at the cell edge, where a modified van Leer limiter is employed to satisfy the condition of total variation diminishing (TVD) \citep{Mignone14}.
The numerical flux is basically calculated by HLLC solver \citep{Toro94}.
For the numerical flux of isotopes, the consistent multi-fluid advection method of \citet{Plewa99} is used. 
The models are computed on a spherical polar coordinate grid with a resolution of 
$n_r \times n_\theta \times n_{\phi} = 512 \times 64 \times 128$ (3D) and
$n_r \times n_\theta = 512 \times 128$ (2D) zones.
The radial grid is logarithmically spaced and covers 
from the center up to the outer boundary of $10^{10}$ cm.
For the polar and azimuthal angle, the grid covers all $4\pi$ steradian.
To focus on the convective activity mainly in the oxygen shell,
the inner 100,000 km is solved in spherical symmetry.
We include self-gravity assuming a spherically symmetric (monopole) gravitational potential. Such a treatment is indispensable for reducing the 
computational time; the non-linear coupling between the core and the surrounding shells 
\citep[e.g.,][]{fuller15} 
is beyond the scope of this study.

We use the ``Helmhotlz" equation of state 
 \citep[EOS:][]{Timmes00}.
  The neutrino cooling is taken into account \citep{Itoh96} as a sink term in the energy equation.
A nuclear reaction network of 21-isotopes (aprox21)\footnote{\url{http://cococubed.asu.edu/code_pages/burn_helium.shtml}} \citep{Paxton11} is implemented, where the inclusion of $^{54}$Fe,$^{56}$Fe and $^{56}$Cr is crucial to treat a low electron fraction $Y_e \ga 0.43$ in the presupernova stage.
The network is as large as that of \citet{Couch15}
and a little larger than 19-isotopes of \citet{bernhard16_prog}.
When the temperature is higher than $5 \times 10^{9}$ K, the chemical composition is assumed to be in nuclear statistical equilibrium (NSE).
To avoid the temperature variations caused by numerical instability, we set an artificial upper bound in our multi-D runs, in such a way that the (absolute) sum of the local energy generation rates by thermonuclear reactions and weak interactions does not exceed 100 times the local neutrino cooling rate. 
To correctly treat the neutronization of heavy elements from
 Si to the iron group and the gradual shift of the nuclear abundances, one needs to use a sufficient number of 
 isotopes ($\sim 100$, \citet{arnett11}),  which is currently computationally and technically very challenging. Since the NSE region appears mainly in the Fe core, this treatment may not significantly 
 affect our results in which we focus on convection in the O layer. 

At a time of $\sim$100 s before the onset of collapse, the 1D evolution models are mapped to our multi-D hydrodynamics code and we follow the 2D and 3D 
evolution for the 
$\sim$100 s until the onset of collapse. 
When we start the multi-D runs, seed perturbations to trigger 
nonspherical motions are imposed to the 1D data
by introducing random perturbations of 
$1\%$ in density 
on the whole computational grid.
We terminate the 2D/3D runs by a
 criterion that the central temperature exceeds $9 \times 10^{9}$ K, because the core is dynamically collapsing at this time.

\section{Results} \label{sec3}

\subsection{1D Stellar Evolution Models} \label{1d}

\begin{figure}[hbpt]
\centering
\includegraphics[width=1.1\linewidth]{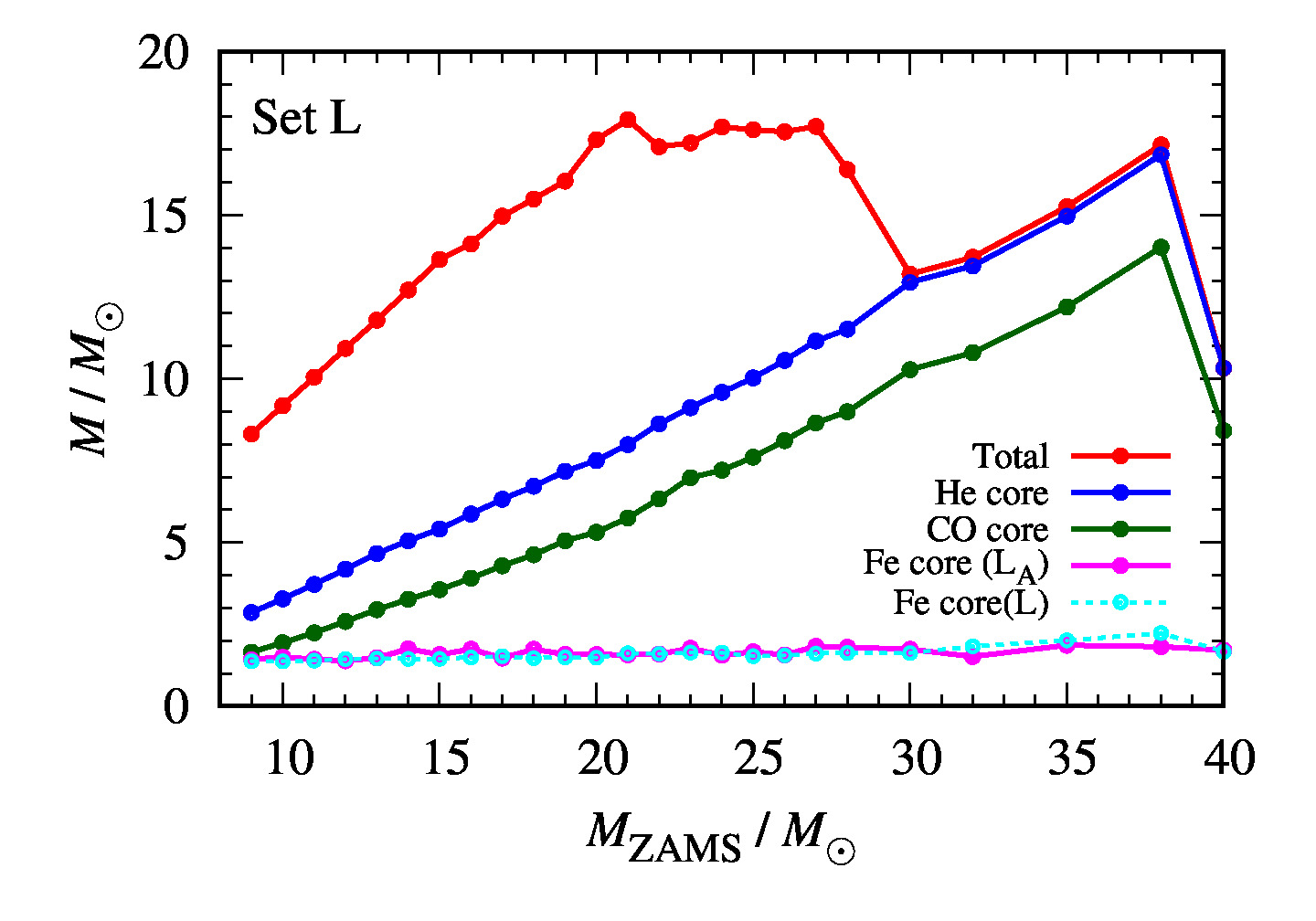}
\includegraphics[width=1.1\linewidth]{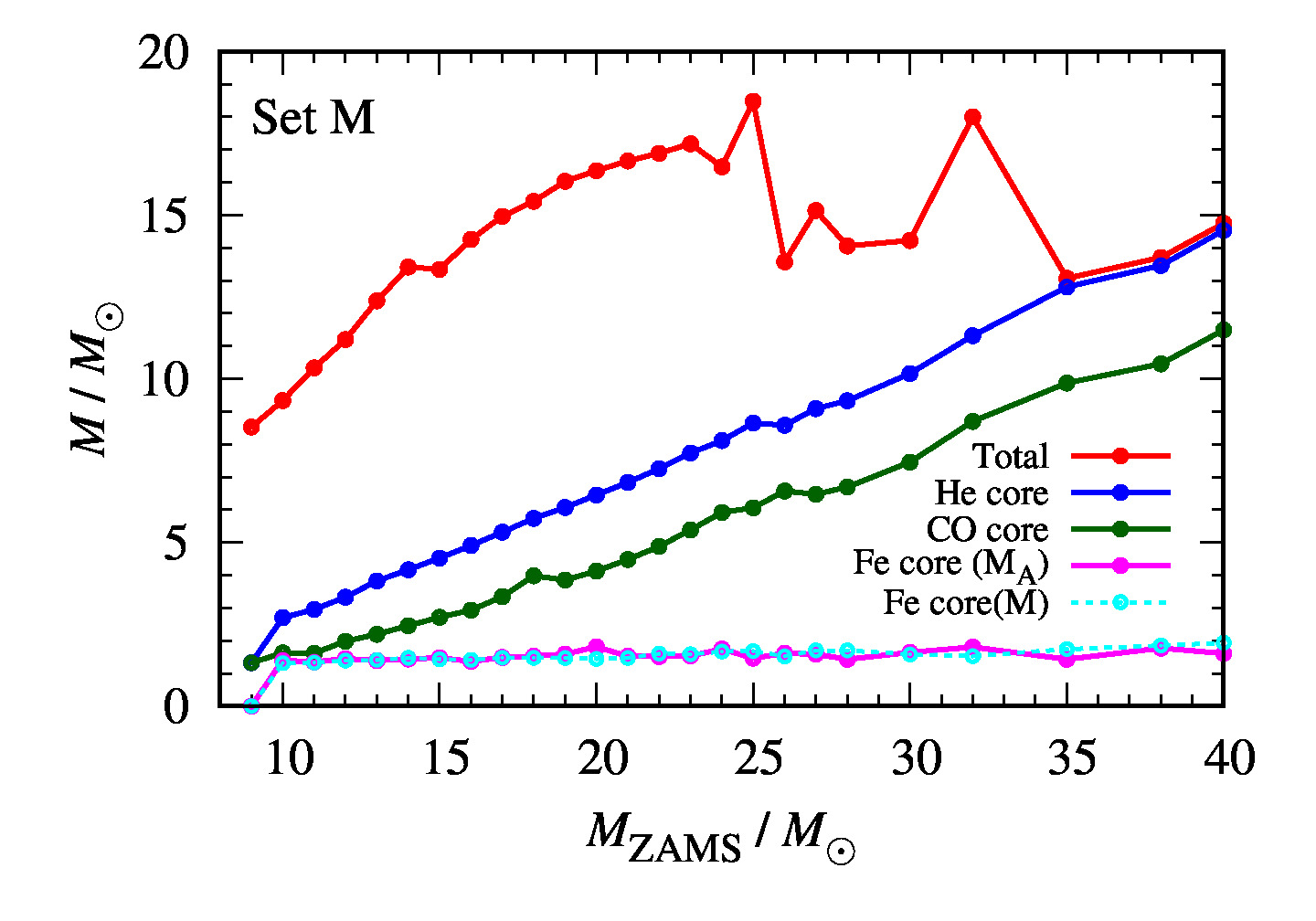}
\caption{The final stellar mass $M$ for Set L (top panel) and Set  M (bottom panel) as a function of the ZAMS mass. Red, blue, green, magenta, and dashed cyan lines correspond to the total mass, He core mass, CO core mass, Fe core mass of Sets L$_{\rm A}$ and M$_{\rm A}$, and Fe core mass of models L and M, respectively.
} 
\label{fig:f1}
\end{figure}

\KT{In total 100 stellar evolution models are calculated in 1D. Four sets of models are constructed, to which different overshoot parameters are applied (Table.1). Each set consists of 25 models to cover the initial mass range of 9--40 $M_\odot$.}
\KT{These 1D models are evolved from the ZAMS stage up to the onset of collapse, which is determined using the threshold central temperature of $T_{\rm C} \sim 10^{9.9}$} K.
Details of our 1D evolution models (e.g., comparison with reference stellar evolution codes) are given in Appendix.

\KT{Figure \ref{fig:f1} shows the total mass (red) and masses of the He (blue), CO (green), and Fe (magenta and cyan) cores at the onset of collapse as a function of the ZAMS mass ($M_{\rm ZAMS}$).}
\KT{The top panel shows results of Sets L and L$_{\rm A}$, while results of Sets M and M$_{\rm A}$ are shown as the bottom panel.
We note that Sets L and L$_{\rm A}$ result in very similar total, He-core, and CO-core masses, and differ only in the Fe-core mass.
This is because the He and CO core masses are mostly determined by the size of the core convection during the H and He burning phases, respectively, and are largely independent of the overshoot during more advanced stages.
The same is true for the Sets M and M$_{\rm A}$.}
\KT{The He core mass is defined as the largest enclosed mass having the hydrogen mass fraction less than $10^{-3}$. Similarly, the CO core mass is defined as the largest enclosed mass with the He mass fraction less than 0.1, and the Fe-core mass is defined as the largest enclosed mass with the sum of the mass fractions of $Z \ge 21$ elements larger than 0.5.}

\KT{The total mass at the collapse is determined by the mass loss history. 
Since the mass loss is relatively weak, the total mass monotonically increases with the ZAMS mass for models below $M_{\rm ZAMS} \lesssim 20$--$25 M_{\odot}$ in both Sets L and M. 
The mass loss rate increases with increasing luminosity, and thus with increasing ZAMS mass. The increasing mass loss rate explains the flat and even decreasing trends seen in 20--30 $M_\odot$ models in Set L. 
At the same time, models in the same mass range show a stochastic trend for Set M. This is caused by the bistability jump of the mass loss rate, which results from the discontinuous rate increase along the decreasing effective temperature \citep[e.g.,][]{Vink00}. 
For more massive models above $M_{\rm ZAMS} \gtrsim 30 M_\odot$, the mass loss rate becomes so efficient to remove most of the H envelope during the He burning phase. 
Therefore, the total masses of these models coincide with their He core masses. This is why the total mass again shows a monotonic increase in this massive end of the ZAMS mass range.
The most massive models (32, 35, and 38 $M_\odot$ models of Set L and 35, 38, and 40 $M_\odot$ models in Set M) finally retain only small amount of hydrogen of 0.26--0.29 $M_\odot$ in their envelopes, which will correspond to be observed as late type WN stars \citep{Crowther07}.}

\KT{As an exception,} model 40L ($M_{\rm ZAMS} = 40 M_\odot$ model in Set L) has lost not only whole the H envelope but also most of the He layer.
\KT{This is due to the even stronger WR wind mass loss during the helium and carbon burning phases.}
The He mass remaining on the surface is 0.24 $M_\odot$.
We apply the mass loss rate of \citet{Nugis00} \KT{for the H-deficient stars}.
\KT{However, there is a large uncertainty in estimation of the WR wind mass loss rate. Especially, among the H-deficient stars,} it has been discussed that the mass loss rate of He-deficient WC stars can be larger than the rate of \citet{Nugis00} \KT{by a factor of $\sim$10} \citep{Yoon2017}. 
\KT{The remaining He mass can be even less if we consider more efficient WR wind mass loss.}
\KT{Therefore, we expect that the star will most probably be observed as a He-deficient WC star, and moreover,}
it will be observed as a Type Ic SN when this star explodes as a SN.

\KT{The features of the distributions of the final stellar mass and He and CO-core masses as a function of ZAMS mass are also seen in the results obtained by previous works using Kepler code  \citep{Woosley07} (e.g., Figure 4 of \citet{Ebinger18} for a concise summary).}

\KT{The He and CO core masses monotonically increase with the ZAMS mass except for the model 40L$_{\rm A}$, which is affected by the strong WR wind during the He-burning phase.
The mass of the helium layer, which is shown as the difference between the He-core and the CO-core masses, also increases with the ZAMS mass.
As mentioned earlier, the He and CO core masses are insensitive to the overshoot parameter after the He burning, $f_{\rm ov,A}$.}
Thus, the difference in the CO core mass between Sets L$_{\rm A}$ and L (and similarly between Sets M$_{\rm A}$ and M) is less than 0.7\%. Furthermore, the difference in the He core mass is less than 0.1\%.
Note that model 18M$_{\rm A}$ exceptionally forms about 3\% larger CO core mass than model 18M. 
This results from an emergence of narrow convection in the outer layer of the CO core, in which small amount of He is contained.
Since this narrow convection is activated only after the core oxygen depletion, the outer structure than the He layer of model 18M$_{\rm A}$ is mostly the same with model 18M like other models.

We will compare Set L with Set M.
Below $M_{\rm ZAMS} \lesssim 25$ $M_\odot$, the final mass is not so sensitive to the overshoot parameter for H and He core convection.
On the other hand, models above 25 $M_\odot$ show a scatter within a factor of $\sim$0.3.
\KT{Set L tends to show larger He and CO core masses than Set M.
This is simply due to the larger overshoot parameter applied during the H and He burning phases. 
For most of the models, the ratio of the He core masses are about a factor of 1.2--1.3, and the CO-core mass ratio is somewhat larger than the He-core mass ratio. 
As an exception, the He core mass ratio reaches 2.16 for the 9 $M_\odot$ models.}
This is due to the merging of He layer to the H envelope by the second dredge-up. 
Larger overshoot for Set L brings about more effective convective mixing to make more massive He and CO cores.
Small core mass ratios for 40 $M_\odot$ models are due to the strong mass loss occurred in the model 40L.

\begin{figure}[htbp]
\includegraphics[width=\linewidth]{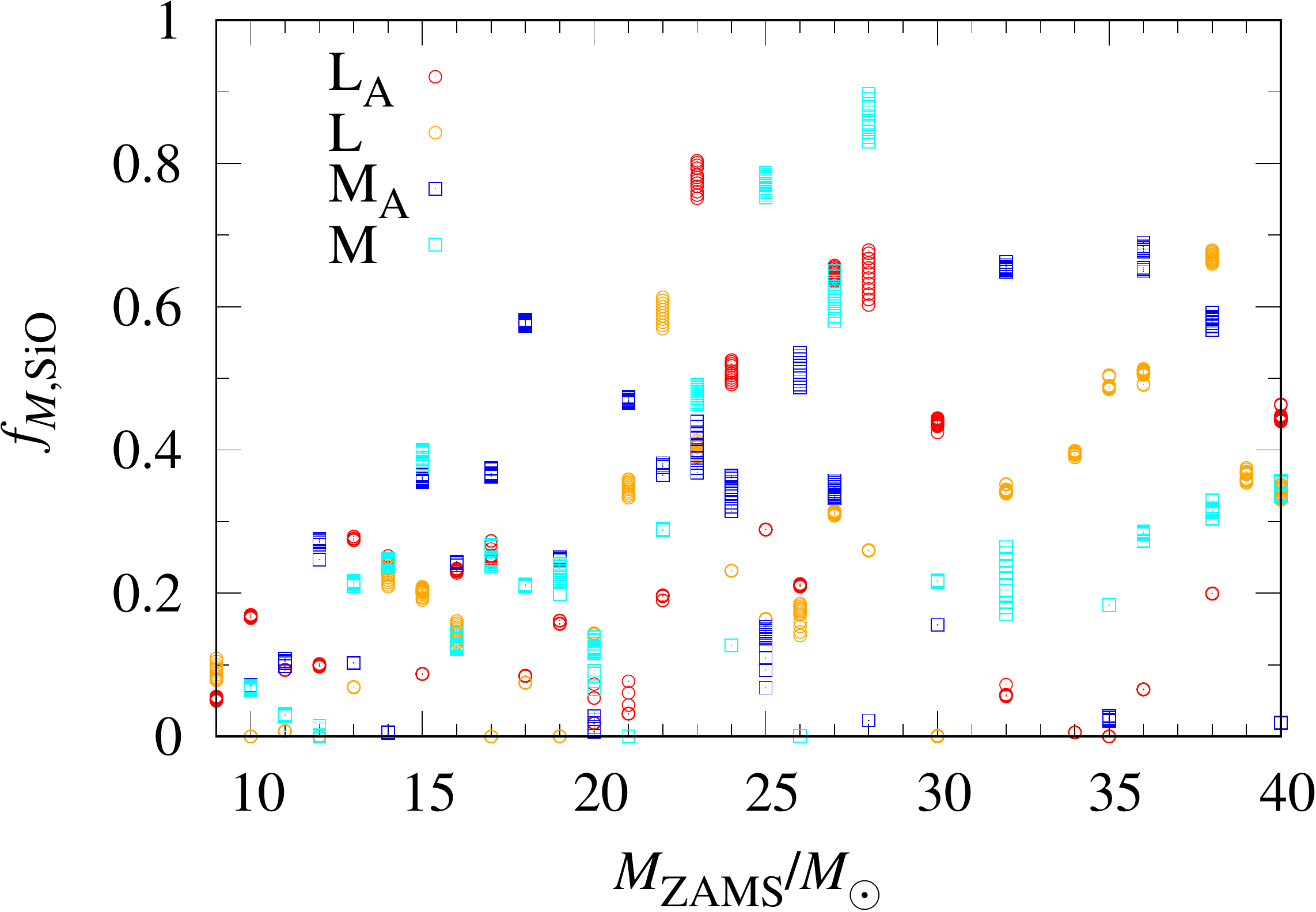}\\
\includegraphics[width=\linewidth]{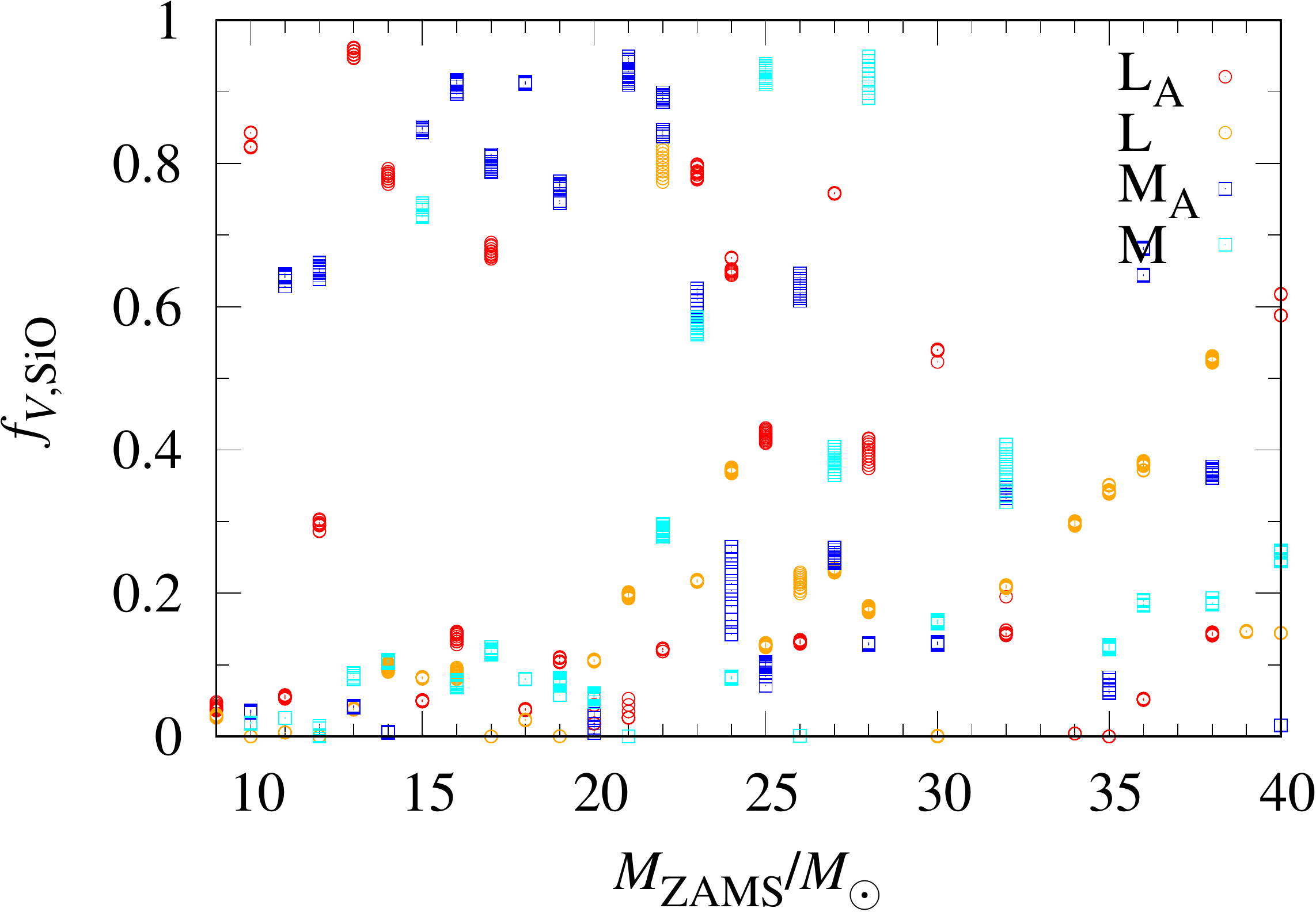}
\caption{SiO-coexistence parameters $f_{M,{\rm SiO}}$ (top panel) and $f_{V,{\rm SiO}}$ (bottom panel).
Red and orange circles represents models in Sets L$_{\rm A}$ and L, respectively.
Blue and cyan squares are for models in Sets M$_{\rm A}$ and M, respectively.}
\label{fig:f3}
\end{figure}

To select models that will show the strong convective activities in the 
SiO-rich layer, we utilize two measures, SiO-coexistence
parameters of $f_{M, {\rm SiO}}$ and $f_{V, {\rm SiO}}$.
Both of them are defined based on mass fraction distributions of $^{16}$O and/or $^{28}$Si.
The $f_{M,{\rm SiO}}$ is a product of the mass fractions of $^{16}$O and $^{28}$Si weighted by the enclosed mass between $10^8$ cm and $10^9$ cm:
\begin{eqnarray}
    f_{M, {\rm SiO}} &=& 
    c_M \int_{1}^{10} X(^{16}{\rm O}) X(^{28}{\rm Si}) \\ 
    &\times& 
    \Theta(X(^{16}{\rm O})-0.1)
    \Theta(X(^{28}{\rm Si})-0.1) 
    \rho r_8^2 d(r_8), \nonumber
\end{eqnarray}
where $c_M$ is a scaling coefficient, $X(^{A}Z)$ is the mass fraction of isotope $^{A}Z$, and $\Theta(x)$ is the step function, which satisfies $\Theta(x) = 1$ for $x \ge 0$ and 0 for $x < 0$, $\rho$ is the density, $r_8$ is the radius in units of $10^{8}$ cm.
\KT{Therefore, the value becomes large in a model that has a layer mainly composed of both oxygen and silicon. 
Such a layer would be the most preferable site to host strong turbulence powered by oxygen shell-burning.}
\KT{This definition of $f_{M,{\rm SiO}}$ has an uncertainty on what power of the local density strength of the turbulence depends. Hence, we also test another indicator, $f_{V,{\rm SiO}}$, in which the product of the mass fraction is weighted not by the enclosed mass but by the enclosed volume instead:}
\begin{eqnarray}
    f_{V, {\rm SiO}} &=& c_V \int_{1}^{10} X(^{16}{\rm O}) X(^{28}{\rm Si}) \\ 
    &\times& \Theta(X(^{16}{\rm O})-0.1)\Theta(X(^{28}{\rm Si})-0.1) r_8^2 d(r_8), \nonumber
\end{eqnarray}
where $c_V$ is a scaling coefficient.
The scaling coefficients are arbitrarily chosen.
We calculate these two measures at every time steps from 120 s to 80 s before the last step of the calculations to see the characteristics at times close to the onset of multidimensional simulations.

The result is shown in Figure \ref{fig:f3}, in which $c_M = 3.2 \times 10^{-10}$ and $c_V = 0.025$ are applied.
We do not see clear dependencies among different treatments of overshoot.
Some models in the ZAMS mass range $\ga 22 M_\odot$ show large ($\ga 0.6$) $f_{M,{\rm SiO}}$ values. 
In the volume weighted case, the ZAMS mass range showing the models having large ($\ga 0.9$) $f_{V,{\rm SiO}}$ values is 13--28 $M_\odot$.
\KT{From this result, we have selected eleven models, in which either $f_{M,{\rm SiO}}$ or $f_{V,{\rm SiO}}$, or possibly both of them, shows a large value.}
Models showing the seven highest $f_{M,{\rm SiO}}$ values are models 28M, 23L$_{\rm A}$, 25M, 28L$_{\rm A}$, 27L$_{\rm A}$, 27M, and 22L.
Models showing the six highest $f_{V,{\rm SiO}}$ values are models 13L$_{\rm A}$, 28M, 21M$_{\rm A}$, 25M, 16M$_{\rm A}$, 18M$_{\rm A}$.
Among them, models 25M and 28M show large values for both f$_{M,{\rm SiO}}$ and $f_{V,{\rm SiO}}$.
The actual values of the parameters are shown in the second and third columns of Table \ref{tab:t2}.

\begin{figure}[htbp]
\includegraphics[width=\linewidth]{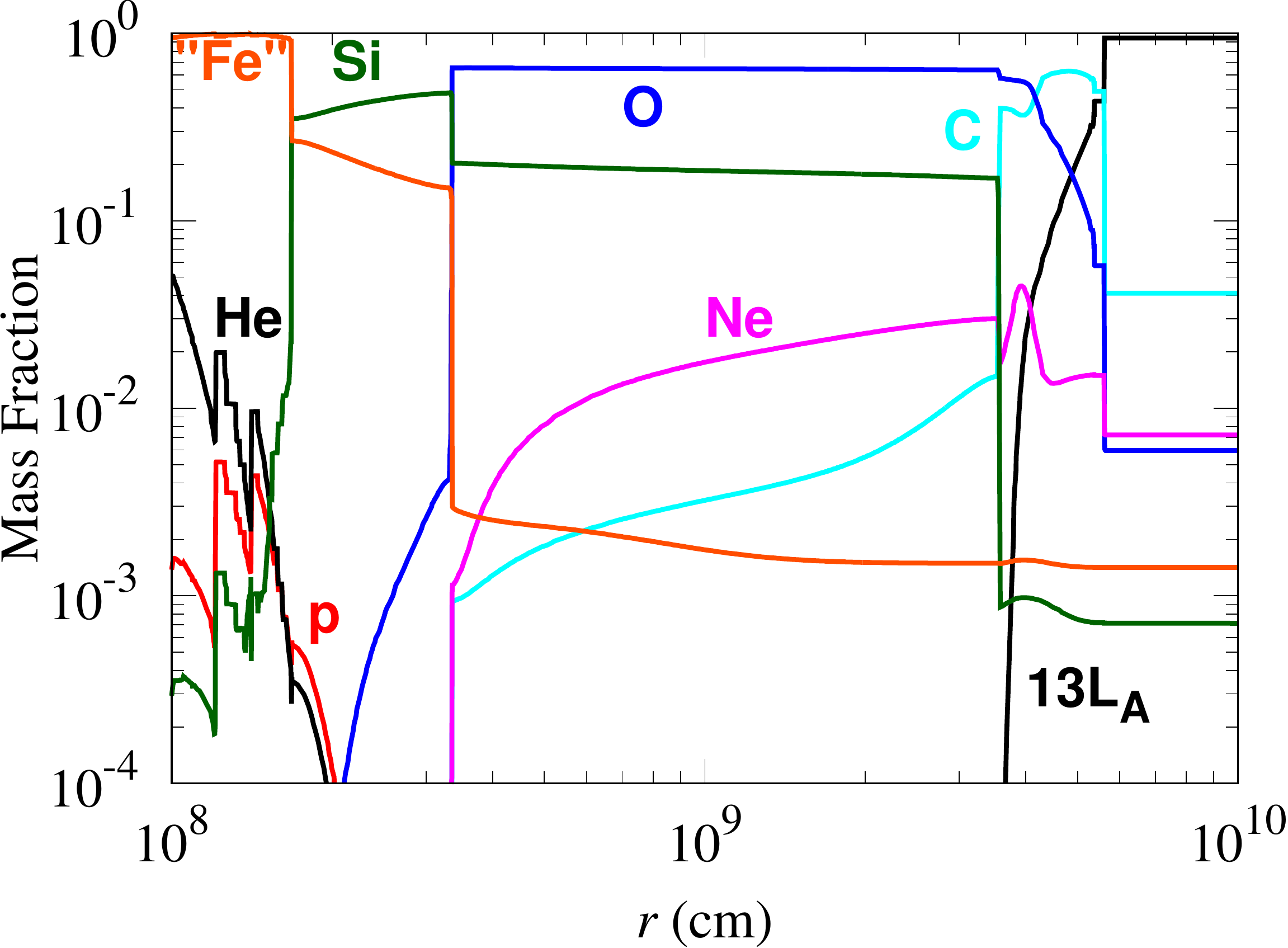}\\
\includegraphics[width=\linewidth]{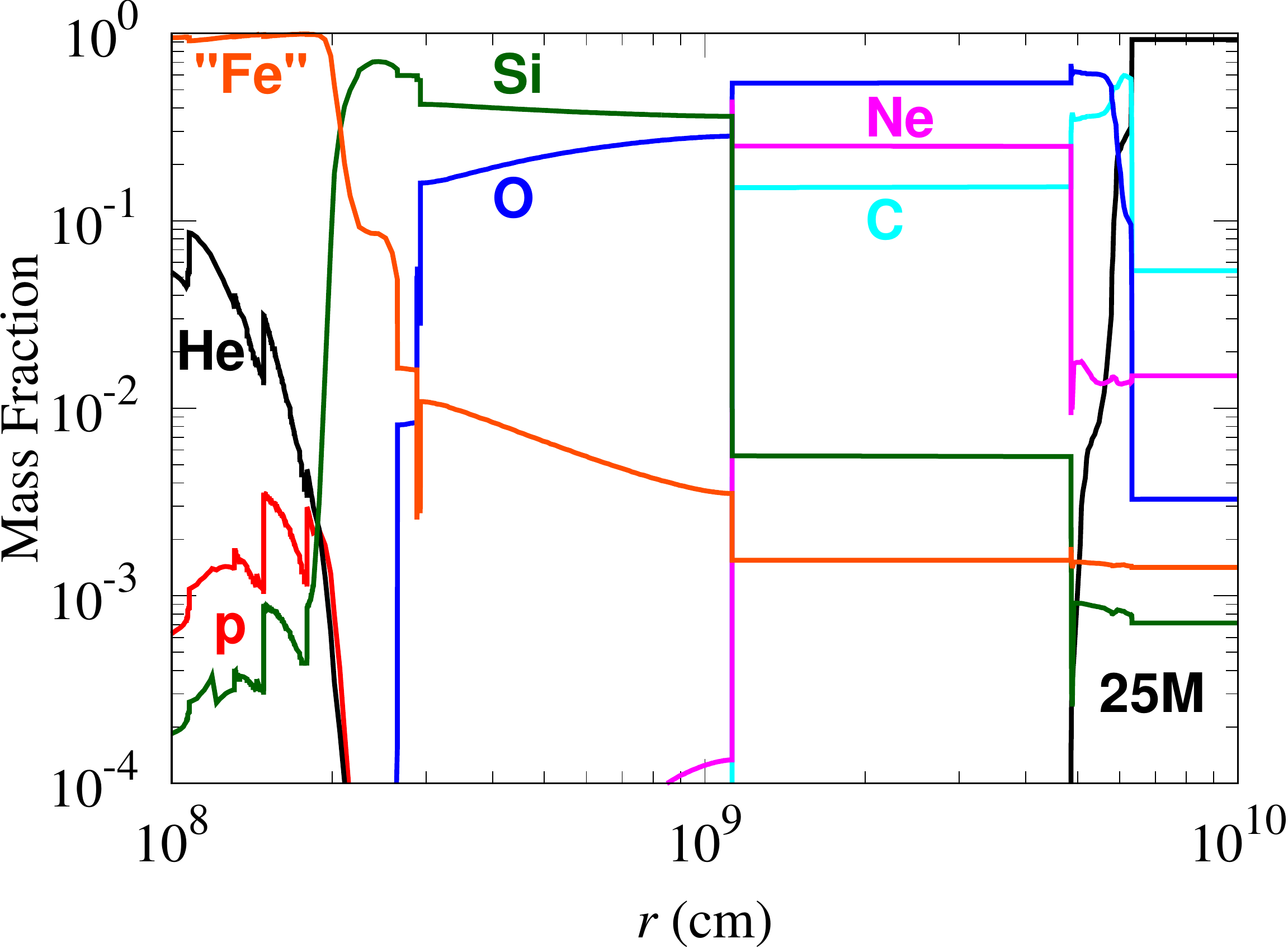}
\caption{Mass fraction distributions of models 13L$_{\rm A}$ (top panel) and 25M (bottom panel) as a function of radius at the last step.
Red, black, cyan, blue, magenta, green, and orange correspond to the mass fractions of p, He, C, O, Ne, Si, and iron-peak elements with $Z \ge 21$ denoted as ^^ ^^ Fe", respectively.
}
\label{fig:mfrad}
\end{figure}

For later convenience, we separate the SiO-rich layer into the ``Si/O" layer and the ``O/Si" layer.
The ``Si/O" layer has larger Si mass fraction than O mass fraction in the layer, i.e., $X(^{28}{\rm Si}) \ge X(^{16}{\rm O})$ and $X(^{16}{\rm O}) \ge 0.1$.
Whereas the ``O/Si" layer has the relation of 
$0.1 \le X(^{28}{\rm Si}) < X(^{16}{\rm O})$.
Then, we may classify these eleven models into two groups having different structures of the SiO-rich layer.
One group has an extended O/Si layer instead of the O/Ne layer above the Si/Fe layer.
The other group has a Si/O layer between the inner Si/Fe layer and the outer O/Ne layer.
The former group consists of models 13L$_{\rm A}$, 16M$_{\rm A}$, 18M$_{\rm A}$, 21M$_{\rm A}$, 23L$_{\rm A}$, 27L$_{\rm A}$, and 28M. 
The top panel of Figure \ref{fig:mfrad} shows the mass fraction distribution of model 13L$_{\rm A}$ as a function of radius.
The radius of the outer boundary of the O/Si layer is $\sim 3 \times 10^{9}$ cm.
This layer was originally formed as an O/Ne layer.
Neon burning has started after the core silicon burning phase, transforming neon into oxygen and silicon.
In case of models 18M$_{\rm A}$, 21M$_{\rm A}$ and 23L$_{\rm A}$, a thin Si/O layer exists between the Si/Fe layer and the O/Si layer with a width of less than 3$\times 10^8$ cm.

Models in the latter group have the layered structure, in which the innermost Fe core is surrounded by the Si/Fe, Si/O, and O/Ne layers.
Models 22L, 25M, 27M, and 28L$_{\rm A}$ comprise this group.
The bottom panel of Figure \ref{fig:mfrad} shows the mass fraction distribution of model 25M, an example of this latter group.
The model has the Si/O layer between $3 \times 10^8$ cm and $1.1 \times 10^9$ cm.
For other models in this group, the width of the Si/O layer is typically several times $10^8$ cm.
We show the mass fraction distributions as a function of radius for models other than models 13L$_{\rm A}$ and 25M in Figure \ref{fig:f18}.

\begin{deluxetable*}{lccccccc}[tbp]
\tablecaption{2D Model Properties and SiO-coexistence parameters. 
SiO-coexistence parameters $f_{M,{\rm SiO}}$ and $f_{V,{\rm SiO}}$ are obtained from the result of 1D evolution simulations.
$\langle Ma^2\rangle^{1/2}_{\rm max}$ represents the maximum convective Mach number obtained at a radius of 
$r$($\langle Ma^2\rangle^{1/2}_{\rm max}$) at the end of the 2D simulations. 
``Layer" represents the composition of the convective region.
$\ell_{\rm  max}$ represents the $\ell$ value at which $c_\ell^2$ has a peak (see equation (\ref{eq:cl2})).
$d_c/H_P$ represents the width of convective region normalized by the local scale height.
 These quantities are all estimated at the last step of the simulations. See the text for more detailed definition.
\label{tab:t2}}
\tablecolumns{8}
\tablewidth{0pt}
\tablehead{
\colhead{Model} &
\colhead{$f_{M,{\rm SiO}}$} &
\colhead{$f_{V,{\rm SiO}}$} &
\colhead{$\langle Ma^2\rangle^{1/2}_{\rm max}$} & 
\colhead{$r$($\langle Ma^2\rangle^{1/2}_{\rm max}$)} &
\colhead{Layer} &
\colhead{$\ell_{\rm max}$} &
\colhead{$d_{\rm c}/H_P$}\\
 &
 &
 &
 &
\colhead{(10$^{8}$ cm)} &
 &
 &
}
\startdata
\multicolumn{8}{c}{ Low-$Ma$ } \\
\hline
13L$_{\rm A}$ & 0.27--0.28 & 0.95--0.96 & 0.018 & 11.6 & O/Si & 12 & 6.22 \\
16M$_{\rm A}$ & 0.24--0.24 & 0.90--0.91 & 0.015 &  3.9 & O/Si &  4 & 3.20 \\
18M$_{\rm A}$ & 0.57--0.58 & 0.91--0.91 & 0.131 &  3.1 & Si/O & 14 & 1.06 \\
21M$_{\rm A}$ & 0.47--0.47 & 0.91--0.95 & 0.134 &  3.0 & Si/O &  8 & 4.42 \\
23L$_{\rm A}$ & 0.75--0.80 & 0.78--0.80 & 0.069 & 11.5 & O/Si &  4 & 5.20 \\
\hline
\multicolumn{8}{c}{ High-$Ma$ } \\
\hline
22L              & 0.57--0.61 & 0.77--0.82 & 0.108 &  9.4 & Si/O &  2 & 2.50 \\
25M              & 0.75--0.79 & 0.91--0.94 & 0.160 & 5.8 & Si/O &  3 & 3.65 \\
27L$_{\rm A}$ & 0.59--0.66 & 0.76--0.76 & 0.179 & 45.0 & O/Si &  2 & 4.56 \\
27M              & 0.58--0.65 & 0.37--0.40 & 0.134 &  4.7 & Si/O & 10 & 2.44 \\
28L$_{\rm A}$ & 0.60--0.68 & 0.37--0.42 & 0.117 &  5.3 & Si/O &  8 & 1.81 \\
28M              & 0.83--0.90 & 0.90--0.95 & 0.369 & 14.6 & O/Si &  2 & 4.08 \\
\enddata
\end{deluxetable*}

\subsection{2D Stellar Hydrodynamics Simulations} \label{2d}
\KT{In order to investigate the convective activities in a multi-dimensional space, we perform 2D hydrodynamics simulations of oxygen shell-burning.}
In the previous subsection, we picked up eleven models that show large 
SiO-coexistence parameters, 
$f_{M,{\rm SiO}}$ and/or $f_{V,{\rm SiO}}$.
\KT{Profiles of these models at $\sim$100 s before the end of the 1D calculations are taken as the initial conditions.
The 2D calculations are proceeded until the central temperature reaches $9\times10^9$ K, by which point the stars have started runaway collapse due to the gravitational instability.}

\KT{Following \citet{bernhard16_prog}, we evaluate the angle-averaged turbulent Mach number as an indicator of the turbulence strength,
\begin{equation}
    \langle Ma^2\rangle^{1/2}(r)
    = \left[ \frac{\int \rho \{ (v_r-\langle v_r \rangle)^2 + v_\theta^2 + v_\phi^2 \} d\Omega}{\int \rho c_s^2 d\Omega} \right]^{1/2},
\end{equation}
where $\rho$ is the density, $v_r$, $v_\theta$, and $v_\phi$ are the radial, tangential, and azimuthal velocities, $\langle v_r \rangle$ is the angle-averaged radial velocity, $c_s$ is the sound velocity, and $\Omega$ is the solid angle.
The maxima of $\langle Ma^2\rangle^{1/2}(r)$ evaluated at the end of the simulations $\langle Ma^2\rangle^{1/2}_{\rm max}$ are shown in the 4th column of Table \ref{tab:t2}, and
}
$r(\langle Ma^2 \rangle ^{1/2}_{\rm max})$ represents the radii where $\langle Ma^2\rangle^{1/2}_{\rm max}$ are obtained.

Based on the Mach number, we divide our 2D models into two groups, either showing ``low-$Ma$" or ``high-$Ma$". 
The criterion of high-$Ma$ is set as $\langle Ma^2 \rangle^{1/2}_{\rm max} \geq 0.1$, because the turbulence with such a high Mach number potentially fosters the perturbation-aided explosion \citep{bernhard15,bernhard16_prog}.
It is noted that models 18\MOVC and 21\MOVC are exceptionally classified into low-$Ma$ in spite of their ``large" Mach numbers. We  discuss this later in this Section.
The column ``Layer" in Table \ref{tab:t2} represents the dominant chemical composition in the convective layer, i.e., the Si/O layer or the O/Si layer. 
We also show the mass fraction distribution of models 13L$_{\rm A}$ and 25M in Figure \ref{fig:mfrad}.
For other models, the mass fraction distributions are shown in Figure \ref{fig:f18} in Appendix.
See the last part of the previous subsection for the definition of the layer.

\begin{figure*}[htb]
\centering
\plottwo{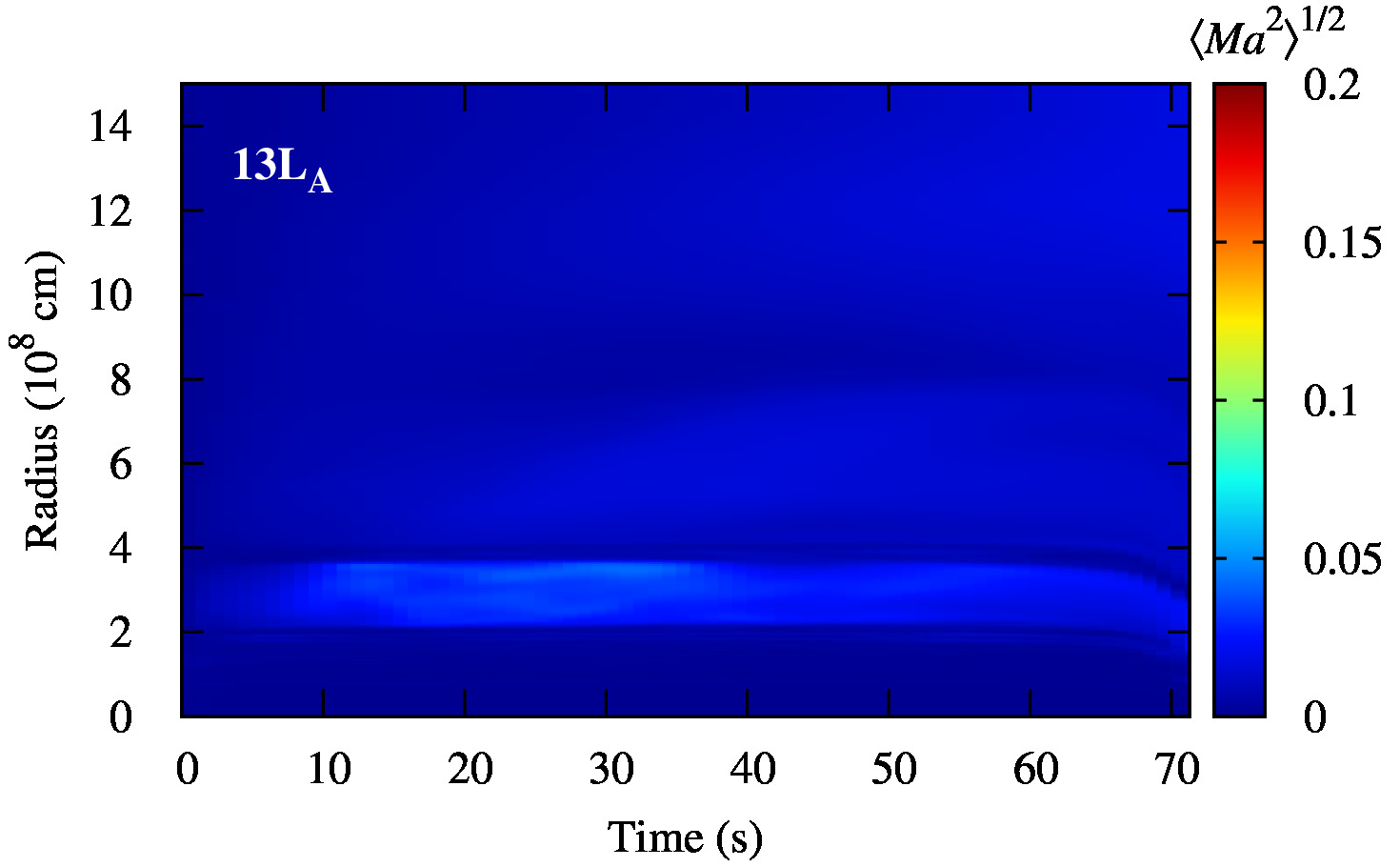}{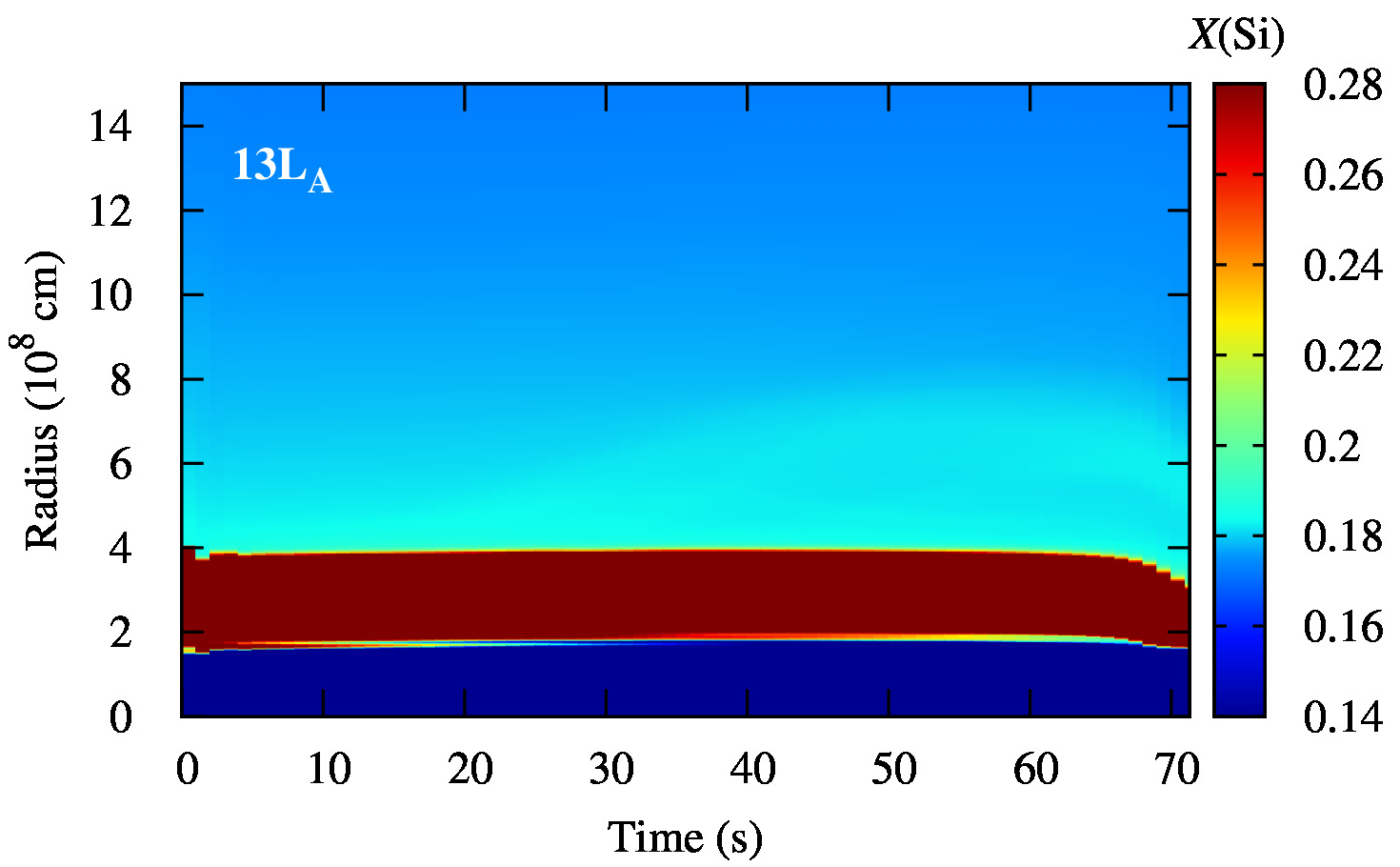}
\plottwo{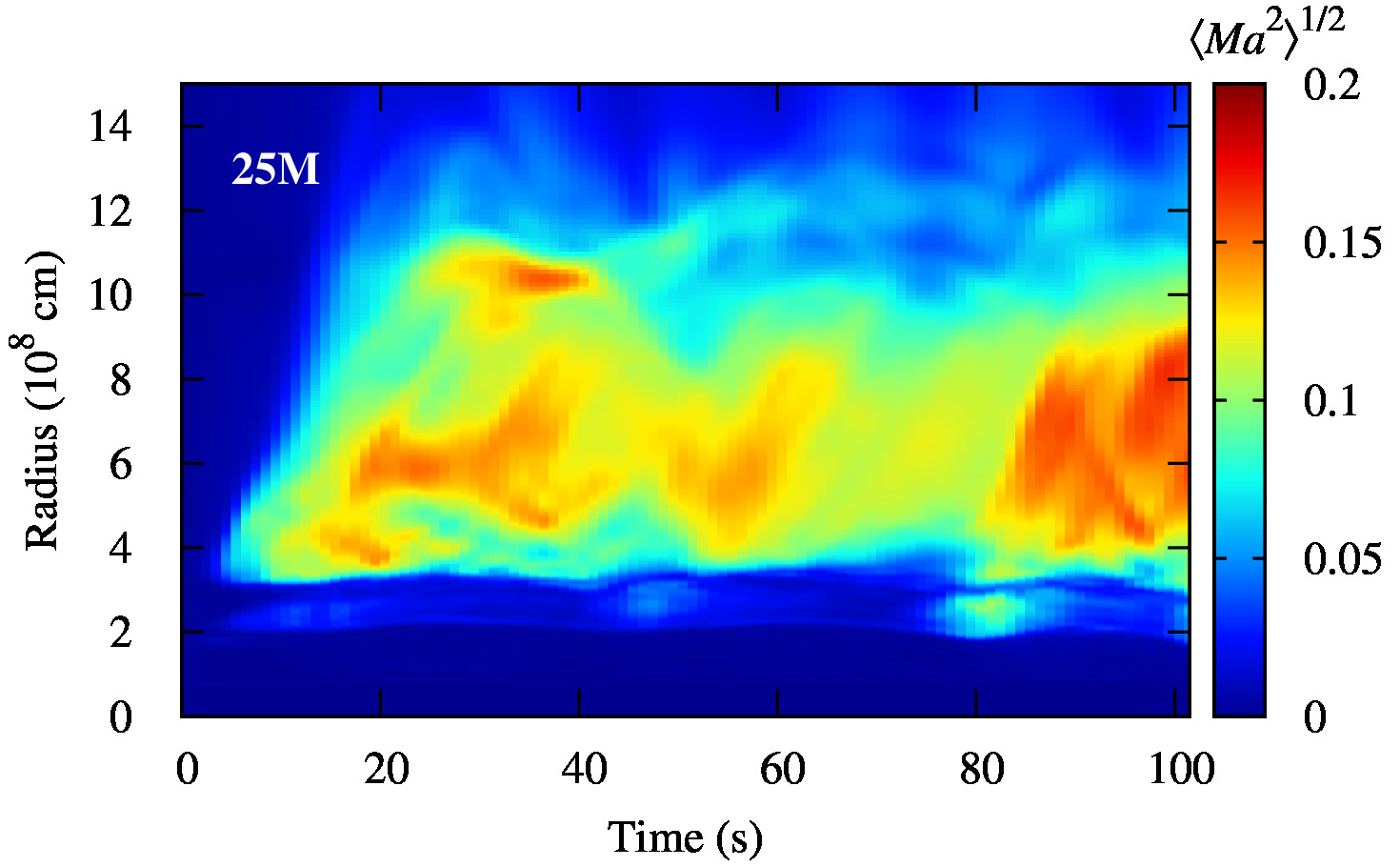}{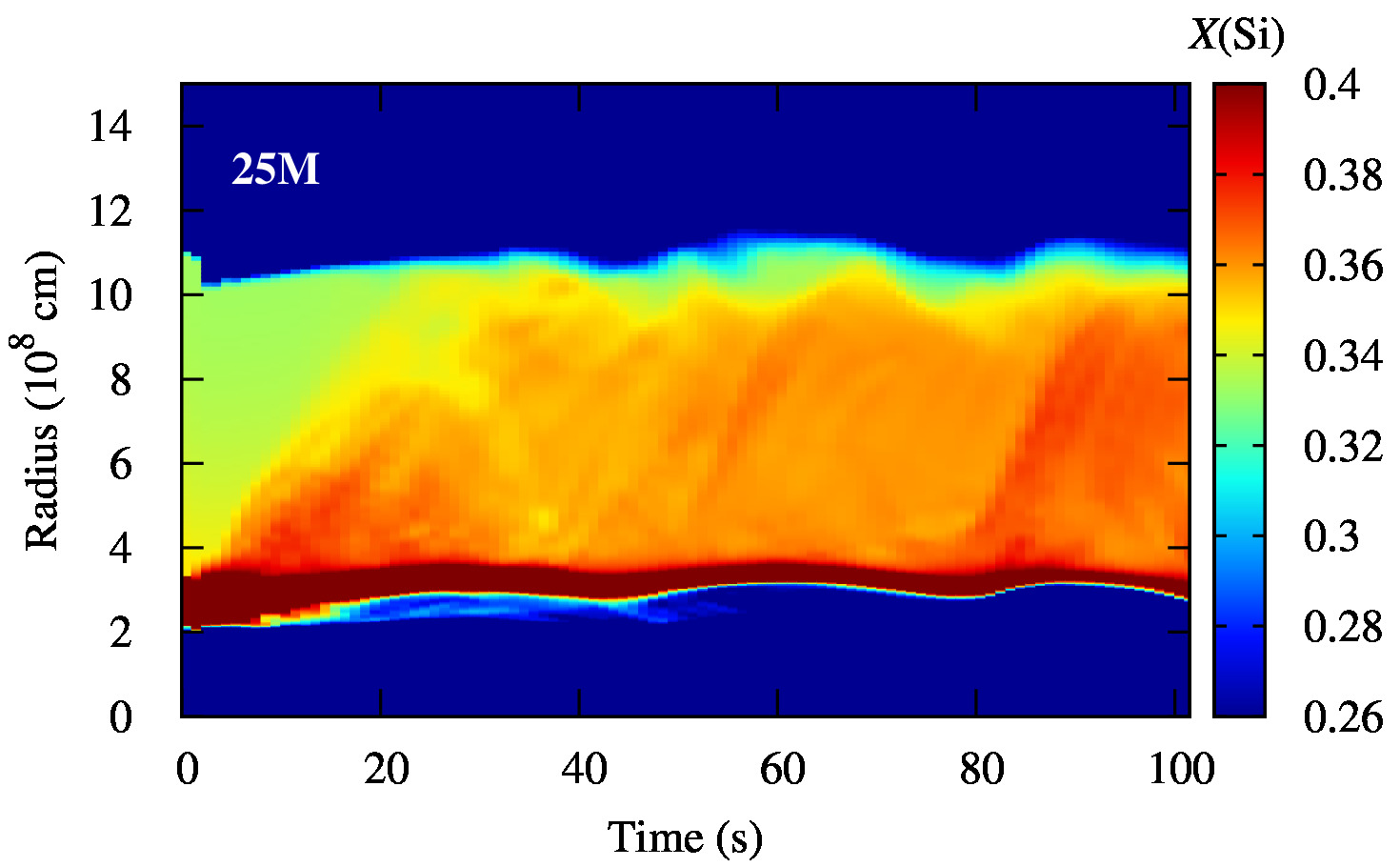}
\plottwo{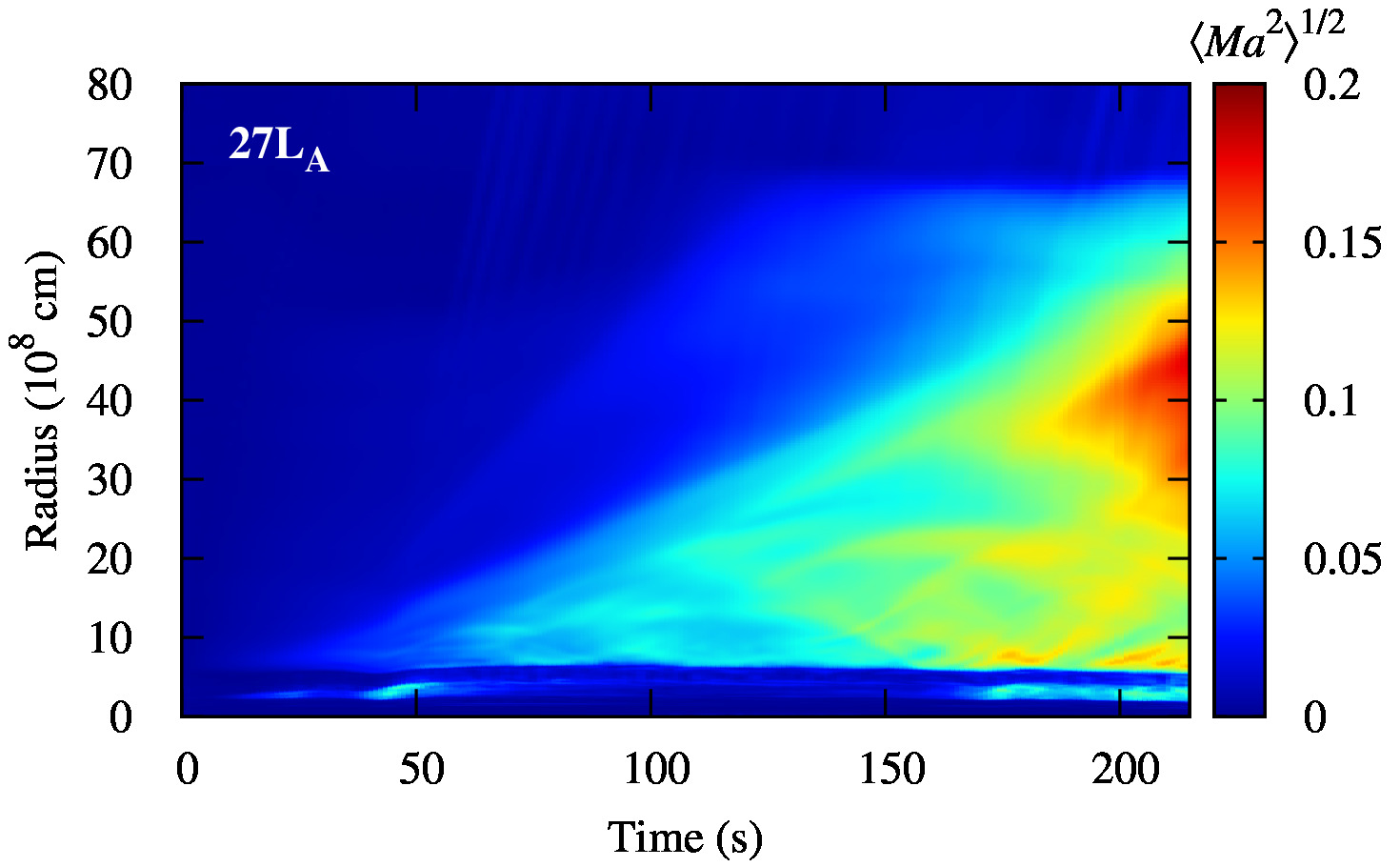}{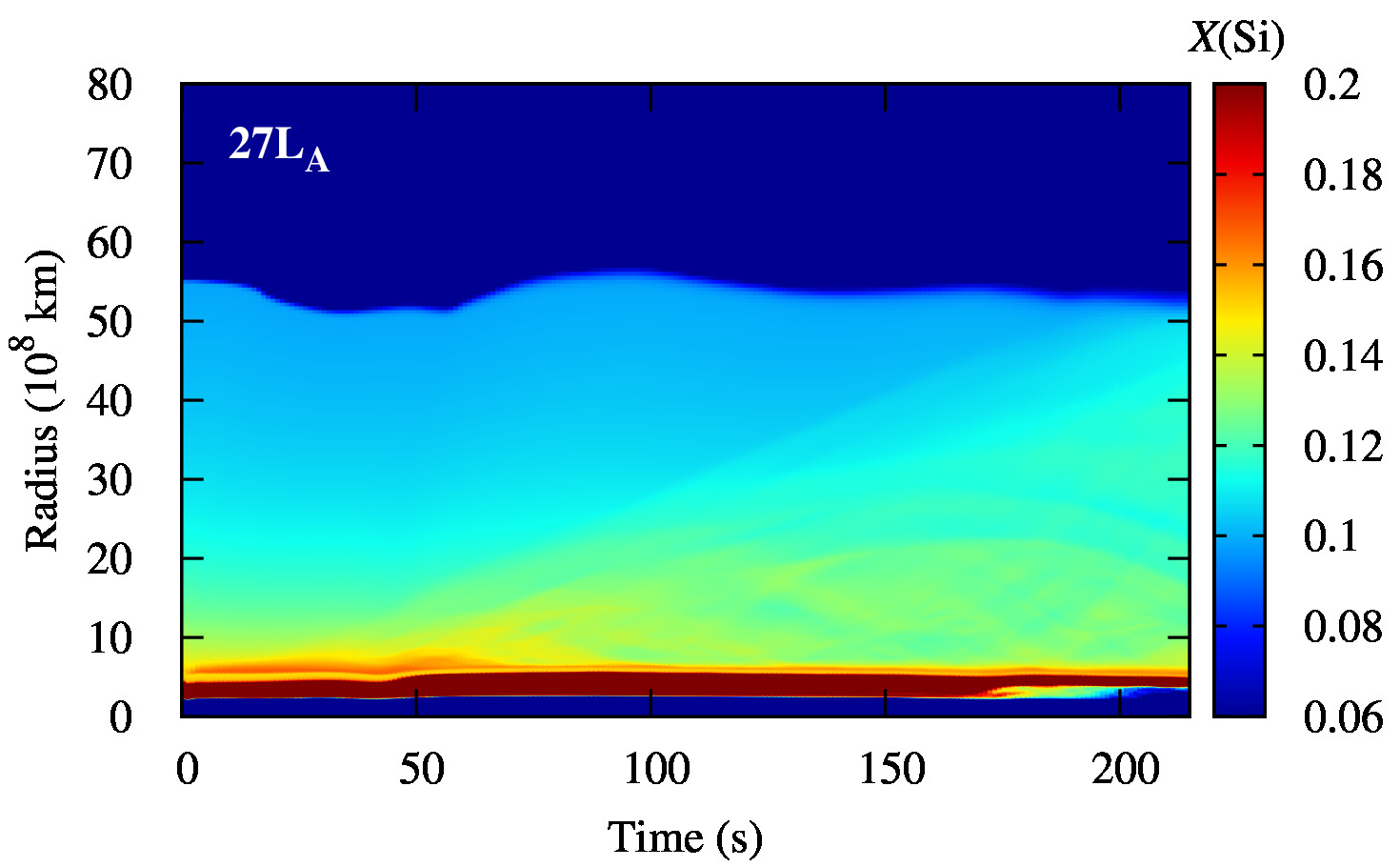}
\caption{The time and radial distributions of the  angular-averaged convective Mach number $\langle Ma^2 \rangle^{1/2}$ (left panels) and the $^{28}$Si mass fraction (right panels).
Top, middle, and bottom panels are for models 13L$_{\rm A}$, 25M, and 27L$_{\rm A}$, respectively.}
\label{fig:vrmftime}
\end{figure*}

\paragraph{Time evolution of convective motion}

In Figure \ref{fig:vrmftime}, we show the time evolution of the turbulent Mach number and the Si mass fraction for representative models 
from low-$Ma$ (13L$_{\rm A}$, top) and high-$Ma$ (25M, middle, and 27L$_{\rm A}$, bottom).
The color visualizes the angle-averaged turbulent  Mach number $\langle Ma^2 \rangle^{1/2}$ (left) and the $^{28}$Si mass fraction $X$(Si) (right).
Note that the outer radial frame of the panels for model 27L$_{\rm A}$ is set to be $8 \times 10^{9}$ cm, in order to show how the outer edge of the convective region keeps moving outward and reaches this radius at the end.

\KT{The model 13L$_{\rm A}$ has no Si/O layer. This star has an Fe core at the central region of $R \la 2 \times 10^8$ cm, which is surrounded by the convective Si/Fe layer ($R \sim$2--4 $\times 10^{8}$ cm) and the convective O/Si layer ($R \ga$4 $\times 10^{8}$ cm). The Si mass fraction at Si/Fe layer is $\sim$0.5 (see the top panel of Figure 
\ref{fig:mfrad}).
As shown in the top right panel of Figure \ref{fig:vrmftime}, the Si mass fraction is small compared to that of 25M model that is shown in the middle right panel of Figure \ref{fig:vrmftime}.
Reflecting this structure, the turbulent Mach number is lower than 0.1 in the inner Si/Fe layer and in the outer O/Si layer throughout the simulation (see the top left panel of Figure \ref{fig:vrmftime}).
However, the oxygen burning slightly enhances the  $^{28}$Si mass fraction in the base region of the O/Si layer of $\sim$4--8 $\times10^8$ cm.
Note that $\langle Ma^2\rangle^{1/2}_{\rm max}$ in Table \ref{tab:t2} is estimated at the end of the simulations, and it does not refer to the peak seen at $\sim 4 \times 10^8$ cm at $\sim$30 sec.}

Readers may be confused by the models of
18M$_{\rm A}$ and 21M$_{\rm A}$ 
since they have Si/O layer in Table \ref{tab:t2} but they are classified into low-$Ma$ group.
\KT{
Actually  models 18M$_{\rm A}$ and 21M$_{\rm A}$ have a thin Si/O layer and show $\langle Ma^2 \rangle^{1/2}$ $\sim 0.13$ in the Si/O layer, but only after the last $\sim$10 s of the simulations.
This is because the turbulence is triggered by the gravitational contraction, which amplifies the temperature at the bottom of the Si/O layer enhancing the oxygen burning rate.
The turbulence powered by the gravitational contraction has too short time to form an extended convective region, which is contrasting to the shell-convection powered by a hydrostatic burning.
This is why we have selected these models as members of  low-$Ma$.
}

Models 22L, 25M, 27L$_{\rm A}$, 27M, 28L$_{\rm A}$, and 28M are categorized into models with high-$Ma$.
Convective motion with such a strong turbulence develops by oxygen burning in the Si/O layer in these models.
We pick out two models in which it is easy to explain typical dynamics of the convection.

\KT{
Model 25M consists of the central Fe core ($R \la 2 \times 10^8$ cm), the Si/Fe layer ($R \sim 2$--$3 \times 10^8$ cm), the Si/O layer ($R \sim 3$--$10 \times 10^8$ cm), and the O/Ne layer ($R \ga 10 \times 10^8$ cm).
It is noteworthy that, despite the Si/Fe layer seems to have a homogeneous chemical composition (see the lower panel of Figure \ref{fig:mfrad}), the outer part of $R \sim 2.5$--$3 \times 10^8$ cm is actually composed of small amount of oxygen with X($^{16}$O)$ < 0.01$.
The oxygen-free region of $R \sim 2$--$2.5 \times 10^8$ cm becomes convective within a short timescale of $\sim$10 s from the start of the simulation, though the shell-convection is not extended further.
At the bottom of the outer Si/O layer, the hydrostatic oxygen shell-burning takes place.
In this case, the nuclear burning drives high turbulent velocity with $\langle Ma^2 \rangle^{1/2} > 0.1$, which is sustained for 20--110 s (see the middle left panel of Figure 4).
Accordingly, turbulent mixing homogenizes the $^{28}$Si mass fraction in the region of $ R =$ 3--10$\times 10^{8}$ cm.
Furthermore, oxygen burning also takes place in the oxygen-containing outer region of the Si/Fe layer.
In spite of the large mean molecular weight, the heating due to the oxygen burning is strong enough to lift the silicon-rich material up into the surrounding Si/O layer.
As a result, the silicon mass fraction in the Si/O layer significantly enhances.
This silicon enhancement repetitively takes place at $\sim$50 and 80 s, which accompanies the enhancement of the convective Mach number as well.
It seems that the repetitive mixing follows the oscillation of the outer edge of the Si/Fe convection.
This will be because, with the small oxygen mass fraction, the temperature fluctuation originally caused by the oscillation is enhanced by the O burning, resulting in a large density fluctuation that triggers convection in the Si/O layer.
Indeed, the temperature rise at the bottom of the Si/O layer where the O mass fraction is $\sim$0.08--0.1 reaches 8 \% at the maximum, which is much higher than the temperature change solely due to the oscillation, less than $\sim$1 \%, measured in outer region of the Si/F and Si/O layers.
}

\KT{In model 27L$_{\rm A}$, the turbulent activity in the O/Si layer starts to increase at $\sim$45 s (see the bottom left panel of Figure 4).
During the simulation time of $\gtrsim$ 200 s, the high Mach number region extends outward, finally reaching $\sim 6 \times 10^8$ cm, which roughly corresponds to the composition jump between the CO-rich and the He-rich layers.
At the same time, the turbulent Mach number grows with time in almost whole region in the convective layer.
As a consequence, the turbulent Mach number exceeds $\sim$0.15 in the wide outer region of $R \sim 30$--$50 \times 10^8$ cm at the end of the simulation.
Initially, the $^{28}$Si mass fraction decreases with the radius in the O/Si layer (see the mass fraction distribution of model 27L$_{\rm A}$ in Figure \ref{fig:f18}).
The convection powered by the oxygen burning mixes material in the slightly silicon-enriched region, which is initially located below $\sim 8 \times 10^{8}$ cm, into the outer slightly silicon-poor region after $\sim$70 s.
However, the convective mixing in the 2D simulation is still not efficient enough to achieve the homogeneous chemical distribution.
This is due to the limitation of the calculation time, because the total time of this simulation covers only about one convection-turnover time.}
Model 28M shows similar convective properties to the model 27L$_{\rm A}$.

\paragraph{High/low Mach number and the chemical distribution}

Models 13L$_{\rm A}$, 16M$_{\rm A}$, 18M$_{\rm A}$, 21M$_{A}$, and 23L$_{A}$ belong to low-$Ma$, namely, they do not show strong turbulence in their convective regions during the simulations.
These low-$Ma$ models have characteristic chemical composition profiles.
A main characteristic is no or thin Si/O layer.
Models 13L$_{\rm A}$, 16M$_{\rm A}$ do not have the Si/O layer and have an extended O/Si layer on the Si layer (see top panel of Fig. \ref{fig:mfrad} and top left panel of Fig. \ref{fig:f18}, respectively).
Model 23L$_{\rm A}$ does not have the Si/O layer at the beginning of the 2D simulation.
The turbulent Mach number of these models in the O/Si layer is low.
Models 18M$_{\rm A}$ and 21M$_{\rm A}$ have a Si/O layer at the beginning of the 2D calculations but the width is less than $\sim 1 \times 10^8$ cm.
Although the turbulent Mach number exceeds 0.1 at the bottom of the Si/O layer for a few seconds before the termination of the simulations, the turbulence in this layer does not develop before this time.

Models 22L, 25M, 27L$_{\rm A}$, 27M, 28L$_{\rm A}$, and 28M belong to high-$Ma$.
The main characteristic of the chemical profiles is an extended Si/O layer.
Models 22L, 25M, and 28M have a Si/O layer with the width of $\sim 8 \times 10^8$ cm.
Note that the Si/O layer of model 28M has merged to the O/Ne layer before the end of the simulation.
Models 27M and 28L$_{\rm A}$ also have a Si/O layer, although their width is thinner than the three models.

Model 27L$_{\rm A}$ is an exception.
This model does not have a Si/O layer but the chemical composition profile is similar to the last step of model 28M.

We briefly discuss the relation to the SiO-coexistence parameters.
All models in high-$Ma$ are selected using a large $f_{M,{\rm SiO}}$ value.
Although models 25M and 28M are also selected using a large $f_{V,{\rm SiO}}$ value, they also have a large $f_{M,{\rm SiO}}$.
The models having a large $f_{M,{\rm SiO}}$ rather than  a large $f_{V,{\rm SiO}}$ are associated with high $Ma$.

\begin{figure*}[htbp]
\plottwo{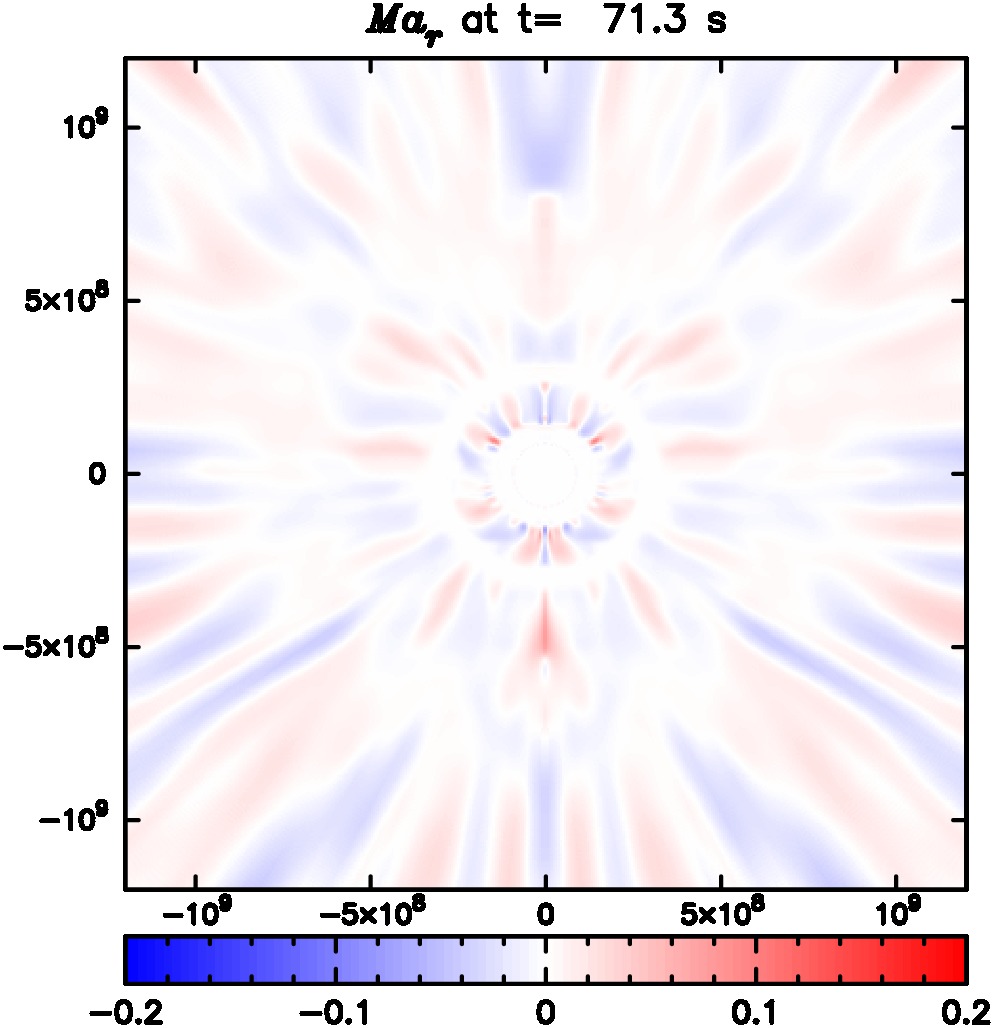}{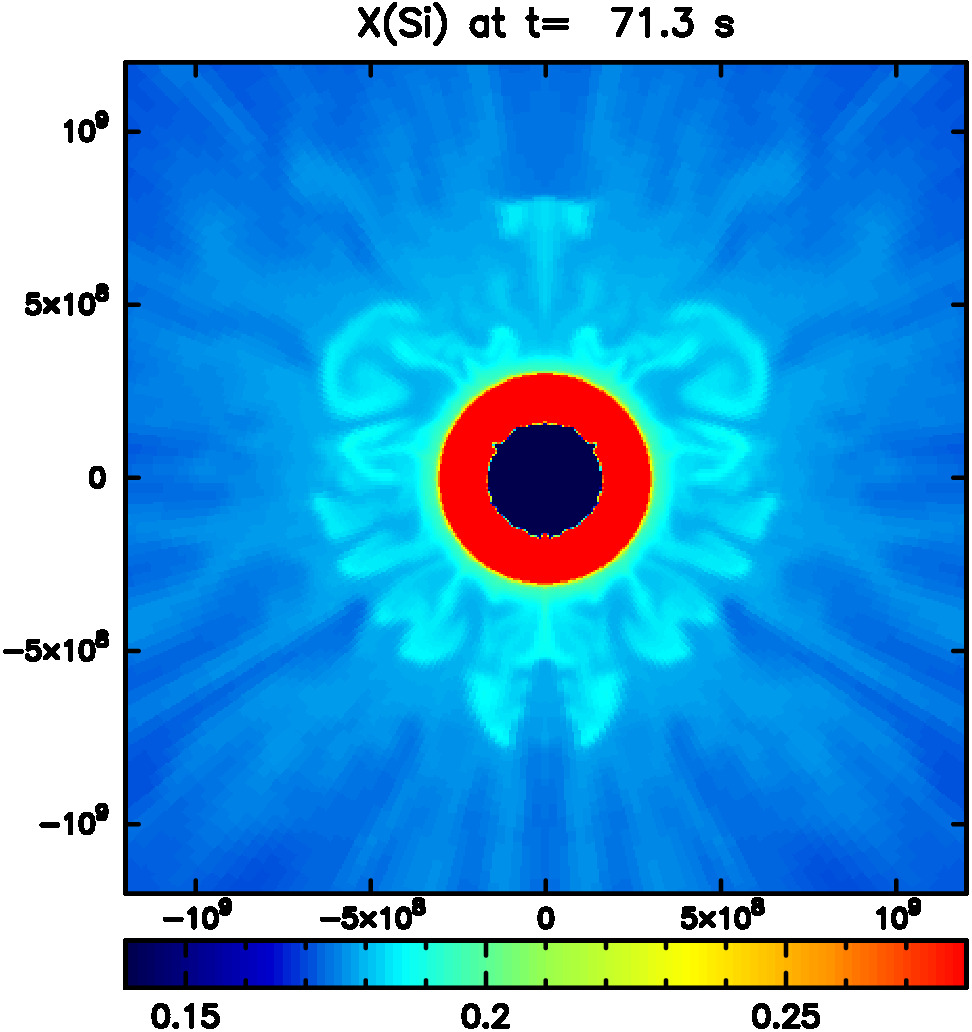}
\plottwo{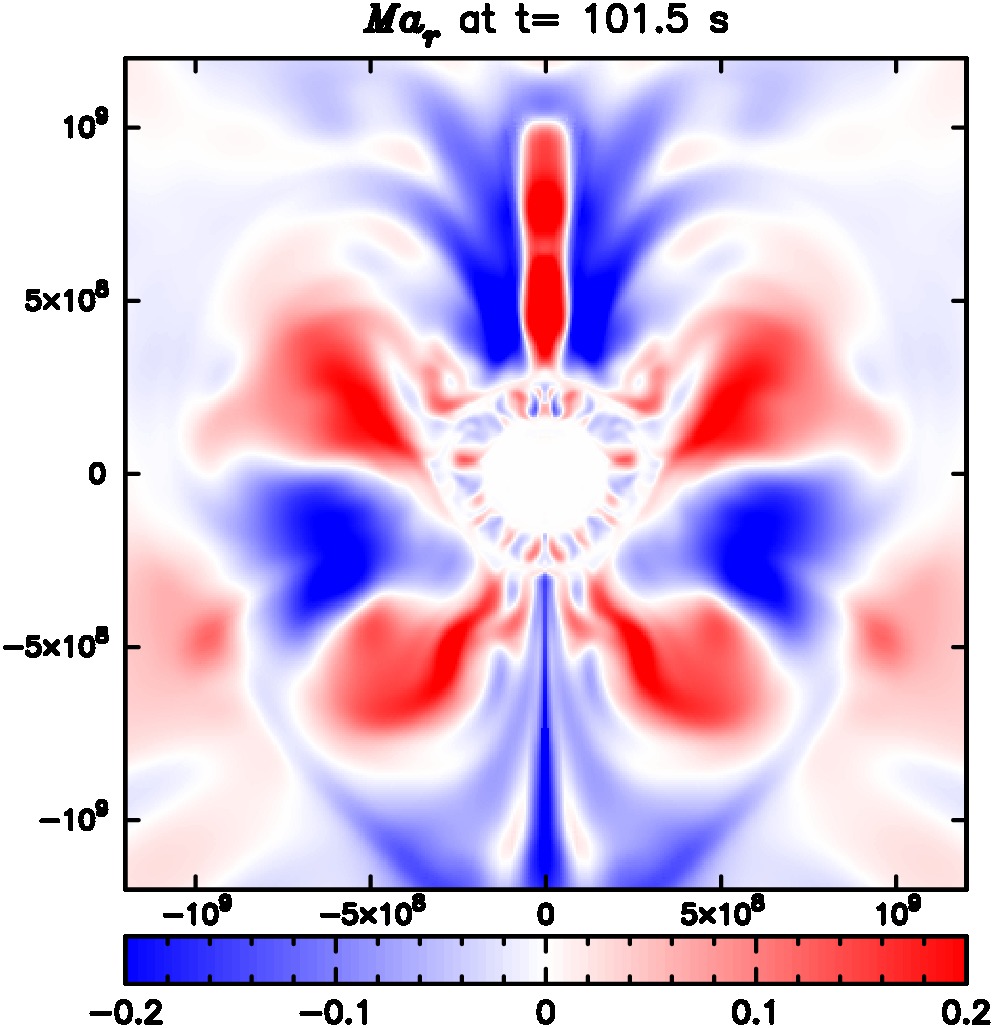}{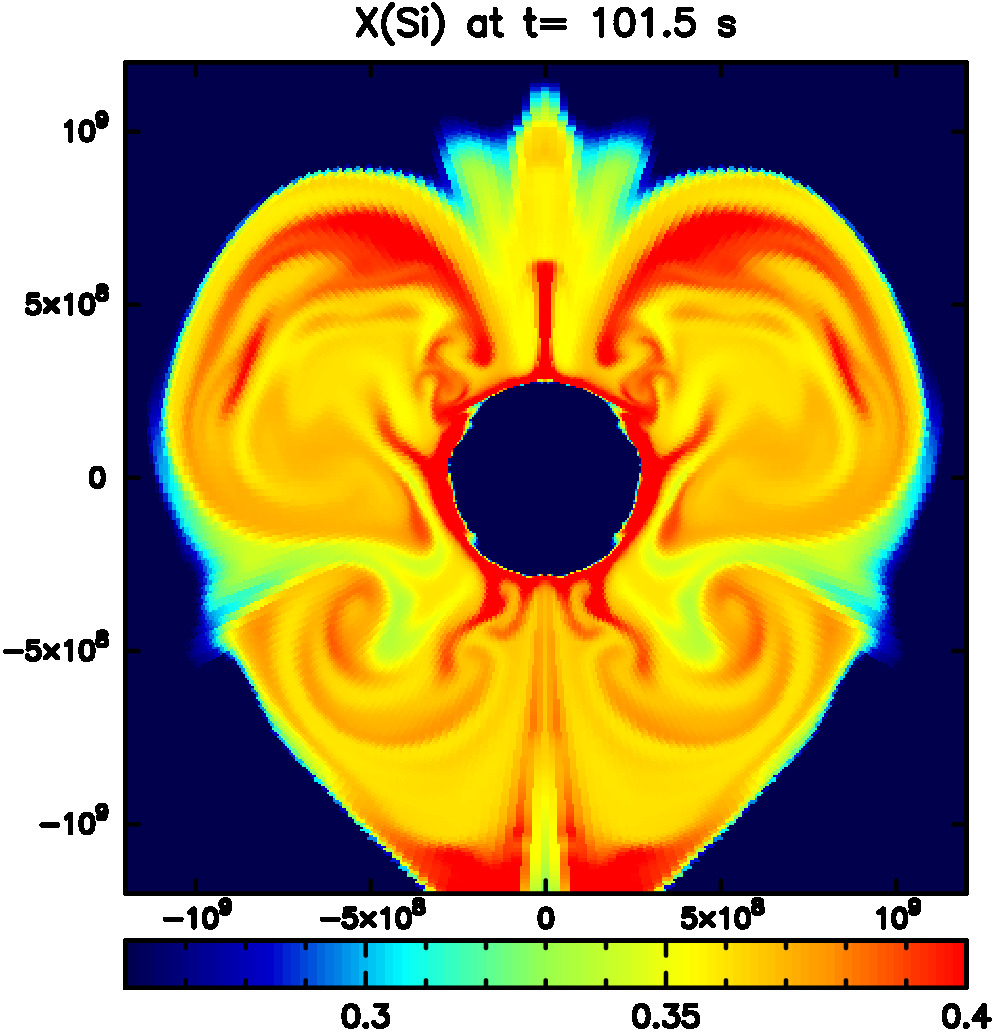}
\plottwo{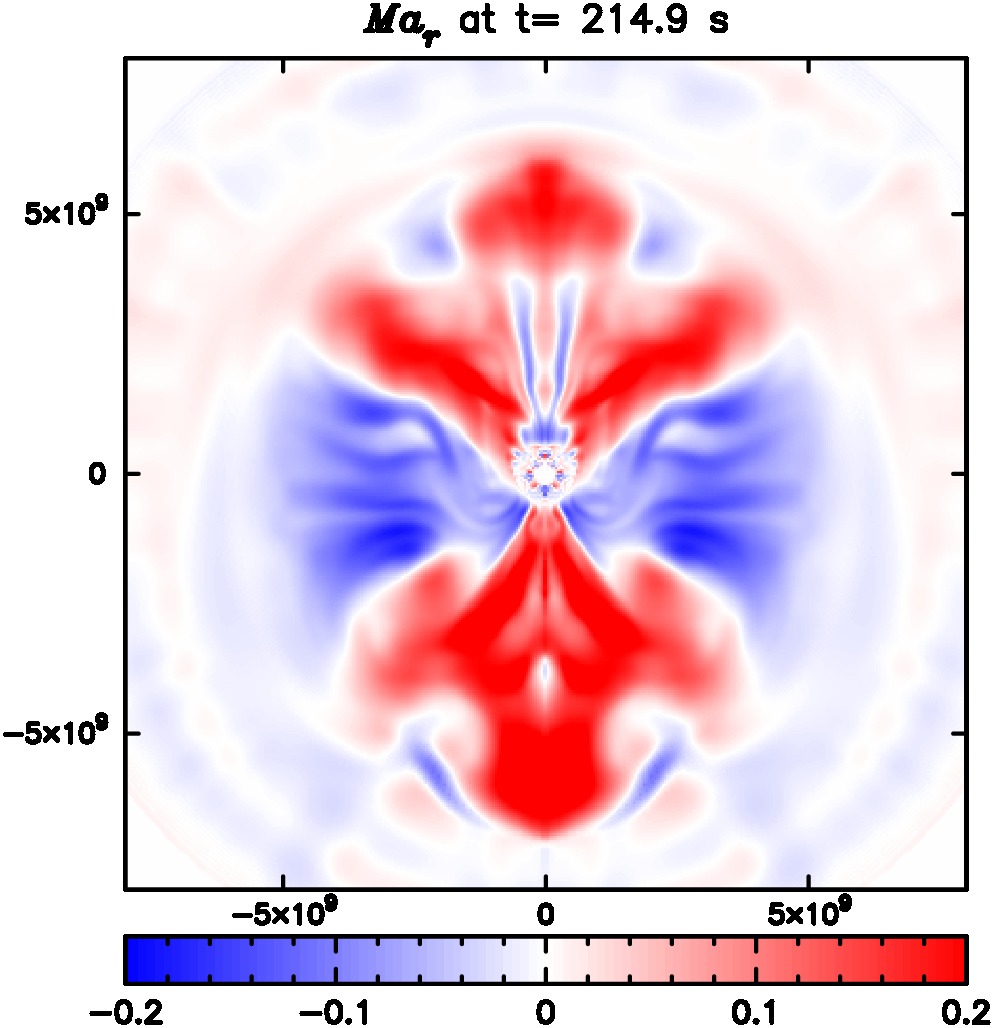}{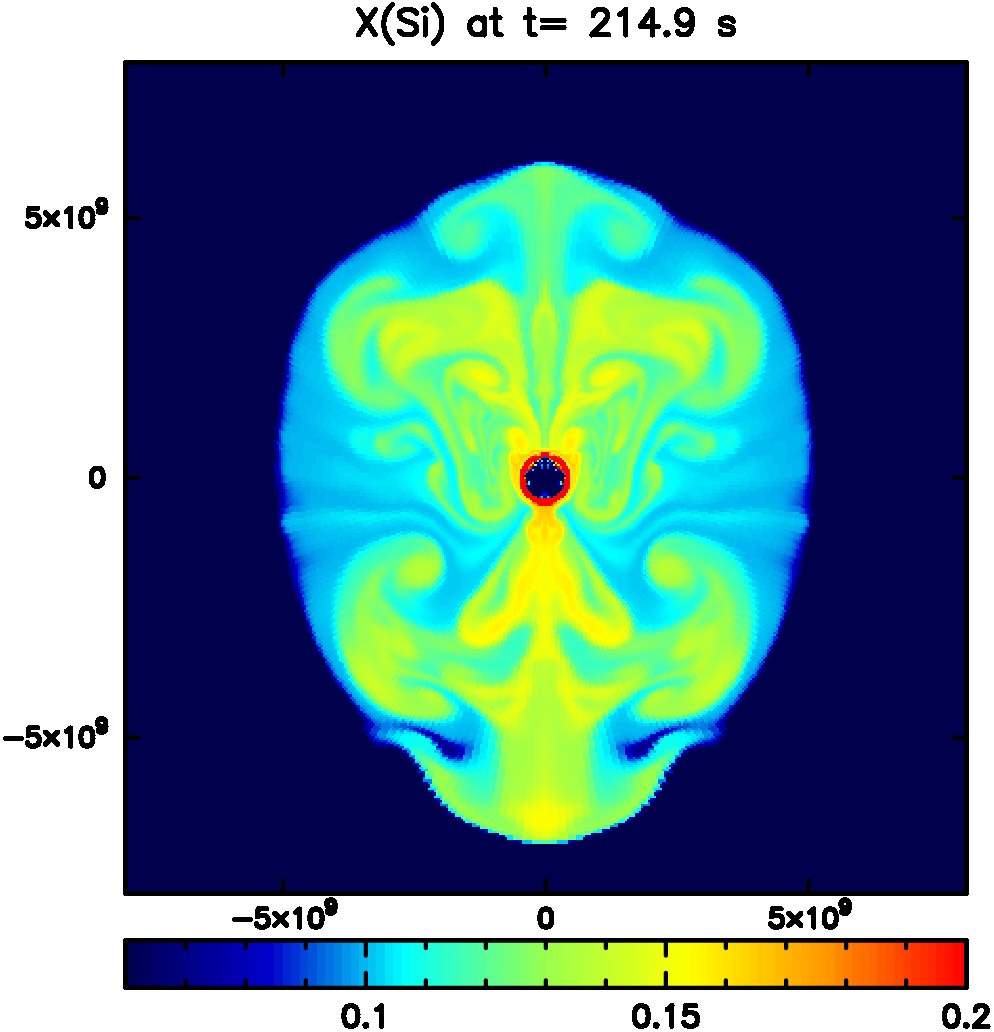}
\caption{2D distributions of the turbulent Mach number of the radial velocity $Ma_r$ (left panels) and $^{28}$Si mass fraction (right panels).
Top, middle, and bottom panels correspond to models 13L$_{\rm A}$, 25M, and 27L$_{\rm A}$, respectively.
}
\label{fig:vrmfsi2d}
\end{figure*}

\paragraph{2D distribution}
\KT{Figure \ref{fig:vrmfsi2d} shows the 2D distributions of the radial turbulent Mach number, $Ma_r$ $= v_r/c_s - \langle v_r \rangle/\langle c_s \rangle$, and the $^{28}$Si mass fraction taken at the last step of the simulations for models 13L$_{\rm A}$, 25M and 27L$_{\rm A}$.}
\KT{The turbulent Mach number of model 13L$_{\rm A}$ (top panels in Figure 5) develops only within the level of $Ma_r$ $\sim$0.01.
The spherical boundary is clearly observed at $r \sim 3 \times 10^{8}$ cm, where the turbulent Mach number becomes almost zero.
Inside the boundary, convection is developed in the Si/Fe layer.
A more extended but even weaker turbulent motion is also developed in the O/Si layer above the boundary.
The outer boundary of the O/Si convection may be defined at $\sim 6 \times 10^{8}$ cm, but there is only a diffuse $Ma_r$ $\sim 0$ region in this case.
Surrounding the inner boundary between the Si/Fe and the O/Si layers, a thin and nearly spherical band with X($^{28}$Si) $\sim$ 0.2 exists.
As a result of the low-velocity turbulence in the O/Si layer, this silicon-rich material is slowly mixed into the inner region of the O/Si layer at $R \la 6 \times 10^{8}$ cm.
}

\KT{Model 25M (middle panels in Figure 5) develops a convective motion with high turbulent Mach number in the Si/O layer ranging from $R \sim 3 \times 10^{8}$ cm to $R \sim 10 \times 10^{8}$ cm.
At the end of the simulation, outflows stream in 3 directions; the northern pole direction, $\sim$45$^\circ$ from the polar axis, and $\sim$135$^\circ$ from the polar axis, and inflows are sandwiched by the outflows.
These convective flows have the turbulent Mach number larger than $\sim$0.1.
The $^{28}$Si mass fraction is roughly homogenized inside the convective region, having the value of 0.3--0.4, though some fluctuations are observed especially near the outer boundary.}

\KT{Model 27L$_{\rm A}$ (bottom panels in Figure 5) has an extended O/Si layer distributed from $R \sim 5 \times 10^{8}$ to $50 \times 10^{8}$ cm.
A large-scale convective motion is developed in this layer; a broad conical outflow with the opening angle of $\sim$45$^\circ$ is formed in the both polar regions, and between them, a thick inflow is formed around the equatorial plane.}
The convective Mach number reaches $\sim$0.12.
\KT{The large scale outflow mixes the silicon rich material into the O/Si layer.
The silicon mass fraction in the most part of this layer is initially $\sim$0.1, while the outflow has a higher fraction of $\sim$0.16.}

\begin{figure}[thbp]
\centering
\includegraphics[width=\linewidth]{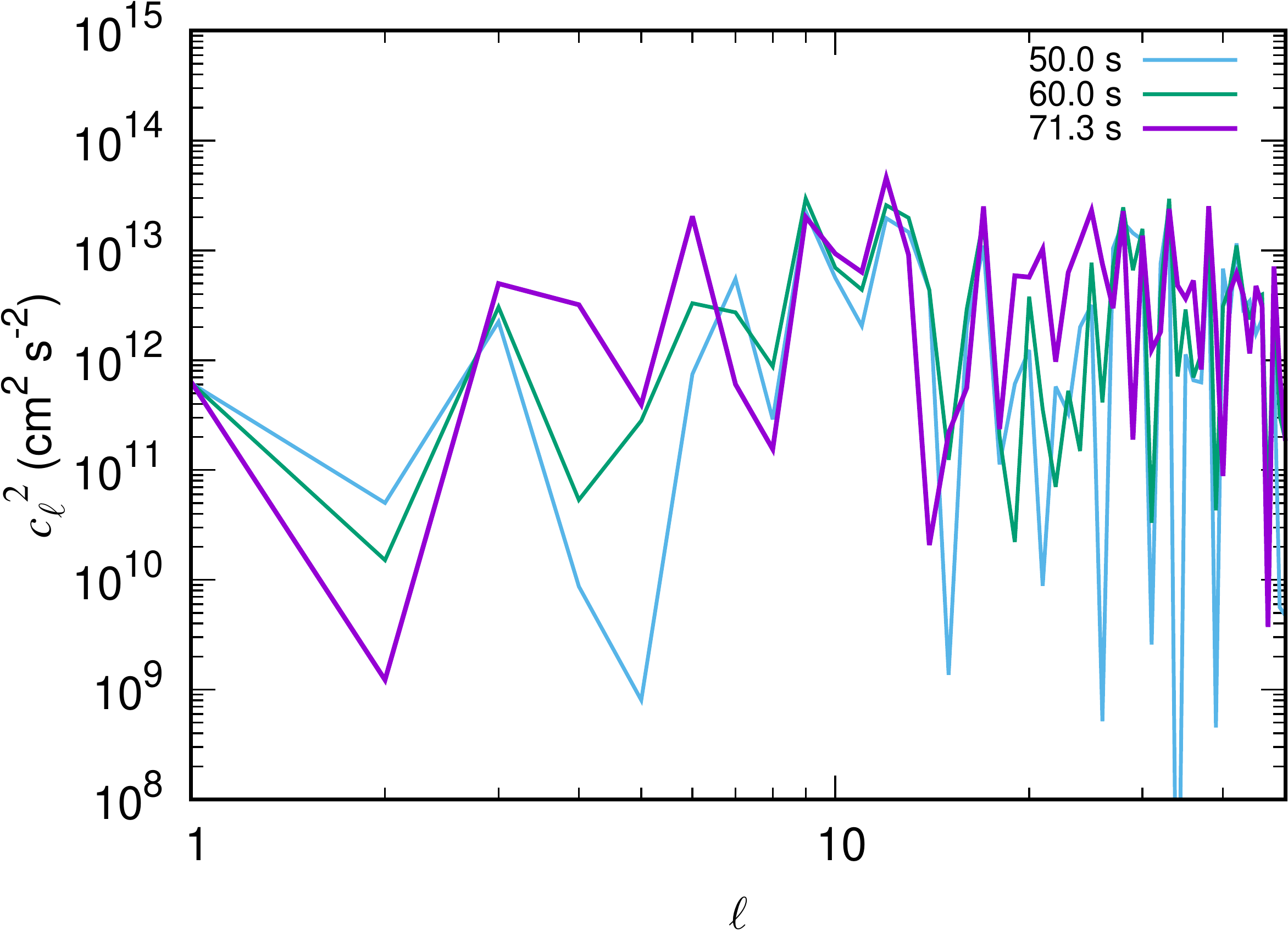}\\
\includegraphics[width=\linewidth]{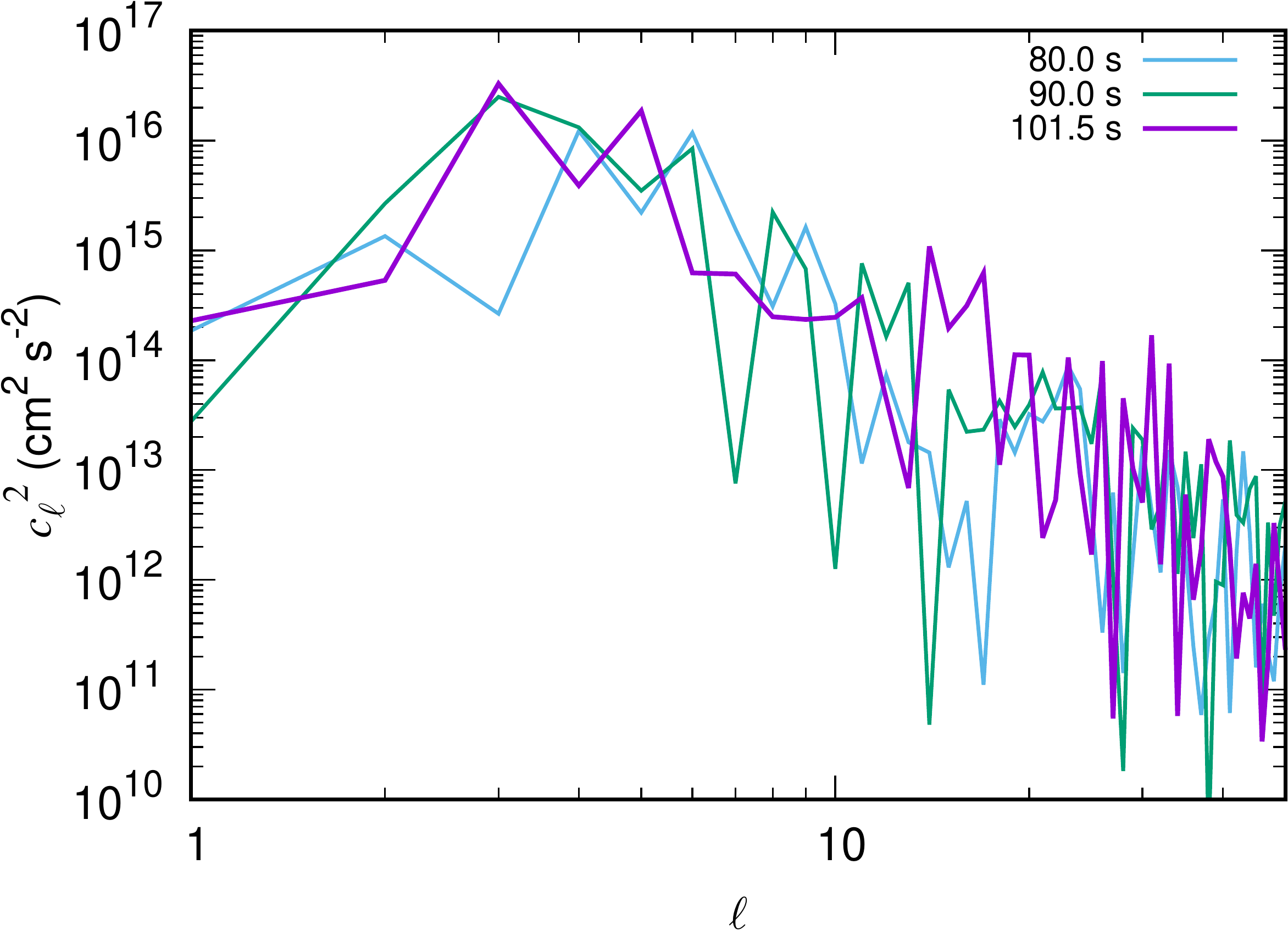}\\
\includegraphics[width=\linewidth]{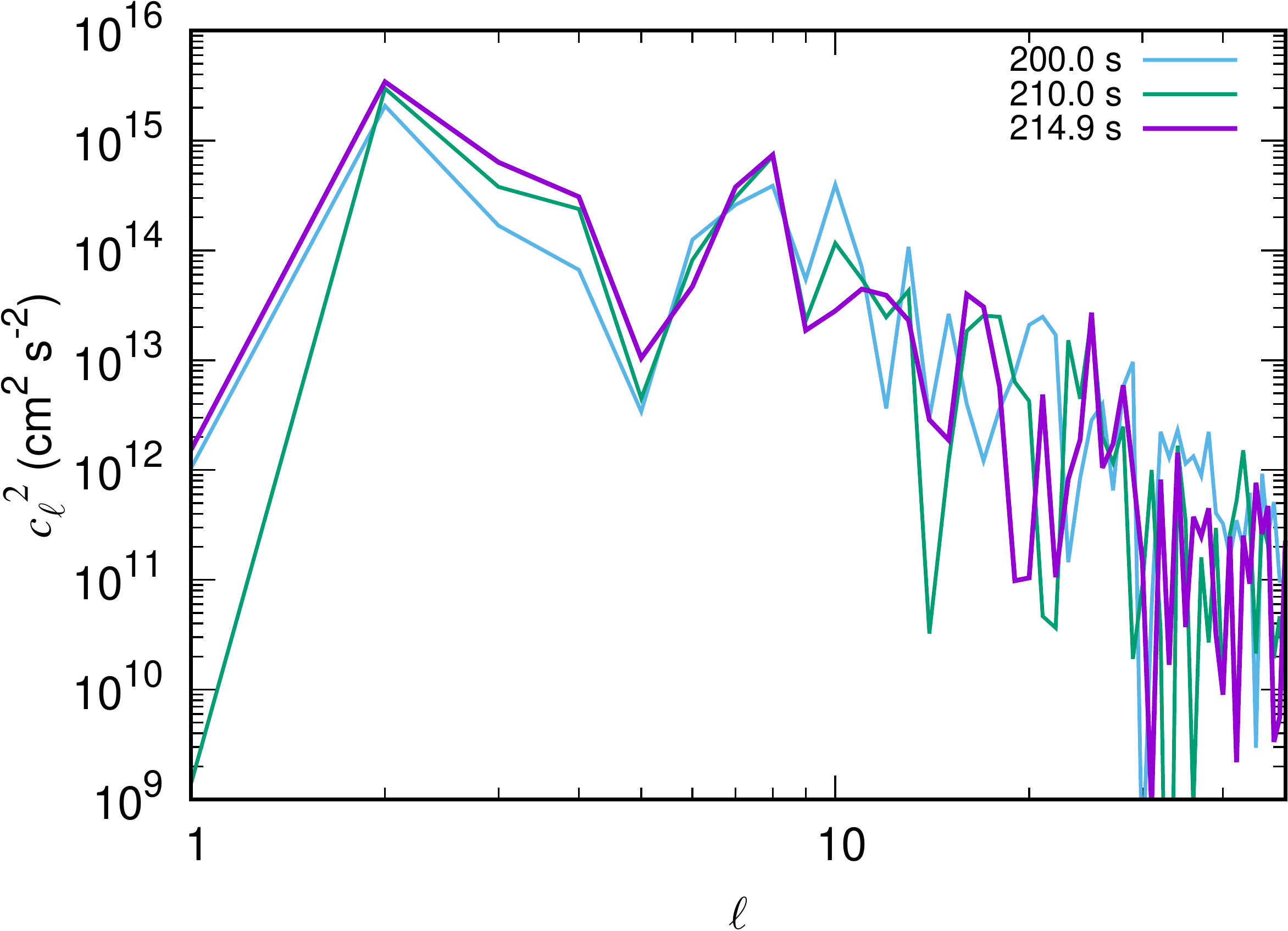}
\caption{Power spectra of the radial velocity dispersion $c_\ell^2$ at $r(\langle Ma^2 \rangle^{1/2})$ for models 13L$_{\rm A}$ (top panel), 25M (middle panel), and 27L$_{\rm A}$ (bottom panel) in the 2D simulations.}
\label{fig:power2d}
\end{figure}

\paragraph{Width of convective region}
We briefly discuss the width of the convective region divided by the local scale height.
We define a convective region as a region having $\langle Ma^2 \rangle^{1/2} > 1/3 \times \langle Ma^2 \rangle^{1/2}_{\rm max}$ including $r(\langle Ma^2 \rangle^{1/2}_{\rm max})$.
We determined the factor 1/3 to avoid including the neighboring convective region because we obtain small turbulent Mach number even at the convection boundary.
The width of the convective region is compared with the pressure scale height $H_P$ at $r(\langle Ma^2 \rangle^{1/2}_{\rm max})$.
This results in 1.8--4.6 for high-$Ma$ models, which are listed in Table 2.

We should note that the above definition does not specify the width of the convective region correctly for low-$Ma$ models.
In models 16M$_{\rm A}$ and 18M$_{\rm A}$, the specified region contains the Si layer inside the O/Si and Si/O layer, respectively.
In model 21M$_{\rm A}$, the calculated region contains a part of the Fe core, the Si and Si/O layer, and a part of the O/Si layer.
The turbulent Mach number at the boundary determined by the abundance distribution of is not small compared with $\langle Ma^2 \rangle^{1/2}_{\rm max}$ enough to specify the boundary of the convective region for these models.

\paragraph{Typical scale of the convection}
In addition to the Mach number, the dominant angular wave number in spherical harmonics also characterizes the convection. This is related to the typical size of the convective flow.
This quantity is important because a large-scale convective flow can amplify the explordability of core-collapse supernovae \citep{bernhard16_prog}.
The power spectrum of the radial turbulent velocity at $r(\langle Ma^2 \rangle ^{1/2}_{\rm max})$ is calculated as
\begin{equation}
    c_\ell^2 = \left| \int (v_r - \langle v_r \rangle) Y_{\ell0}^{*}(\theta) d\Omega \right|^{2}, \label{eq:cl2}
\end{equation}
where $Y_{\ell m}(\theta)$ is the spherical harmonics function of degree $\ell$ and order $m$.
$\ell_{\rm max}$ in the table represents $\ell$ value, at which $c_\ell^2$ has a peak.

Figure \ref{fig:power2d} shows the power spectrum $c_\ell^2$ at three different times for models 13L$_{\rm A}$ (top), 25M (middle), and 27L$_{\rm A}$ (bottom).
For model 13L$_{\rm A}$, $c_\ell^2$ has a maximum at $\ell = 12$, but the spectrum is rather flat and less energetic.
The radius of the highest Mach number in the O/Si layer is $1.16 \times 10^9$ cm.
Although the convective mixing occurs in the inner region of $R \la 6 \times 10^8$ cm, the turbulent velocity  is lower than the outer region at the last step.
Large scale convection in the O/Si layer is not developed probably because of small turbulent motion.
For model 16M$_{\rm A}$, $\ell_{\rm max}$ is equal to 4 and the trend of the power spectrum is similar to model 13L$_{\rm A}$.

For other low-$Ma$ models, models 18M$_{\rm A}$ and 21M$_{\rm A}$ 
show large $\ell_{\rm max}$ (see Table \ref{tab:t2})
and they have a thin Si/O layer.
This trend is roughly consistent with the analysis of convective eddy scale relating to the scale of the convective layer and the typical radius of the layer \citep[e.g.,][]{bernhard16_prog}.
In these models, the SiO-rich layer is thin compared to the radius of the layer.

In models 25M and 27L$_{\rm A}$, high-$Ma$ models, the power spectrum peaks at $\ell_{\rm max} = 3$ and 2, respectively, and the spectrum decreases with increasing $\ell$ above that.
Models 22L and 28M show a similar power spectrum to model 27L$_{\rm A}$.
For these models, large-scale convective eddies have been developed.
On the other hand, models 27M and 28L$_{\rm A}$ indicate larger $\ell$ values probably owing to thin SiO-rich layer.
Indeed, the former three models show larger width of the convective region normalized by the scale height compared to the latter two models (see Table 2).
Note that these models develop shell-convection in the Si/Fe layer as well. 
However, the convective region is always confined inside the layer.
This will be because the timescale of the silicon burning is shorter than the convective turn-over time, so that the mean molecular weight of the convective blob soon increases, suppressing the convective motion.

From the results shown above, it is discerned that 
$f_{M,{\rm SiO}}$ will be a more suitable measure than $f_{V,{\rm SiO}}$
to discriminate a model that develops convection with high turbulent velocity and a small $\ell_{\rm max}$.
First of all, we have shown that high-$Ma$ models are selected based on the high $f_{M,{\rm SiO}}$ values.
Moreover, models showing small $\ell_{\rm max} \leq 3$ are all selected based on $f_{M,{\rm SiO}}$
(models 22L, 25M, 27L$_{\rm A}$, and 28M), 
and only one of the two models showing $\ell_{\rm max} = 4$ (model 16M$_{\rm A}$) is selected based on $f_{V,{\rm SiO}}$.

\begin{figure*}[th]
\begin{center}
\includegraphics[width=7.5cm]{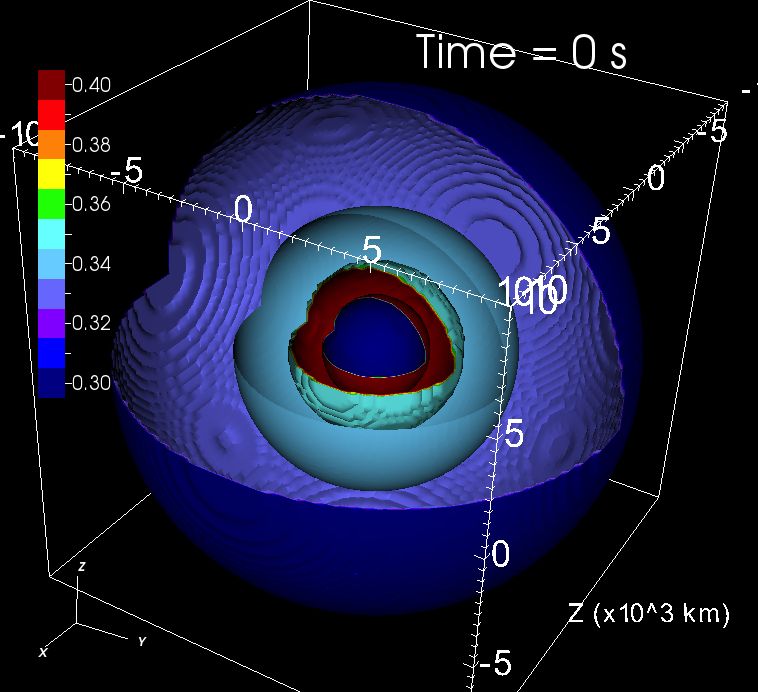}
\includegraphics[width=7.5cm]{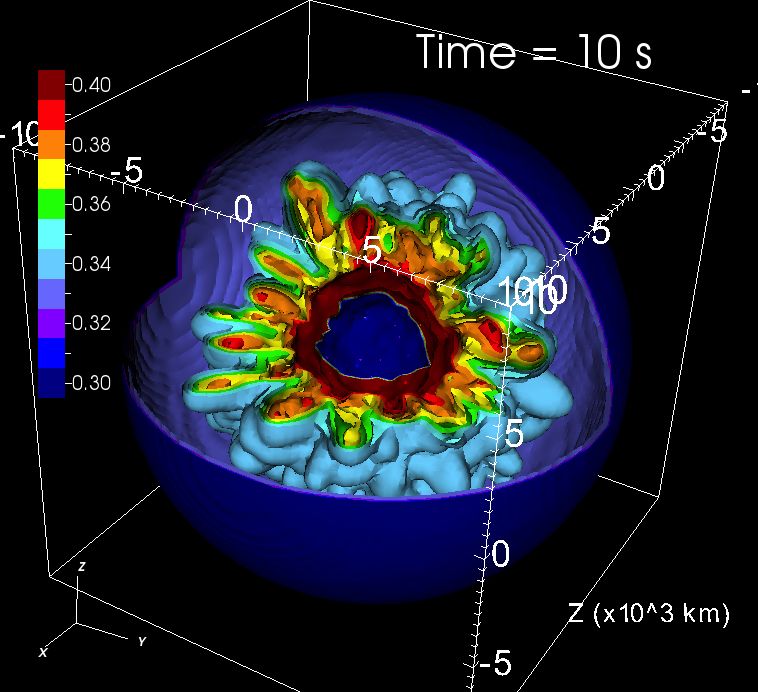}
\includegraphics[width=7.5cm]{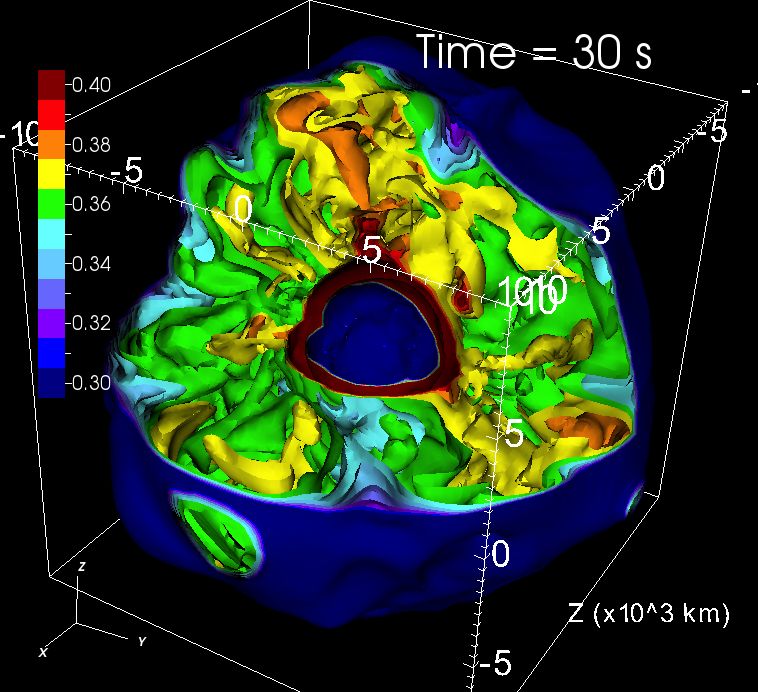}
\includegraphics[width=7.5cm]{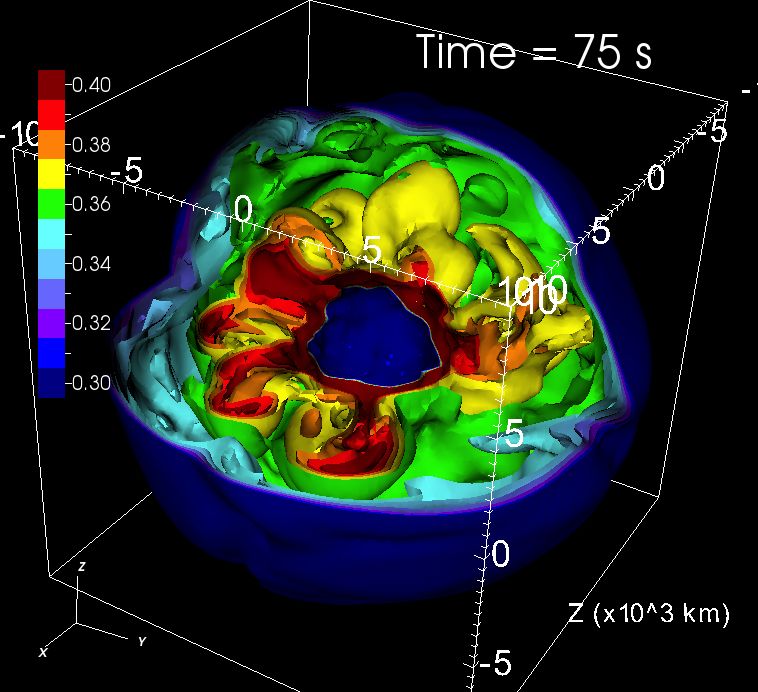}
\includegraphics[width=7.5cm]{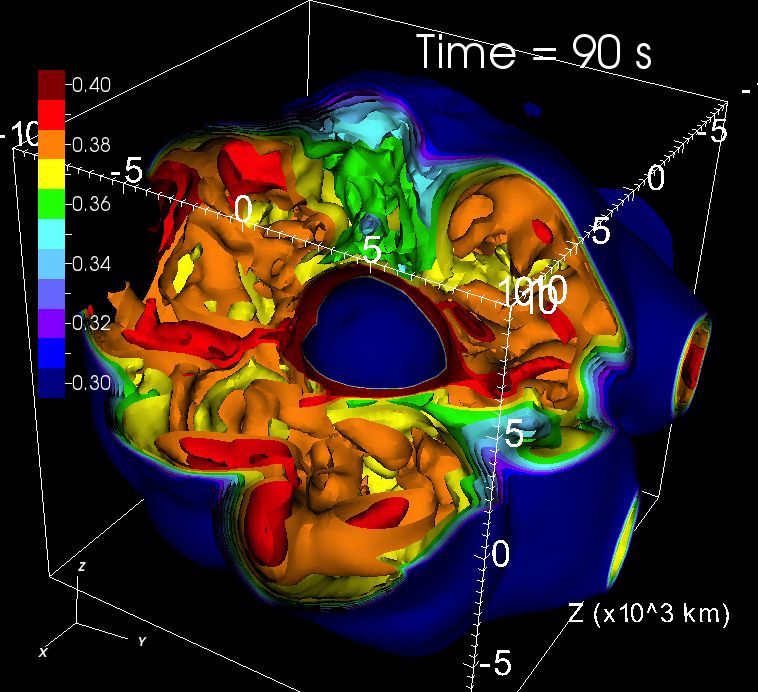}
\includegraphics[width=7.5cm]{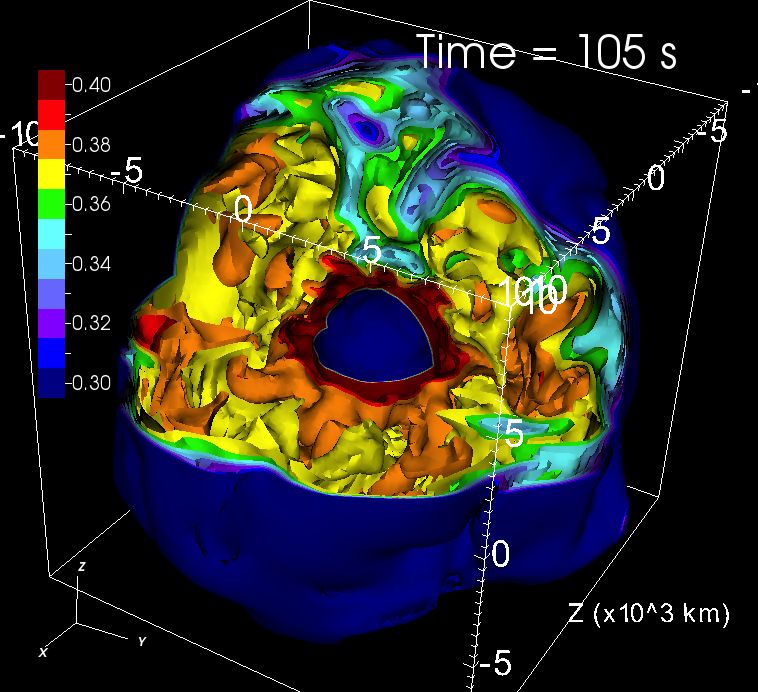}
\end{center}
\caption{The time variation of the $^{28}$Si mass-fraction distribution of model 25M at $t=$ 0 s (top left), 10 s (top right), 30 s (middle left), 75 s (middle right), 90 s (bottom left), and 105 s (bottom right).
}
\label{fig:f7}
\end{figure*}

\subsection{3D Stellar Hydrodynamics Simulation} \label{3d}

A 3D hydrodynamics simulation is conducted using the 1D model 25M as the initial condition.
The size of the convective region of model 25M would be suitable for investigating multidimensional effects of the structure of presupernova star to the SN explosion\footnote{Note that strong turbulent activity still continues for model 27L$_{\rm A}$ at the last step of the simulation.
 We shall perform a 3D hydrodynamics simulation for other models including model 27L$_{\rm A}$ in the future work}.
The hydrodynamical evolution is followed $\sim$100 s until the central temperature reaches $9 \times 10^9$ K, at which point the Fe core is unstable enough to collapse.

Figure \ref{fig:f7} shows the time evolution of the $^{28}$Si mass fraction distribution of model 25M.
The initial distribution of the $^{28}$Si mass fraction is spherically symmetric (top left panel).
After the start of the simulation, the convection in the Si/O layer develops from the inner region.
We see that Si-enriched plumes go up into the Si/O layer (top right panel).
The convective motion reaches a steady flow by $\sim$20 s.
The inhomogeneous $^{28}$Si mass fraction distribution introduced by the convection at 30 s is shown in the middle left panel.
After a while, the turbulent velocity becomes small.
We will discuss the mechanism of this weakening later.

At $\sim$70 s, the Si/O layer gradually contracts, triggering the strong oxygen shell-burning at the bottom of the Si/O layer.
This strong burning drives  high-velocity turbulence and expands the Si/O layer.
We see some $^{28}$Si enriched plumes flow from the inner region of the Si/O layer at 75 s (see red region in the middle right panel).
As a result, the  high-velocity turbulent flow mixes with the surroundings and increases the Si mass fraction in the whole Si/O layer.
The $^{28}$Si mass fraction in the Si/O layer slightly increases from about 0.36 at 30 s to 0.38 at 90 s (bottom left panel).
The convective motion in the Si/O layer continues until the last step of the simulation.
We see inhomogeneous $^{28}$Si mass fraction distribution at the last step (bottom right panel).

\begin{figure}
\begin{center}
\includegraphics[width=8.5cm]{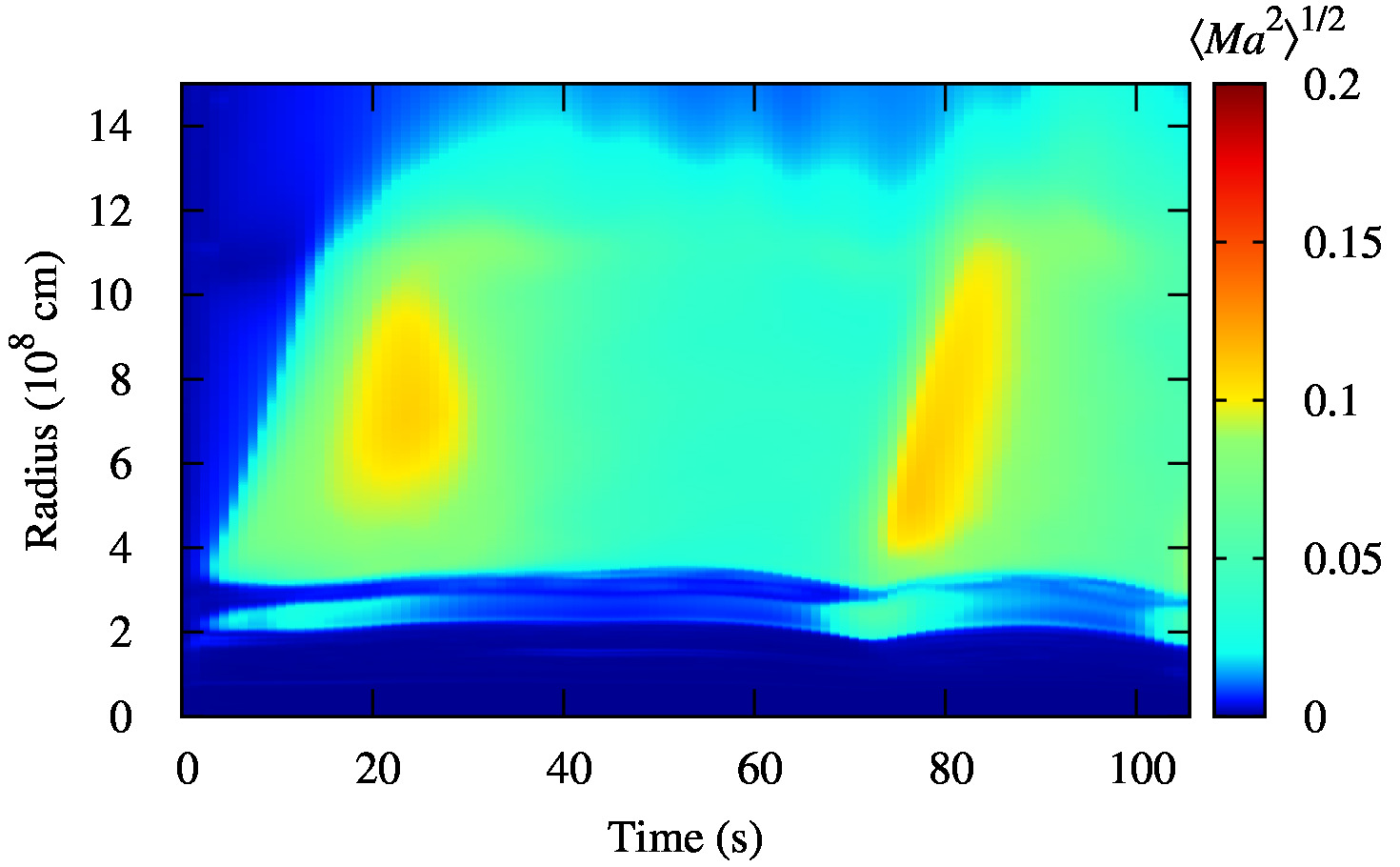}
\includegraphics[width=8.5cm]{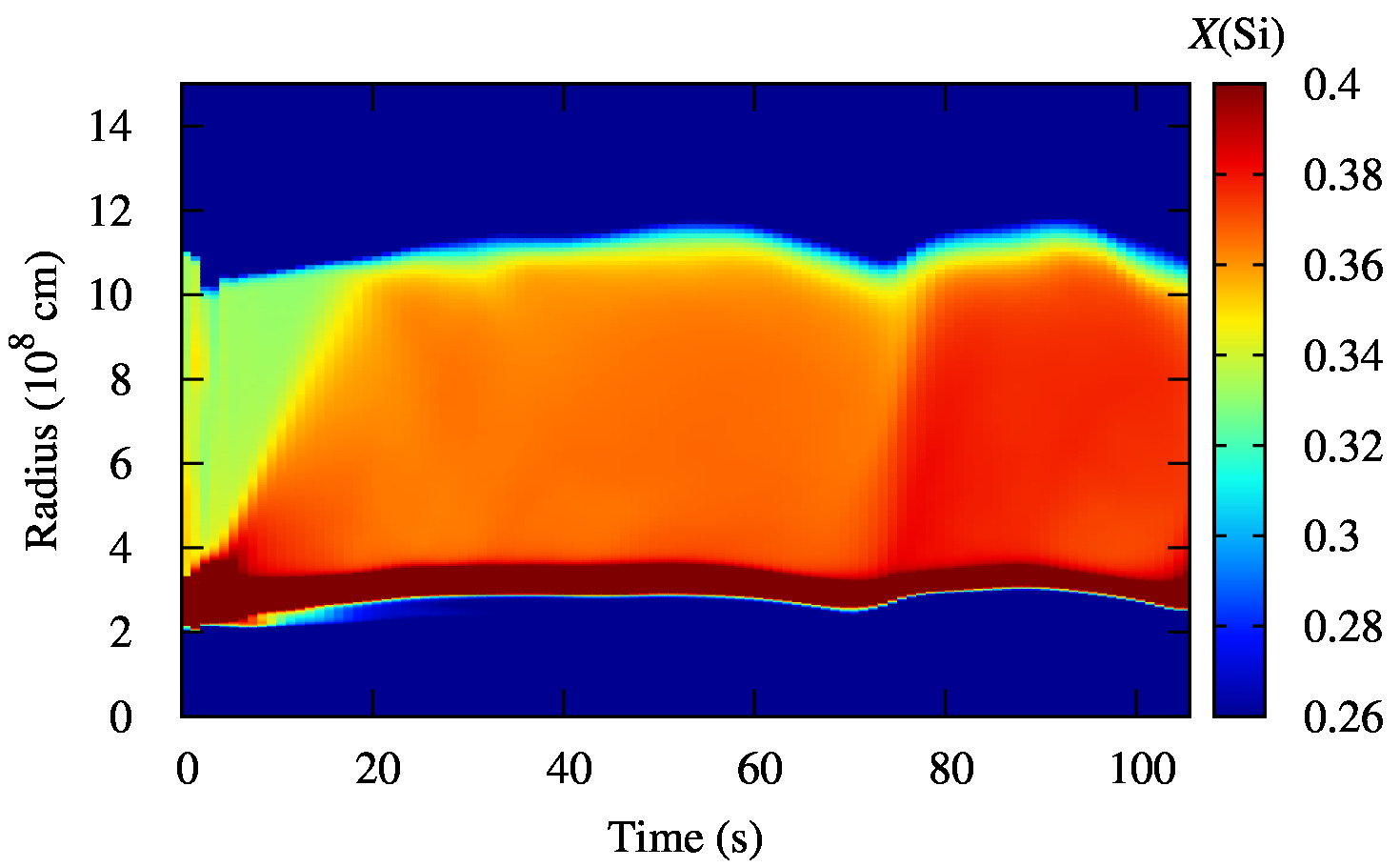}
\end{center}
\caption{The angle-averaged radial and time distribution of the Mach number $\langle Ma^2 \rangle^{1/2}$ (top) and the $^{28}$Si mass fraction (bottom) of model 25M.}
\label{fig:f8}
\end{figure}

In Figure \ref{fig:f8}, the time evolution of $\langle Ma^2 \rangle ^{1/2}$ and the $^{28}$Si mass fraction obtained from the 3D simulation is shown.
As shown by the left yellow region in the top panel, the convective motion with $\langle Ma^2 \rangle ^{1/2} \sim 0.1$ is obtained in the Si/O layer at $\sim$20 s.
The bottom panel shows that the $^{28}$Si mass fraction is enhanced in this layer by that time.
After a while, the Si/O layer slightly expands and the convective motion weakens.
The averaged Si mass fraction in the Si/O layer does not change significantly from 20 s to 70 s.

By $\sim$70 s, the Fe core contracts and silicon shell burning as well as oxygen shell burning is enhanced.
Because of this, the turbulent Mach number becomes large not only at $R \sim$ 3--12 $\times 10^{8}$ cm but also at $R \sim 2 \times 10^{8}$ cm.
In addition, the Si-rich material at the bottom of the Si/O layer is carried into the Si/O layer through this convective motion and the $^{28}$Si mass fraction in this region increases during $\sim$70--80 s.

From $\sim$90 s, the convective motion becomes weak again.
The whole Si/O layer gradually contracts towards the core-collapse.
\KT{Meanwhile, the base temperature of the Si/O layer increases, boosting the oxygen shell burning.
This results in the enhancement of the convective motion at the bottom of the Si/O layer.}
The Si/O layer at the end of the simulation is distributed from 3.0$\times 10^{8}$ cm to 10.5$\times 10^{8}$ cm.
The maximum radial convective Mach number is $\langle Ma^2 \rangle^{1/2} = 0.087$, which is obtained at $R = 3.6 \times 10^{8}$ cm. 

\begin{figure}
\includegraphics[width=\linewidth]{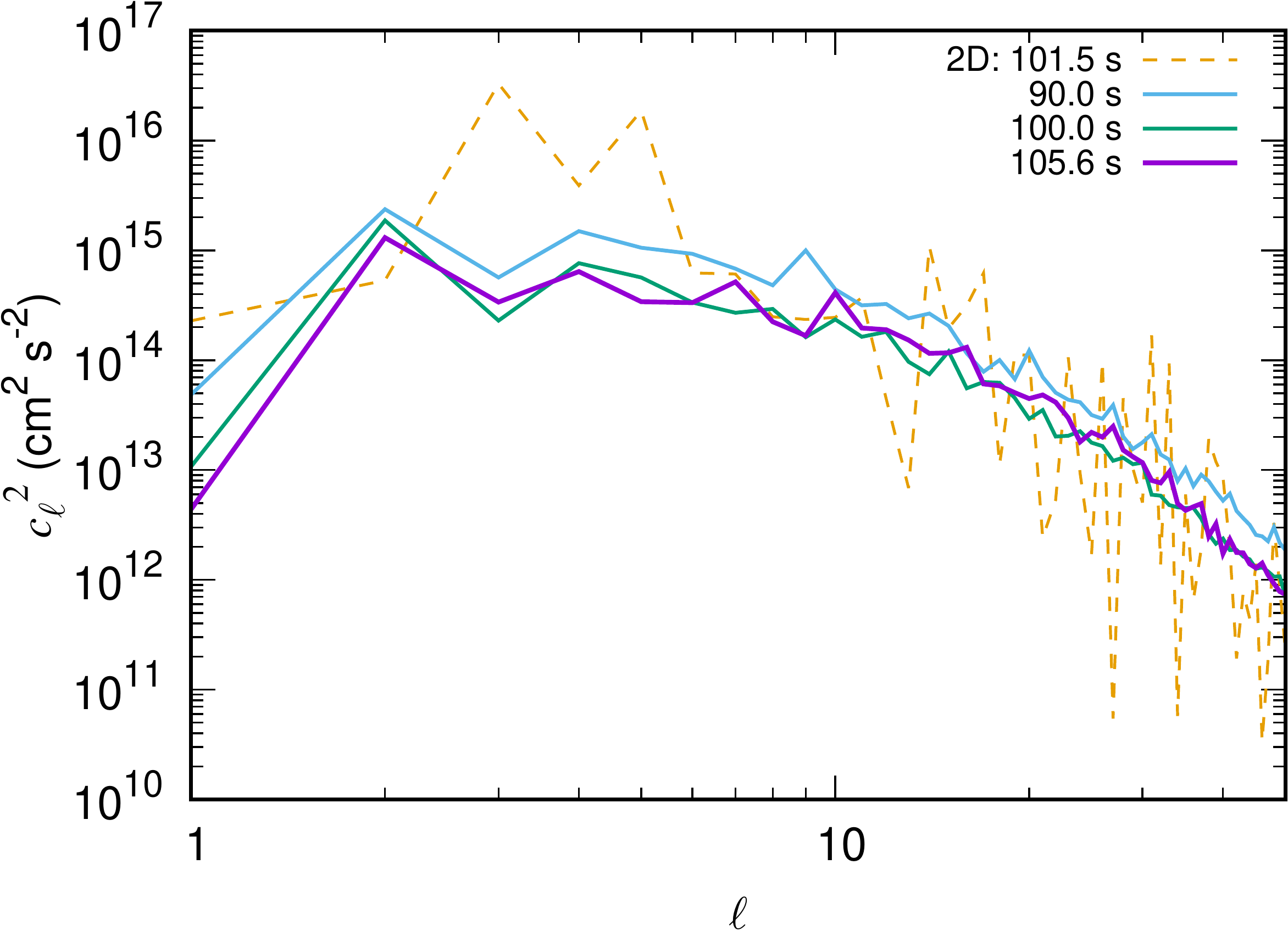}
\caption{The power spectrum of the convective radial velocity $c_\ell^2$ at $R = 5.8 \times 10^8$ cm for model 25M taken at three different times of 90.0, 100.0, and 105.6 s in the 3D simulation.}
\label{fig:f9}
\end{figure}

Figure \ref{fig:f9} shows the power spectrum of the radial convective velocity $c_\ell^2$ at $R = 5.8 \times 10^{8}$ cm taken at three different times.
The radius, which is located in the middle of the convective Si/O layer, is determined from the radius where the maximum $\langle Ma^2 \rangle^{1/2}$ is obtained in the 2D simulation.
Similar to the 2D case, the power spectrum in the 3D case is calculated as
\begin{equation}
    c_\ell^2 = \sum_{m = -\ell}^{\ell} \left| \int (v_r - \langle v_r \rangle) Y_{\ell m}^{*}(\theta,\phi) d\Omega \right|^{2}. \label{eq:cl2-3d}
\end{equation}
The power spectrum $c_\ell^2$ in the 3D simulation peaks at $\ell = 2$. 
This is consistent with the finding in \citet{bernhard16_prog} for the $18 M_{\odot}$ star.
The small maximum mode means that the convective motion is dominated by a large-scale flow.
The main difference between 2D and 3D power spectra is the weaker turbulence in the 3D simulation. Note that stronger turbulence in 2D than in 3D is not surprising because the turbulent energy cascade could occur artificially, as previously identified, from small to large scales along the coordinate symmetry axis, this is mainly due to a reduced degree of freedom in which the energy can dissipate.

\begin{figure}
\includegraphics[width=\linewidth]{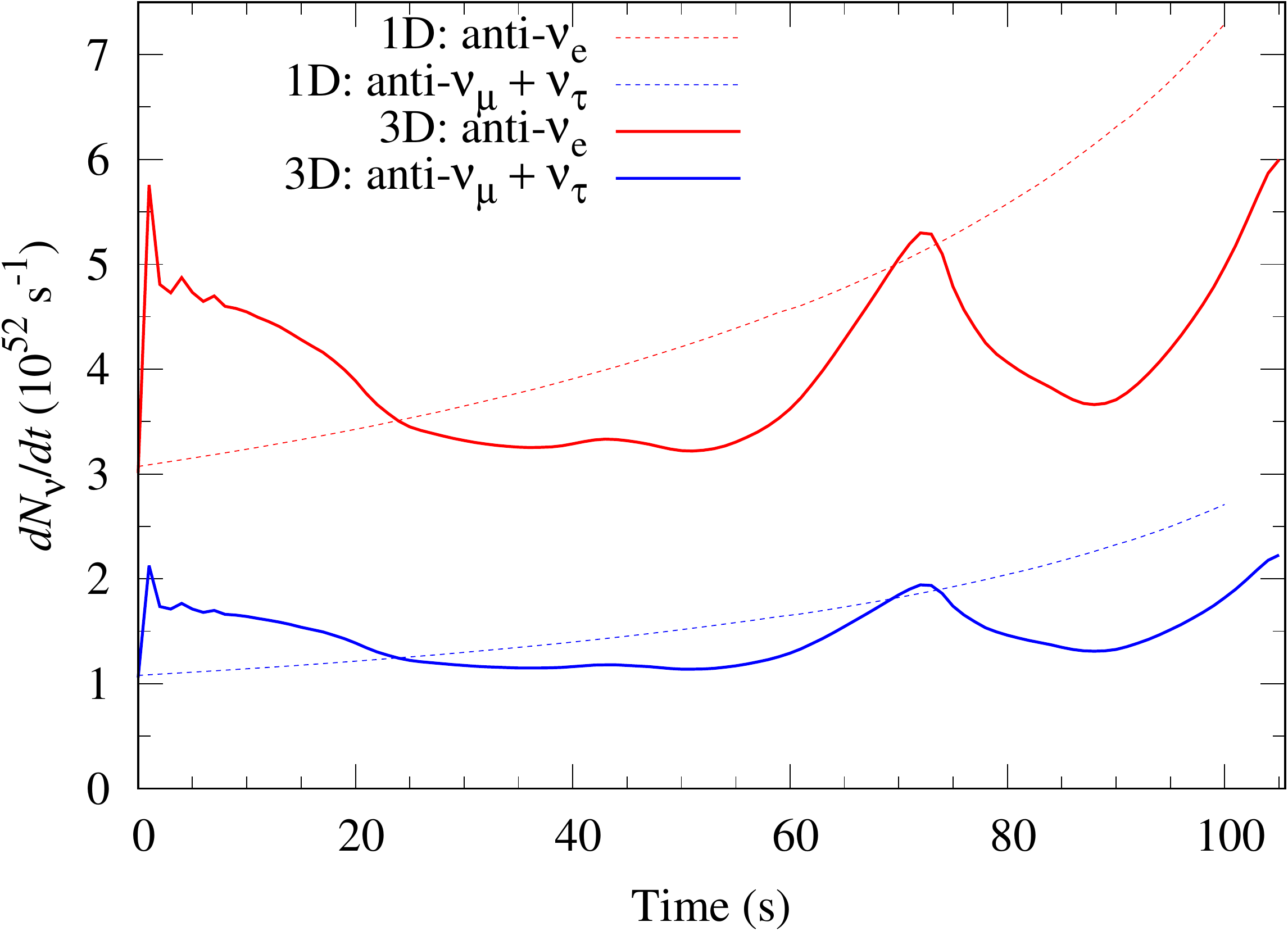}
\includegraphics[width=\linewidth]{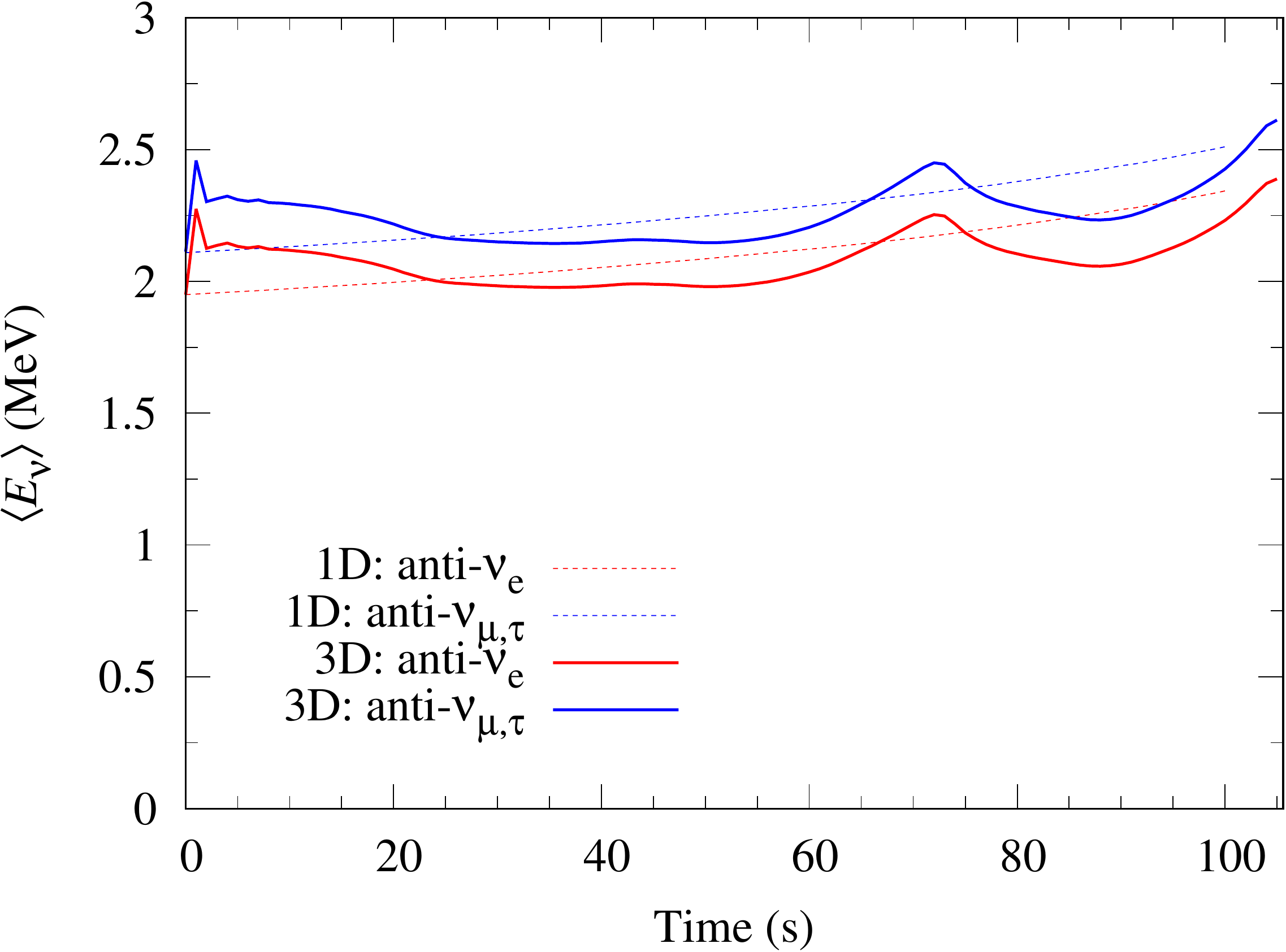}
\caption{Time variation of the emission rate (top panel) and average energy (bottom panel) of anti-neutrinos produced through pair-neutrino process for model 25M.
Thick solid and thin dashed curves correspond to the results of 3D and 1D simulations, respectively.
Red and blue curves correspond to $\bar{\nu}_e$ and $\bar{\nu}_{\mu,\tau}$}.
\label{fig:f10}
\end{figure}

Finally, we briefly present the result of the neutrino emission at the precollapse stage of the 3D simulation.
Using the same method in \citet{Yoshida16}, we calculate the time evolution of the luminosity and the spectrum of neutrino emitted via pair-neutrino process for model 25M.
In Figure \ref{fig:f10}, the emission rate and the average energy of neutrino obtained from the 1D and the 3D simulations are compared.

The overall features of neutrino spectra for the 1D and 3D simulations are in common.
The neutrino emission rate and the average temperature increase with time towards the core-collapse after $\sim$30 s in both of the simulations.
The emission rates of $\nu_e$ and $\bar{\nu}_e$ are larger than that of $\nu_{\mu,\tau}$ and $\bar{\nu}_{\mu,\tau}$.
The average energy of $\bar{\nu}_e$ is slightly smaller than that of $\bar{\nu}_{\mu,\tau}$.
In the 3D simulation, however, the decrease in the emission rate and the average energy of neutrinos is observed in 70--90 s.

\begin{figure}
\includegraphics[width=\linewidth]{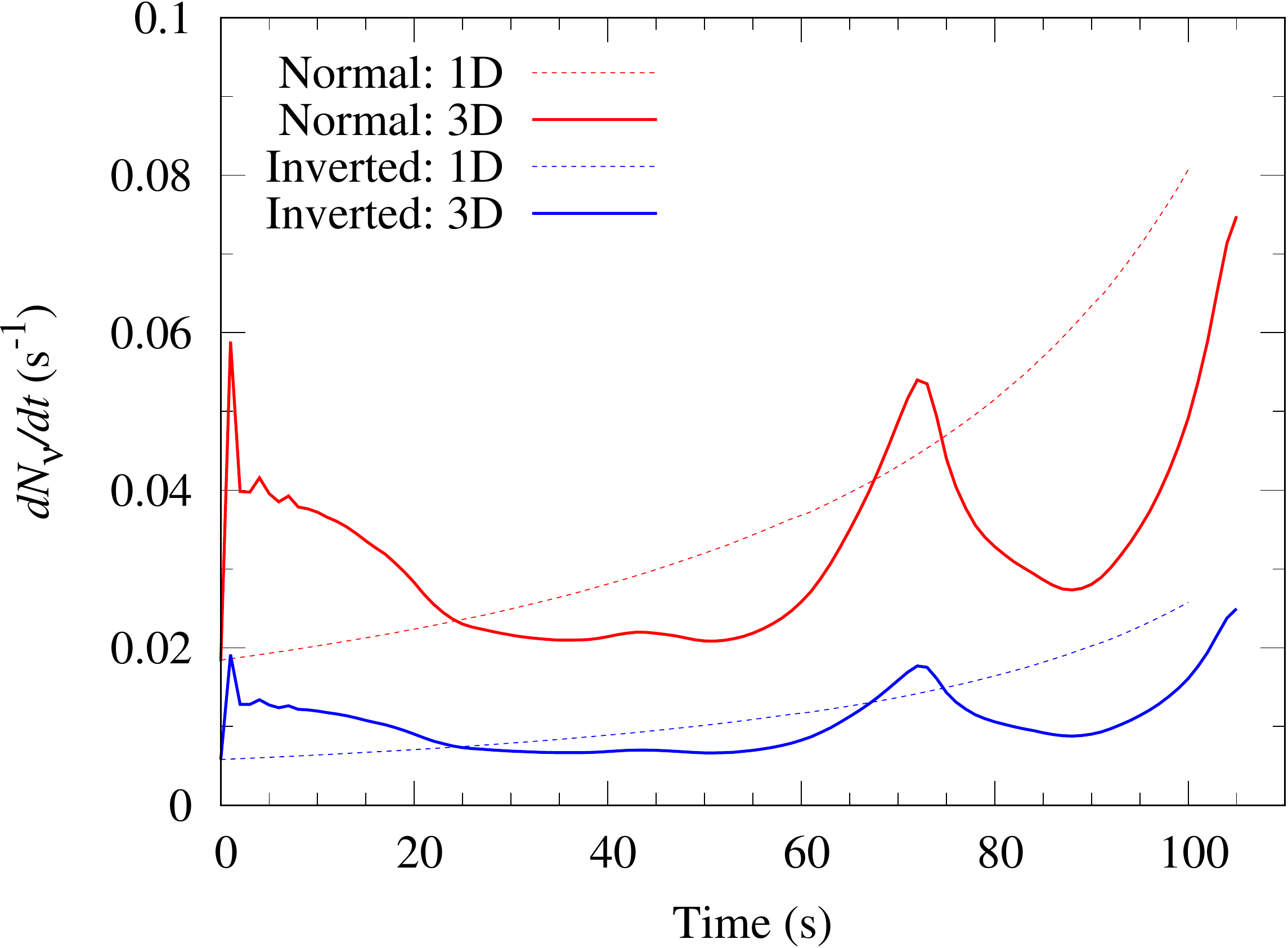}
\caption{The evolution of neutrino event rate of model 25M by KamLAND.
Thick solid and thin dashed curves correspond to the results of 3D and 1D simulations, respectively.
Red and blue curves correspond to the normal and inverted orderings.}
\label{fig:f11}
\end{figure}

Furthermore, we evaluate the time evolution of the event rate of $\bar{\nu}_e$ by KamLAND \cite[e.g.,][]{Gando13}, assuming that the initial mass of 25 $M_\odot$ at the distance of 200 pc is at the ongoing core collapse phase, emitting neutrinos via pair-neutrino process.

KamLAND is a one-kton size liquid-scintillation type neutrino detector \cite[e.g.,][]{Gando13}.
We take the neutrino oscillation into account in a simple manner: the survival probability of $\bar{\nu}_e$ is set to be 0.675 and 0.024 in the normal and inverted mass ordering, respectively \citep{Yoshida16}.
The livetime-to-runtime ratio and the total detection efficiency are set to be 0.903 and 0.64 \citep{Yoshida16}.
Figure \ref{fig:f11} shows the evolution of the neutrino event rate by KamLAND.
As an overall feature, the rate increases with time independent of mass ordering.
In the 3D simulation, however, the decrease in the event rate due to the oxygen shell burning at the bottom of the Si/O layer is seen at 70--90 s.
The time variation in the precollapse neutrino detection thus can be used as the indicator of the multidimensional convective activity deep inside the star, although it is practically impossible to detect such a time variation by current one kilo-ton size neutrino detectors.

Next, we examine the detectability of precollapse neutrinos by Hyper-Kamiokande.
Hyper-Kamiokande is a planned Water-Cherenkov type neutrino detector, which has a huge fiducial mass of 190 kton \citep{HKPC18}.
If a supernova explodes in the vicinity of the earth, high quality data of supernova neutrinos with well resolved time and energy bins are expected to be obtained \citep[e.g.,][]{Takahashi01,Mirizzi16}.
However, since the threshold energy of Hyper-Kamiokande is expected to be the same level of Super-Kamiokande ($\sim$4.79 MeV corresponding to 3.5 MeV for recoil electrons \citep{Sekiya13}) and is higher than that of liquid scintillation type detector, Hyper-Kamiokande is less preferable for the observations of neutrinos produced through pair-neutrino process during the precollapse stage.
\citet{Yoshida16} discussed a possibility of the neutrino observations using delayed $\gamma$-rays from Gd in Gd-loaded Hyper-Kamiokande \citep{Beacom04}, because the energy threshold will be reduced to 1.8 MeV, a similar energy to the case of KamLAND, in this case.
Considering a moderate detection efficiency of 50\%, the detection rate is about 178 times as large as the rate of KamLAND.
When the enhancement factor of 178 for the detection rate is applied, the event rate is expected to be 3--14 s$^{-1}$ in the normal mass ordering.
So, the time variation due to the convective motion with a time scale of seconds could be observed by Hyper-Kamiokande.

We should note that we consider only pair-neutrino process for presupernova neutrinos in this study.
For a few minutes before the core-collapse, the main source of $\bar{\nu}_e$ will be $\beta^-$-decays rather than the pair-neutrino process \citep{kato17}.
$\beta^-$-decays mainly take place in the innermost Fe core and the $\bar{\nu}_e$ energy is similar to pair-neutrinos.
We expect that the neutrino event rate by $\beta^-$-decays also decreases for 70--90 s because the central temperature and density decrease during this period.
So, even when we take into account the neutrino event rate by $\beta^-$-decays, the time variation of presupernova neutrino events would give us the information about dynamics in the SiO-rich layer or a collapsing Fe core. 

\section{Summary and Discussions} \label{sec4}

We have performed 1D, 2D, and 3D simulations of the oxygen shell-burning just before the core-collapse of massive stars.
First, we have calculated the 1D evolution of solar metallicity massive stars with ZAMS mass of 9--40 $M_\odot$.
We have considered four cases of the convective overshoot parameters in hydrogen and helium burning and the following stages.

From these, we have searched for the massive stars having O and Si-rich layers located in the range of $10^8$--10$^{9}$ cm, because the oxygen burning shell is
expected to give rise to large asphericity in the mass accretion rates onto the proto-neutron star, favoring the onset of neutrino-driven explosions.
For the enclosed-mass weighted SiO-coexistence parameter $f_{M,{\rm SiO}}$, the stars with $M_{\rm ZAMS} \sim$20--30 $M_\odot$ have SiO-enriched environment in 10$^8$--10$^9$ cm.
For the volume weighted SiO-coexistence parameter $f_{V,{\rm SiO}}$, the mass range indicating SiO-rich layer increases to 13--30 $M_\odot$.

We have selected eleven models showing a large SiO-rich layer from the result of the massive star evolution and performed 2D hydrodynamics simulations of the evolution for $\sim$100 s until the central temperature reaches $9 \times 10^9$ K.
We have investigated time evolution of the Mach number of convective velocity and analyzed the power spectrum of the radial convective velocity.
Based upon the analysis, we have classified the eleven models into five models with low turbulent Mach number (low-$Ma$) and six models with high turbulent Mach number (high-$Ma$).

High-velocity turbulence is obtained in four models having a Si/O layer with the range up to $\sim 1 \times 10^9$ cm and two models having an extended O/Si layer.
These models commonly have large $f_{M,{\rm SiO}}$ values.
All models with large $f_{M,{\rm SiO}}$ except for 23L$_{\rm A}$ exhibit high-velocity turbulence in the 2D simulations.
In addition, most of the small $\ell_{\rm max}$ models have been selected based on $f_{M,{\rm SiO}}$. 
Our results suggest that high density in the SiO-rich layer could be conducive to producing 
vigorous convection just before the core-collapse
(see the difference in Equations (2) and (3)) and that $f_{M,{\rm SiO}}$ could be a more suitable measure for convection with high velocity turbulence and large scale eddies than $f_{V,{\rm SiO}}$.

We have performed a 3D hydrodynamics simulation for model 25M for $\sim$100 s until the central temperature reaches $9 \times 10^9$ K.
The time evolution of the convection properties is qualitatively similar between 2D and 3D simulations.
The convection is dominated by large-scale flows with either $\ell_{\rm max}=2$ in the 3D case or $\ell_{\rm max}=3$ in the 2D case.
The main difference is smaller turbulent velocity in the 3D simulation.

Using the 3D result of the hydrodynamics, we have evaluated the time evolution of the neutrino emission through electron-positron pair annihilation.
The emission rate and the average energy of neutrino evolve similarly in 1D and 3D simulations.
However, in the 3D model, the neutrino emission rate shows significant variation due to the strong oxygen shell burning at $\sim$70 s.
The multi-D effect of the convective burning would be observed in presupernova neutrino events by the present and future neutrino detectors such as KamLAND and Hyper-Kamiokande.

In this study, we were only able to compute 
a single 3D model of the $25 M_{\odot}$ star due to the high numerical cost. However, more systematic and long-term 3D simulations employing a variety of the progenitors are needed 
\citep[e.g.,][]{bernhard18_prog}.
In order to accurately deal with neutronization of heavy elements to the iron-group nuclei, we need to not only implement bigger nuclear network but also a complete set of neutrino opacities \citep{kato17} 
in our multi-D models, again albeit computationally expensive. Employing the 3D progenitor models in core-collapse supernova simulations is also mandatory  to see how much the precollapse inhomogeneities could foster the onset of neutrino-driven explosions \citep{Couch15,bernhard18_prog}. Impact of rotation (not to mention magnetic fields!)
 on the burning shells during the last stage of massive stars is yet to be studied in 3D (see \citet{chatz16} for 2D evolution models with rotation). All in all, this study is only the very first step toward the more sophisticated and systematic multi-D presupernova modeling.

\acknowledgments

We acknowledge anonymous referee for reading our manuscript carefully and giving us many 
useful comments to improve this manuscript.
We thank C. Nagele for proofreading.
This study was supported in part by the Grants-in-Aid for the Scientific Research of Japan Society for the Promotion of Science (JSPS, Nos. 
JP26707013, 
JP26870823, 
JP16K17668, 
JP17H01130, 
JP17K14306, 
JP18H01212  
), 
the Ministry of Education, Science and Culture of Japan (MEXT, Nos. 
26104007,   
JP15H00789, 
JP15H01039, 
JP15KK0173, 
JP17H05205, 
JP17H05206 
JP17H06357, 
JP17H06364, 
JP17H06365, 
JP24103001 
JP24103006 
JP26104001, 
JP26104007), 
by the Central Research Institute of Explosive Stellar Phenomena (REISEP) of Fukuoka University and the associated research projects (Nos.171042,177103), and by JICFuS as a priority issue to be tackled by using Post `K' Computer.
T.T was supported by the NINS program for cross-disciplinary
study (Grant Numbers 01321802 and 01311904) on Turbulence, Transport,
and Heating Dynamics in Laboratory and Solar/Astrophysical Plasmas: SoLaBo-X.
K. T. was supported by Japan Society for the Promotion of Science (JSPS) Overseas Research Fellowships.
The numerical simulations have been done using XC50 at Center for Computational Astrophysics at National Astronomical Obervatory of Japan and 
XC40 at Yukawa Institute of Theoretical Physics.

\software{HOSHI \citep{Takahashi16,Takahashi18}, 3DnSNe \citep{Takiwaki16,Nakamura16,kotake18}}

%



\appendix


We show some evolution properties of the massive star models.
\begin{figure*}
\centering
\includegraphics[width=8cm]{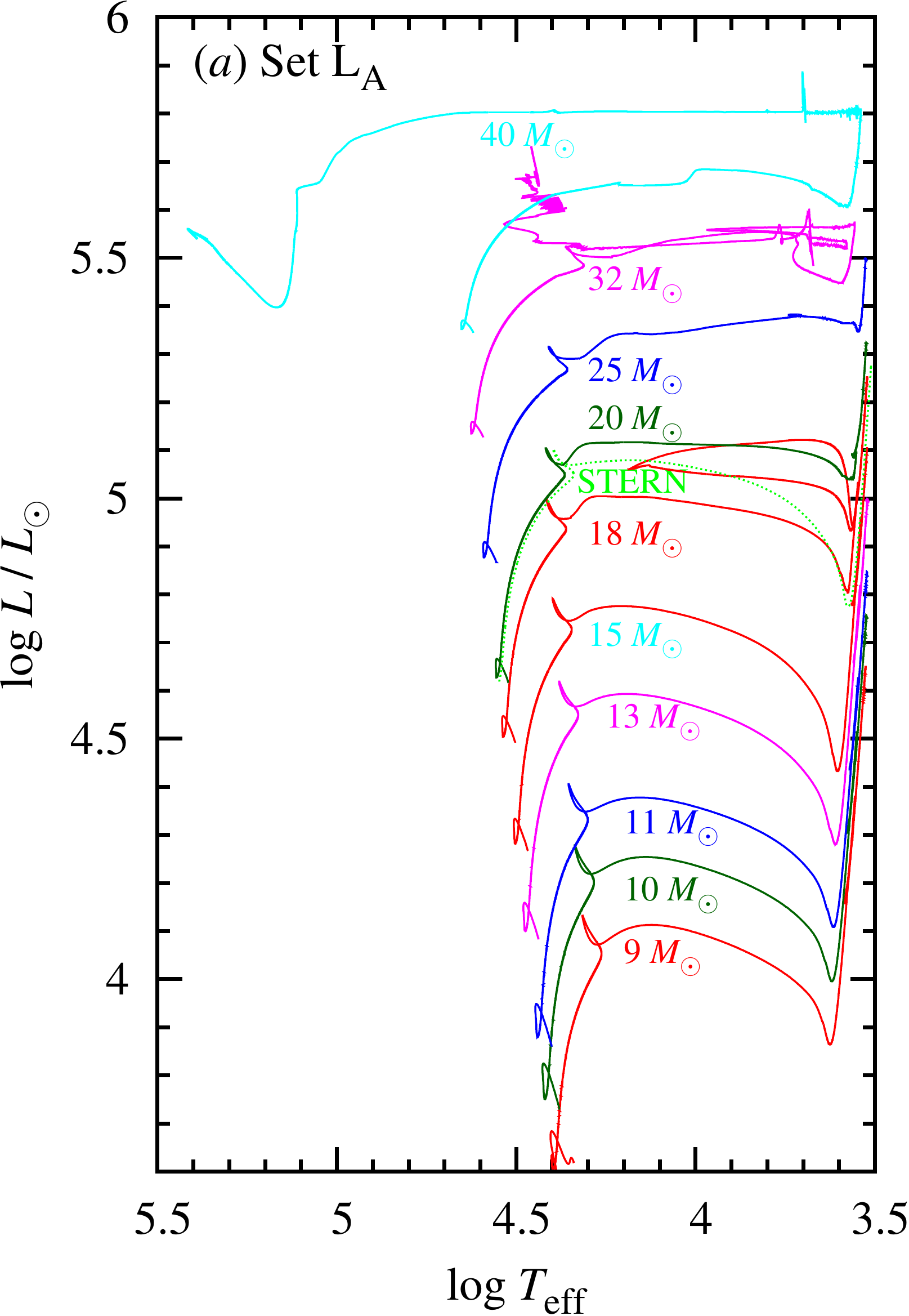}
\includegraphics[width=8cm]{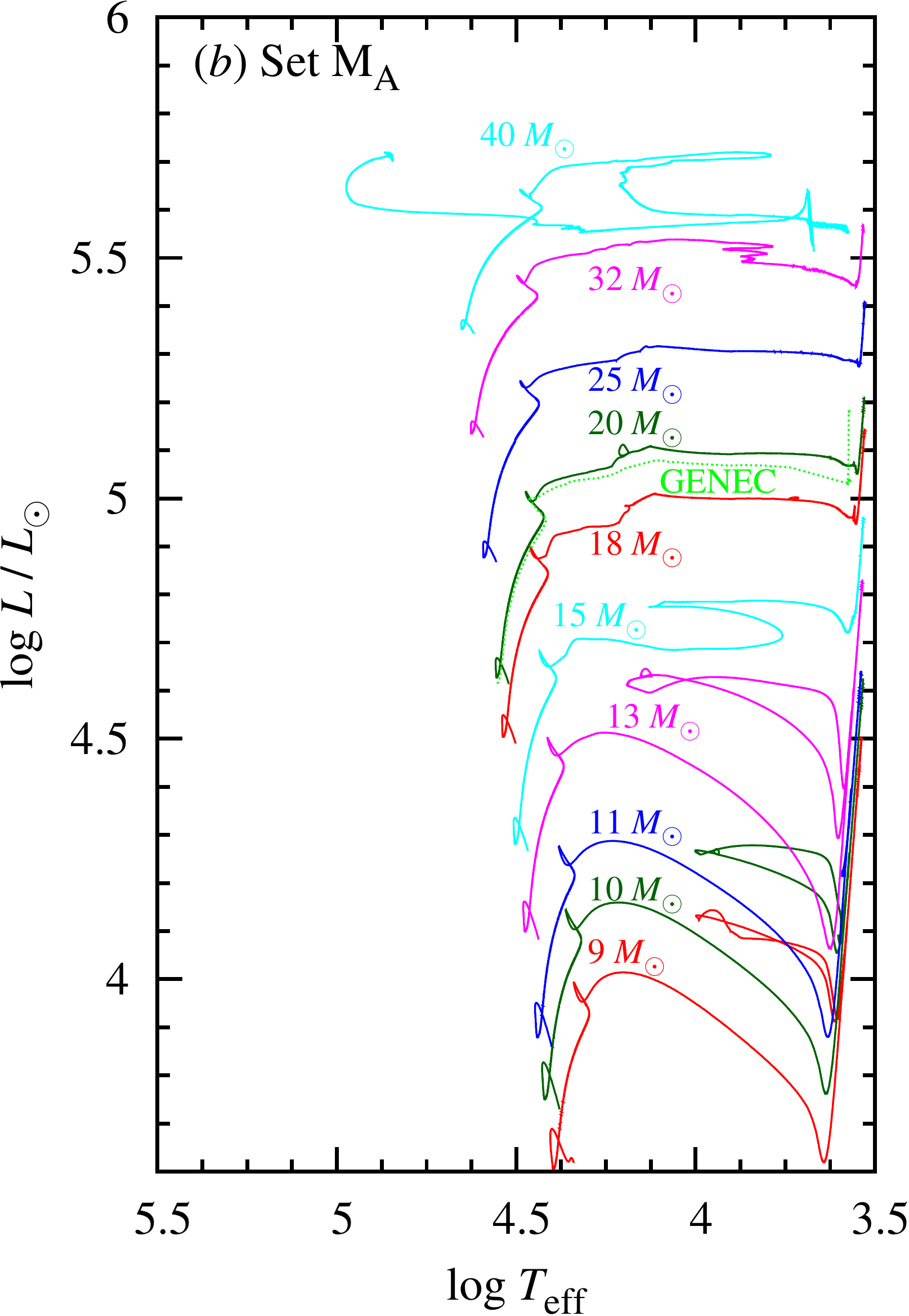}
\caption{HR diagram of Sets L$_{\rm A}$ (panel ($a$)) and M$_{\rm A}$ (panel ($b$)).
Green-dotted lines in panels ($a$) and ($b$) are 20 $M_\odot$ models of Stern \citep{Brott11} and GENEC \citep{Ekstroem12}, respectively.
}
\label{fig:HRdiag}
\end{figure*}
We show the HR diagram of ten different ZAMS mass models for Sets \LOVC  and \MOVC in Figure \ref{fig:HRdiag}.
Stars with $\MZAMS <$ 30 $M_\odot$ for Set  \LOVC and with $\MZAMS <$ 35 $M_\odot$ for Set \MOVC end as a red supergiant.
Stars with heavier $M_{\rm ZAMS}$  evolve to a Wolf-Rayet star.

As shown in Section 2, we considered four sets of models with different overshoot parameters.
The overshoot parameter during the hydrogen and helium burning affects the evolution on the Hertzsprung-Russell (HR) diagram.
We compare the HR diagram of models 20L$_{\rm A}$ and 20M$_{\rm A}$ with that of the 20 $M_\odot$ star models of Stern \citep{Brott11} and GENEC \citep{Ekstroem12}, respectively (see green-lines in Figure \ref{fig:HRdiag}).
The main difference in the HR diagram between model 20L$_{\rm A}$ and model 20M$_{\rm A}$ is the main-sequence (MS) band width, i.e., the difference of the effective temperature between ZAMS and the hydrogen burning termination.
The MS band width of model 20L$_{\rm A}$ is almost identical to the Stern model.
On the other hand, model 20M$_{\rm A}$ is almost identical to GENEC.
As explained in Section 2, the overshoot parameters of Sets L$_{\rm A}$ and M$_{\rm A}$ are determined based on the calibrations to early-B type stars in the LMC similar to the Stern model and the MS width observed for AB stars in open-clusters of the MW Galaxy similar to the GENEC model, respectively.
Except for the MS width, we do not see any large differences in the HR diagram among these models.
The evolution of 20 $M_\odot$ star models of Stern, GENEC, MESA \citet{Paxton11,Martins13}, and Starevol \citep{Decressin09} are compared in \citet{Martins13}.
The HR diagram of model 20M$_{\rm A}$ is also similar to that of MESA and Starevol.

\begin{figure}
\includegraphics[width=0.5\linewidth]{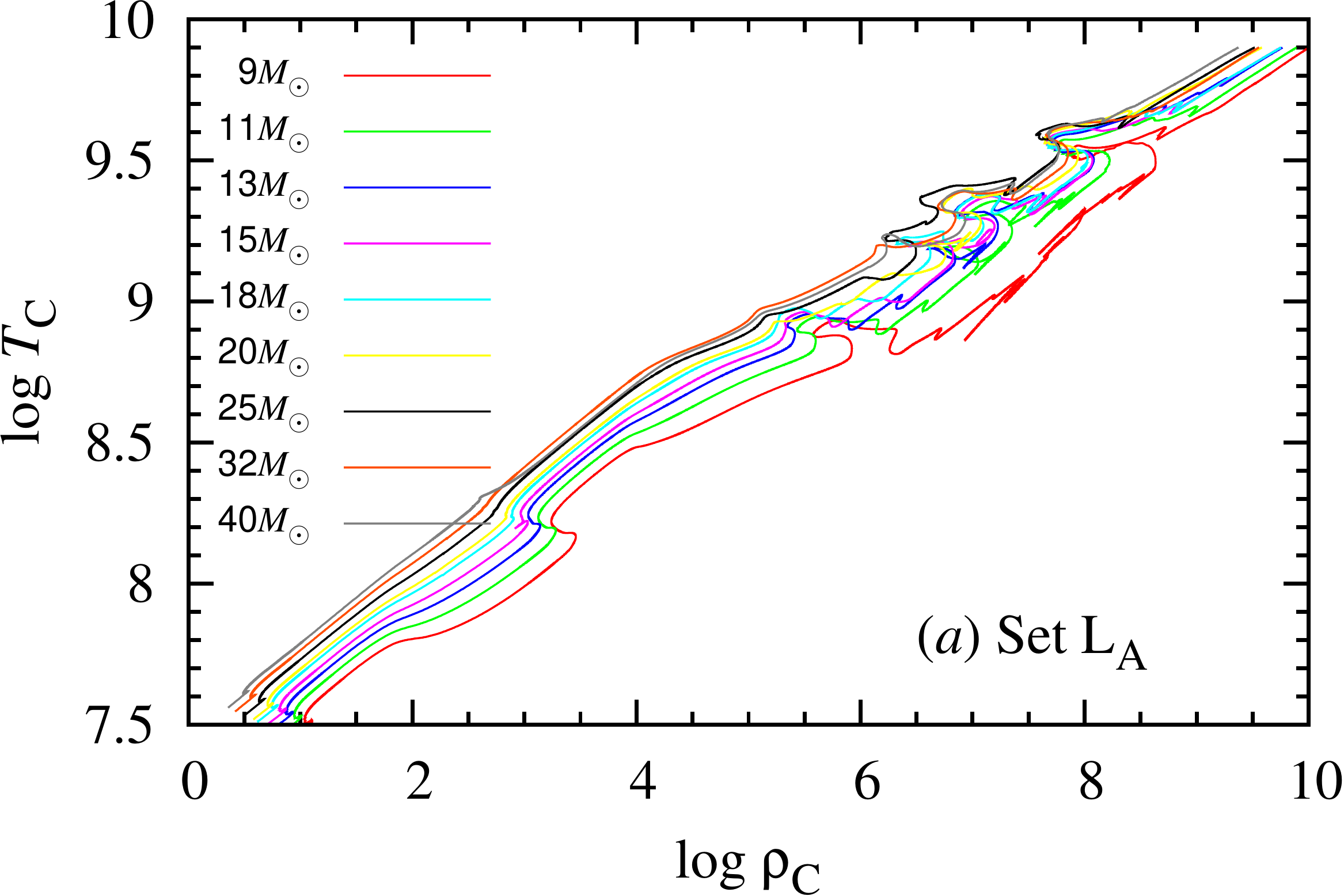}
\includegraphics[width=0.5\linewidth]{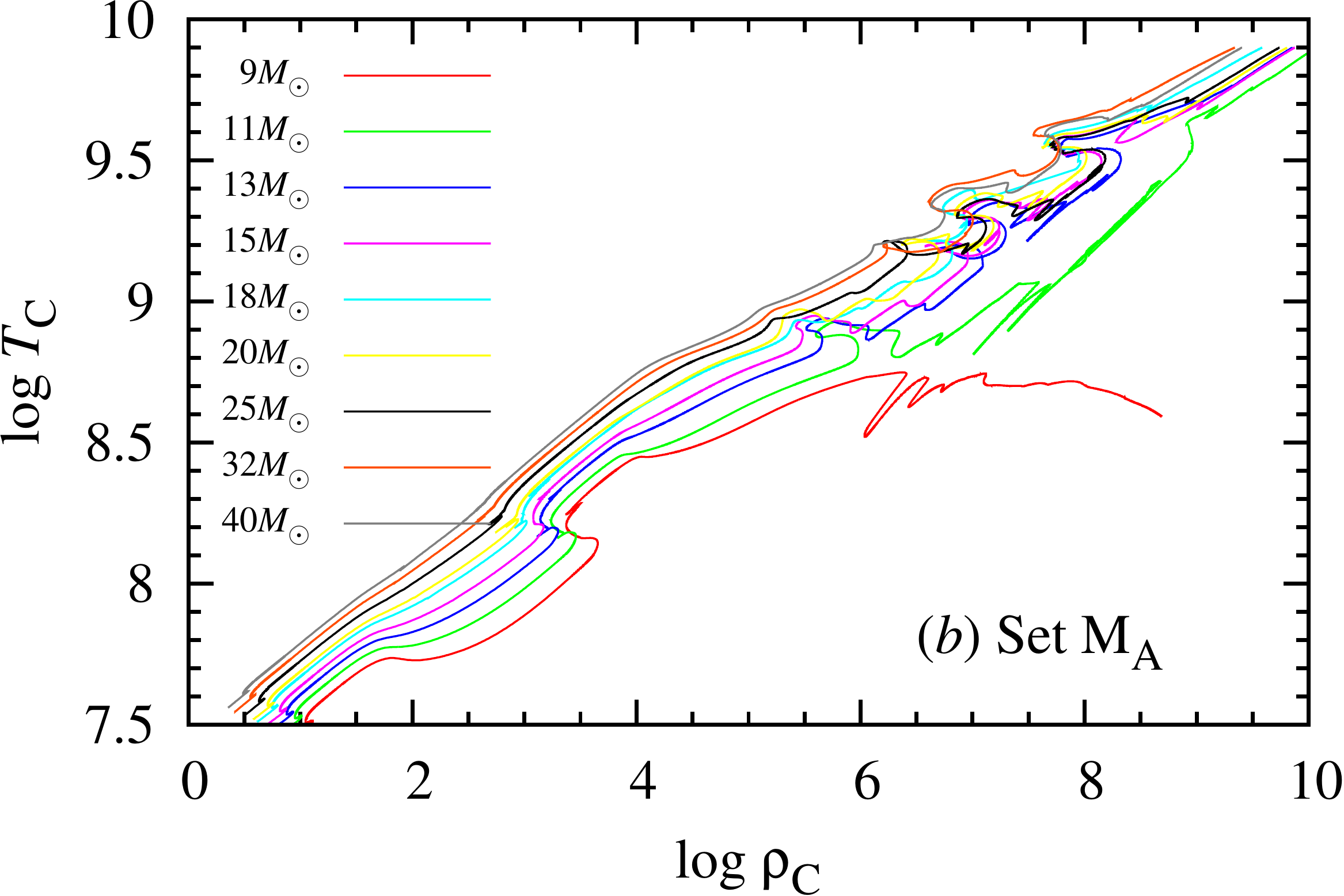}
\caption{The evolution of the central density $\log \rho_{\rm C}$ and the central temperature $\log T_{\rm C}$ of Sets L$_{\rm A}$ (panel ($a$)) and M$_{\rm A}$ (panel ($b$)).}
\label{fig:tcrhoc}
\end{figure}

We show the evolution of the relation between the central density and the central temperature in Figure \ref{fig:tcrhoc}.
All stars except for the ZAMS mass of 9 $M_\odot$ in Sets M$_{\rm A}$ and M end as the core-collapse of an Fe core.
The 9 $M_\odot$ star in Set M$_{\rm A}$ does not bring about Ne ignition (panel ($b$): red line).
We also confirmed no Ne ignition in the 9 $M_\odot$ star in Set M until the central density becomes $\log \rho_{\rm C} = 8.5$.
These stars are expected to end as an ONe white dwarf or an electron-capture SN.
The off-center Ne ignition occurs for the 9 $M_\odot$ star in Set L$_{\rm A}$ (panel ($a$): red line) and the 11 $M_\odot$ star in Set M$_{\rm A}$ (panel ($b$): green line).
The similar off-center neon burning is also seen in a 10 $M_\odot$ star in Sets M$_{\rm A}$ and M.
The low-mass M$_{\rm A}$ and M models ignite silicon at an off-center region.

Whether the carbon core burning occurs convectively or radiatively depends on the stellar mass.
The convective carbon-core burning occurs in the stars with $M_{\rm ZAMS} \le 18 M_\odot$ in Sets L$_{\rm A}$ and L and with $M_{\rm ZAMS} \le 21 M_\odot$ in Sets  M$_{\rm A}$ and M.
We suggest that the maximum CO-core mass for the convective carbon burning is between 4.6 and 4.9 $M_\odot$ for our models.

\begin{figure*}
\centering
\plottwo{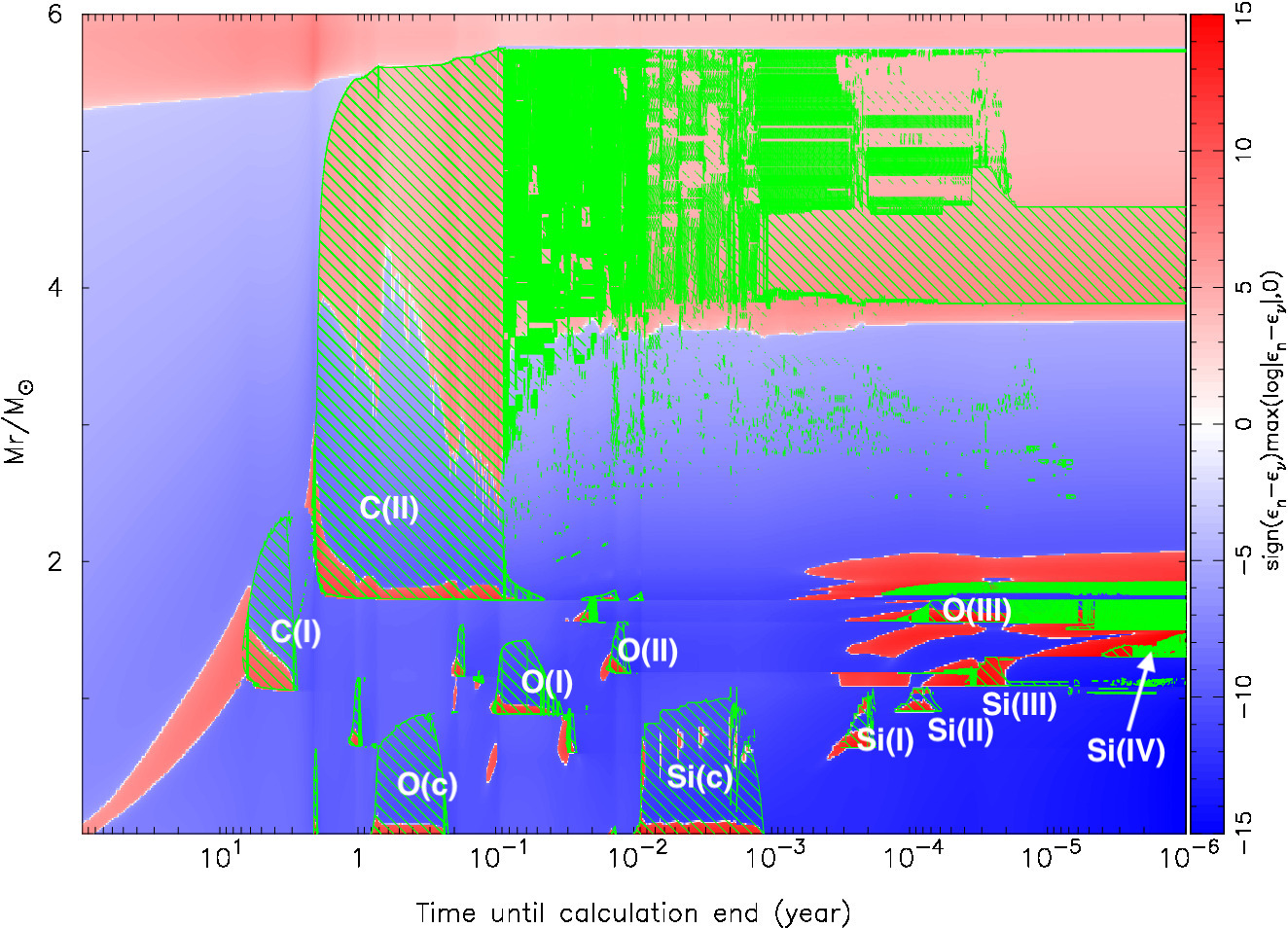}{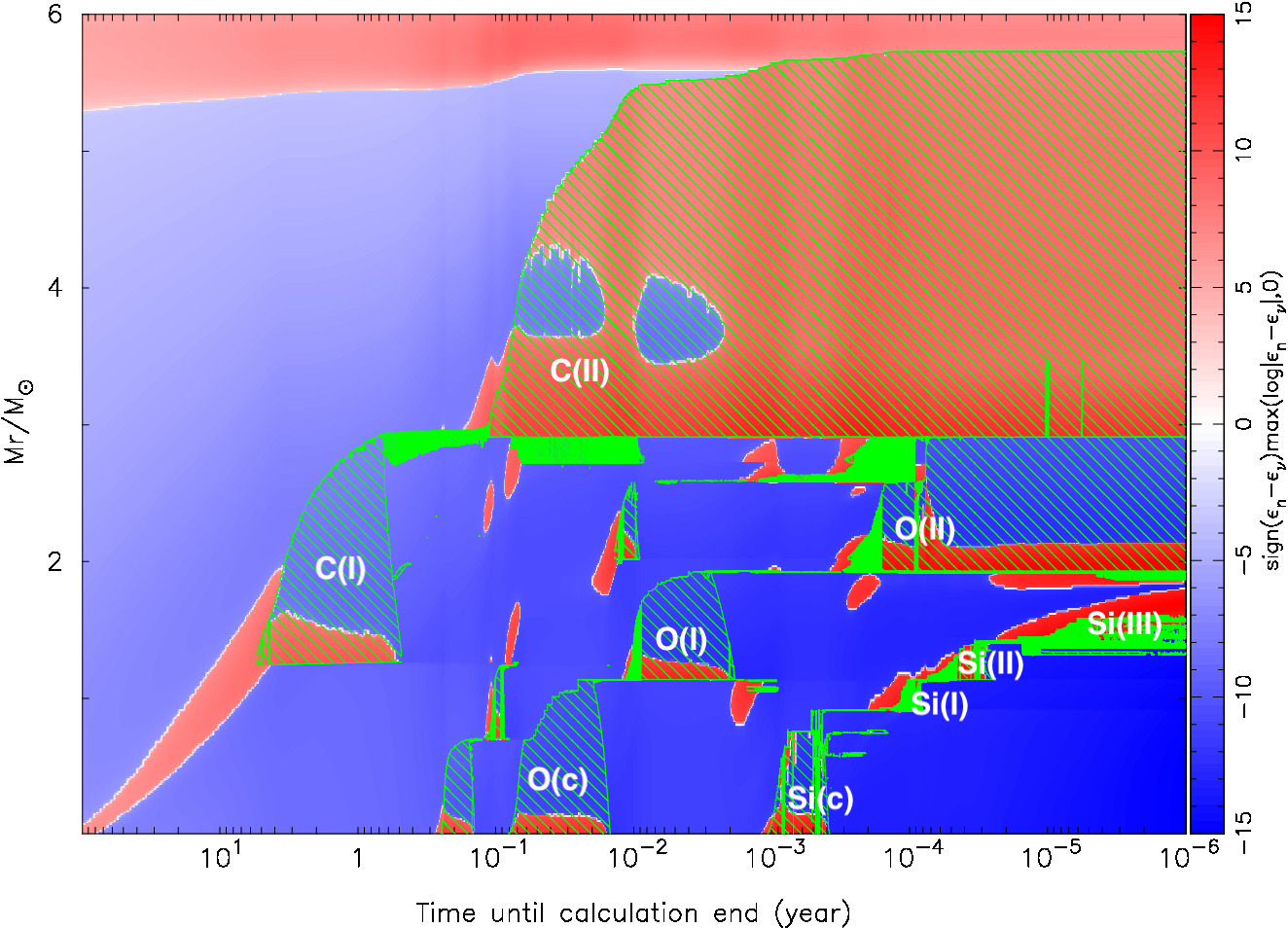}
\caption{Kippenhahn diagram of models 25M$_{\rm A}$ (left panel) and 25M (right panel) from the ignition of the central carbon burning until the calculation termination.}
\label{fig:Kip25}
\end{figure*}

\begin{figure*}
\centering
\includegraphics[width=0.48\linewidth]{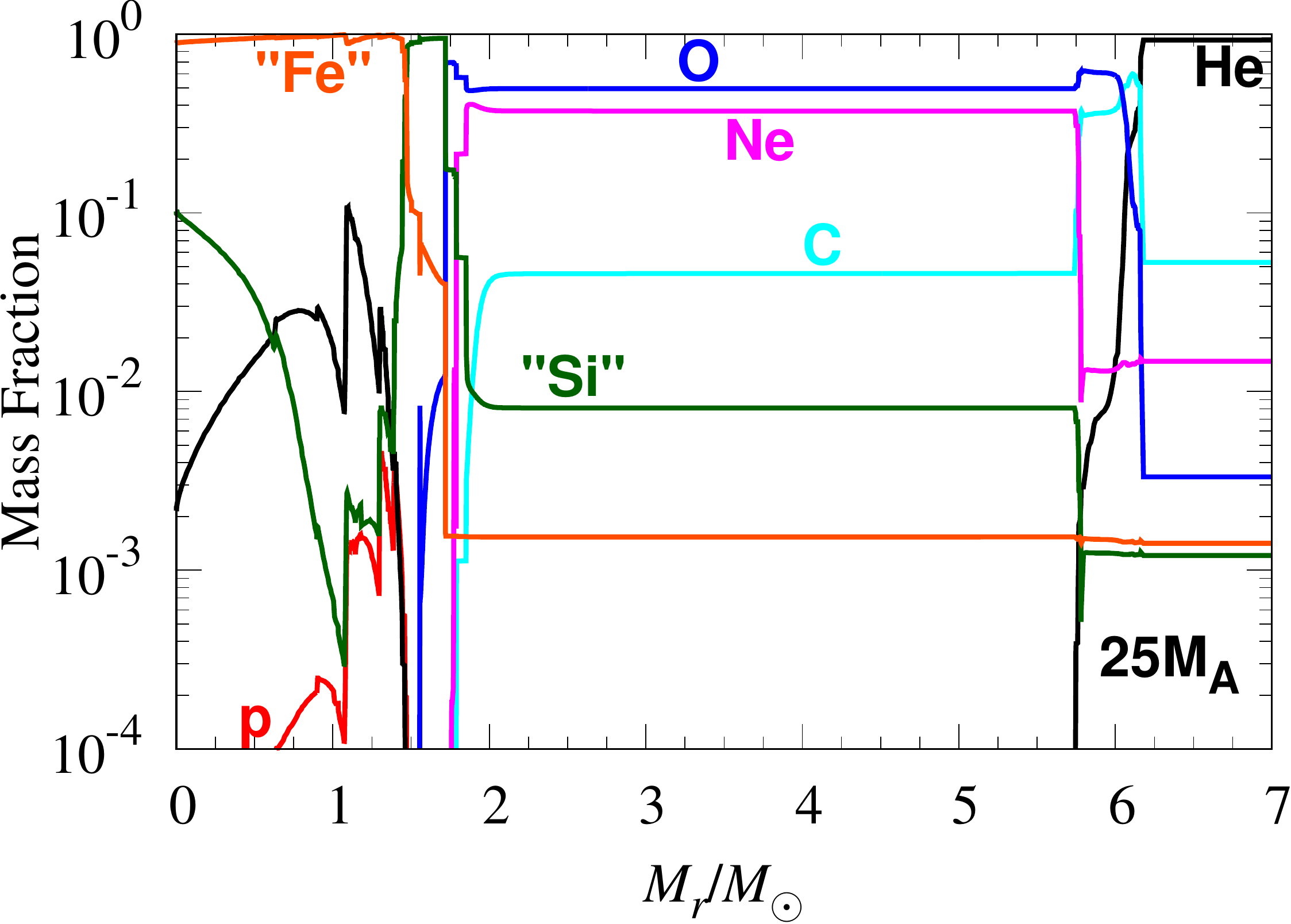}
\includegraphics[width=0.48\linewidth]{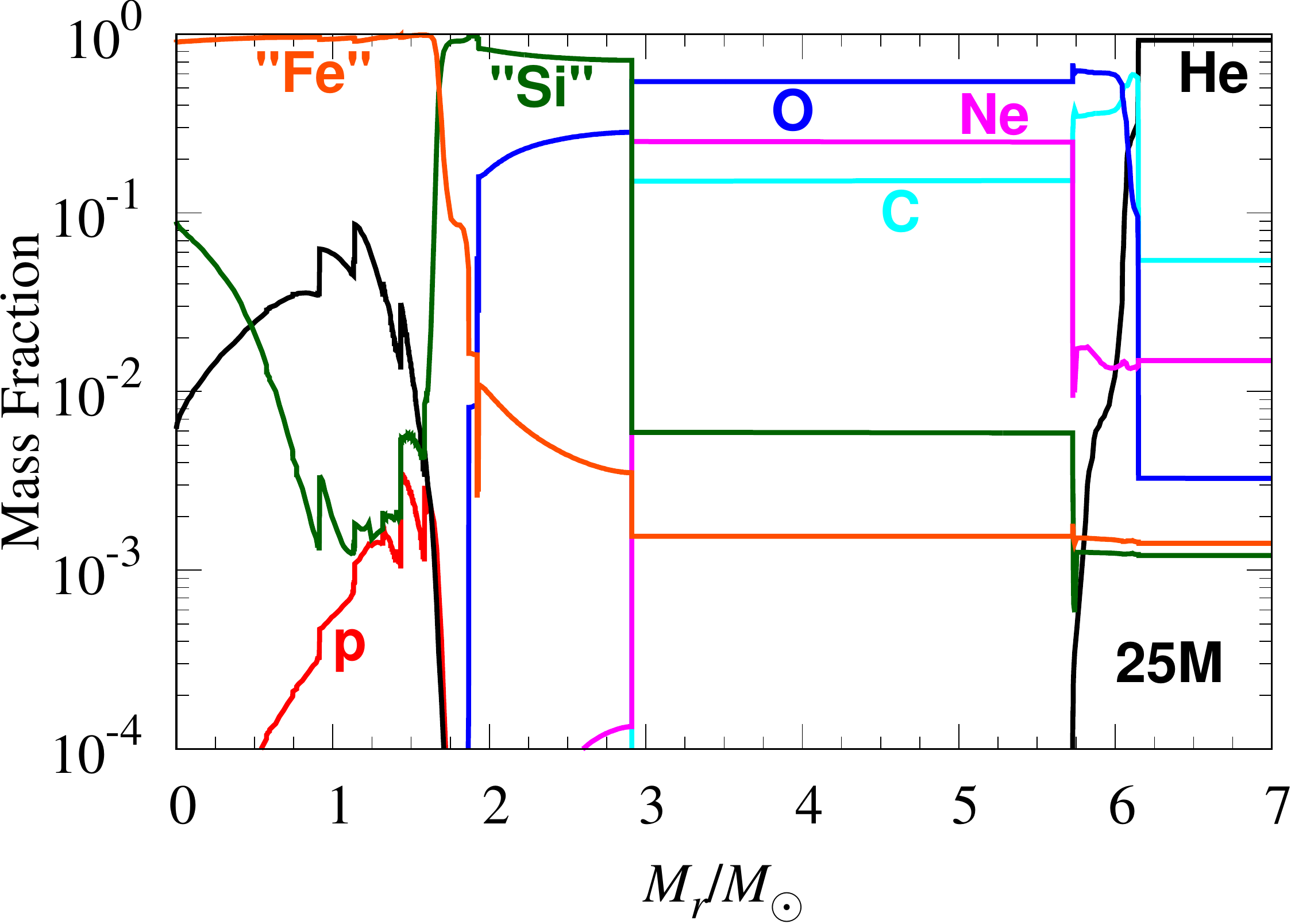}
\caption{Mass fraction distributions of the last time step of models 25M$_{\rm A}$ (left panel) and 25M (right panel).
Red, black, cyan, blue, magenta,green, and orange lines correspond to the mass fractions of p, He, C, O, Ne, intermediate elements with $Z=14$--20 denoted as ^^ ^^ Si", and iron-peak elements with $Z \ge 21$ denoted as ^^ ^^ Fe", respectively.
}
\label{fig:mf25}
\end{figure*}

We calculate the massive star evolution taking account of a weak overshoot after the helium burning for models in Sets L$_{\rm A}$ and M$_{\rm A}$.
Despite the small value compared with the hydrogen and helium burning, the overshoot in the advanced stage affects advanced evolution in a complicated way.
We show Kippenhahn diagram of models 25M$_{\rm A}$ and 25M in Figure \ref{fig:Kip25} for comparison.
In these models, the carbon core burning occurs radiatively and the first carbon shell burning (C(I)) ignites at $M_r \sim 1.3 M_\odot$.
In model 25M$_{\rm A}$ the convective region of the C(I) burning extends both inward and outward with the help of overshoot.
The convective C layer extends in the range 1.0--2.3 $M_\odot$.
Then, the second carbon shell burning (C(II)) ignites at $\sim 2.1 M_\odot$ and the convective region extends again inward and outward.
The inner convection boundary of the second carbon shell burning reaches 1.74 $M_\odot$.
This inner boundary restricts the region of the following burning.
The Si core is formed through the O core burning (O(c)) and the first (O(I)) and second (O(II)) oxygen shell burning.
The third oxygen shell burning (O(III)) extends the Si layer up to 1.67 $M_\odot$ but the boundary is still below the inner boundary of the second carbon shell burning.
The Si core (Si(c)) and the following four silicon shell burning (Si(I)-(IV)) form an Fe core of $\sim$1.56 $M_\odot$.
The left panel of Figure \ref{fig:mf25} shows the mass fraction distribution of model 25M$_{\rm A}$.
There is a SiO-rich layer between the Si layer and the O/Ne layer but it is very thin (the width is about $10^8$ cm).

Model 25M indicates the evolution different from model 25M$_{\rm A}$ from the first carbon shell burning (C(I)).
The first carbon shell burning ignites at $\sim$1.3 $M_\odot$ and the convective region extends outward to 2.9 $M_\odot$.
The convection does not extend inward in this model.
Then, the second carbon shell burning (C(II)) starts at 2.9 $M_\odot$.
Since the second carbon burning shell starts at a large radius, the O core burning (O(c)) and the first oxygen shell burning (O(I)) extend more effectively outward and form a larger Si core.
The Si core (Si(c)) and the following three silicon shell burning (Si(I)-(III)) form an Fe core of 1.68 $M_\odot$.
The second oxygen shell burning (O(II)) makes a large Si/O layer between the top of first oxygen shell burning (1.9 $M_\odot$) and the bottom of the second carbon shell burning (2.9 $M_\odot$).
The mass fraction distribution of model 25M is shown in the right panel of Figure \ref{fig:mf25}.
We see the large Si/O layer in this figure.

\begin{figure}[htbp]
\includegraphics[width=0.5\linewidth]{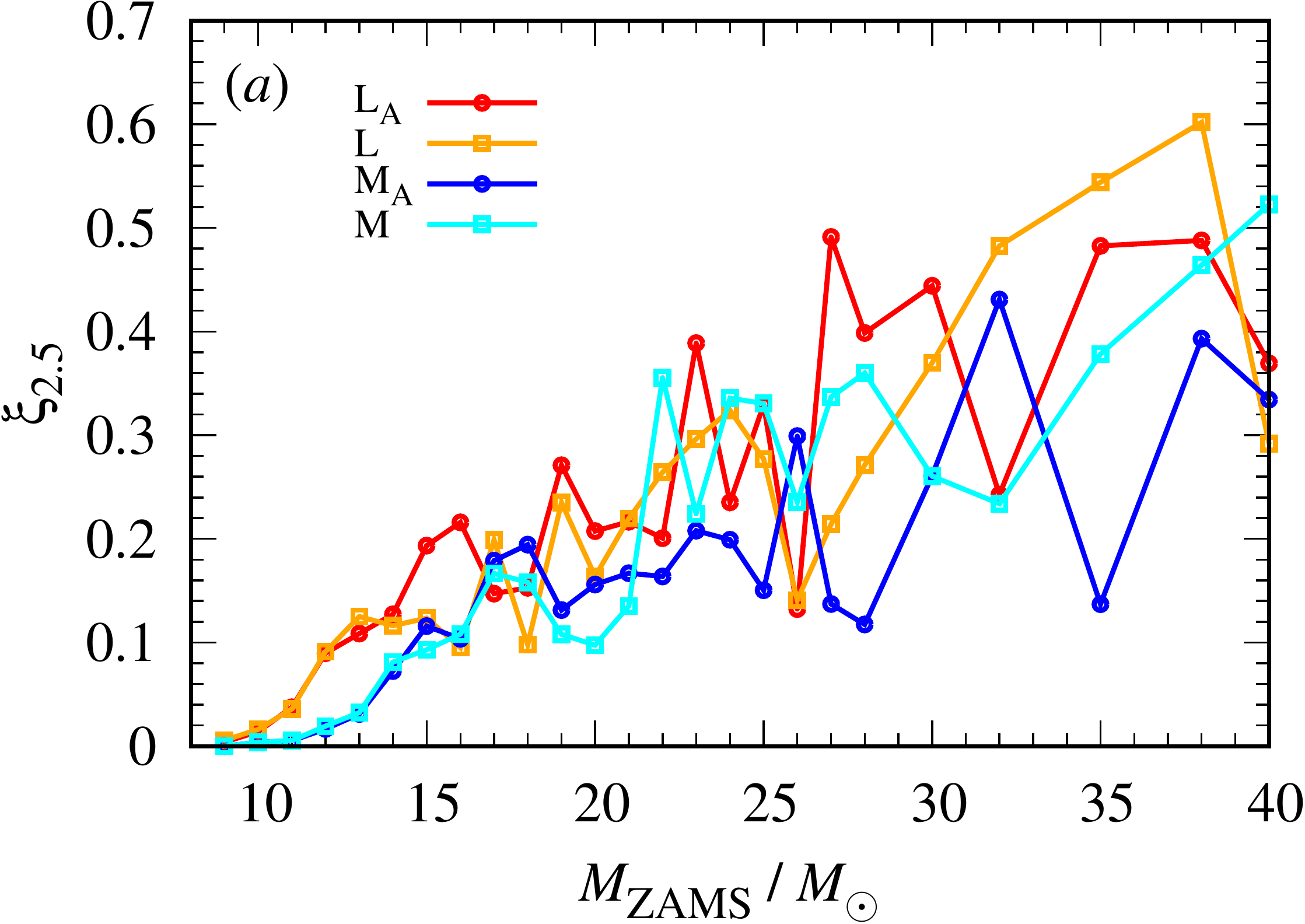}
\includegraphics[width=0.5\linewidth]{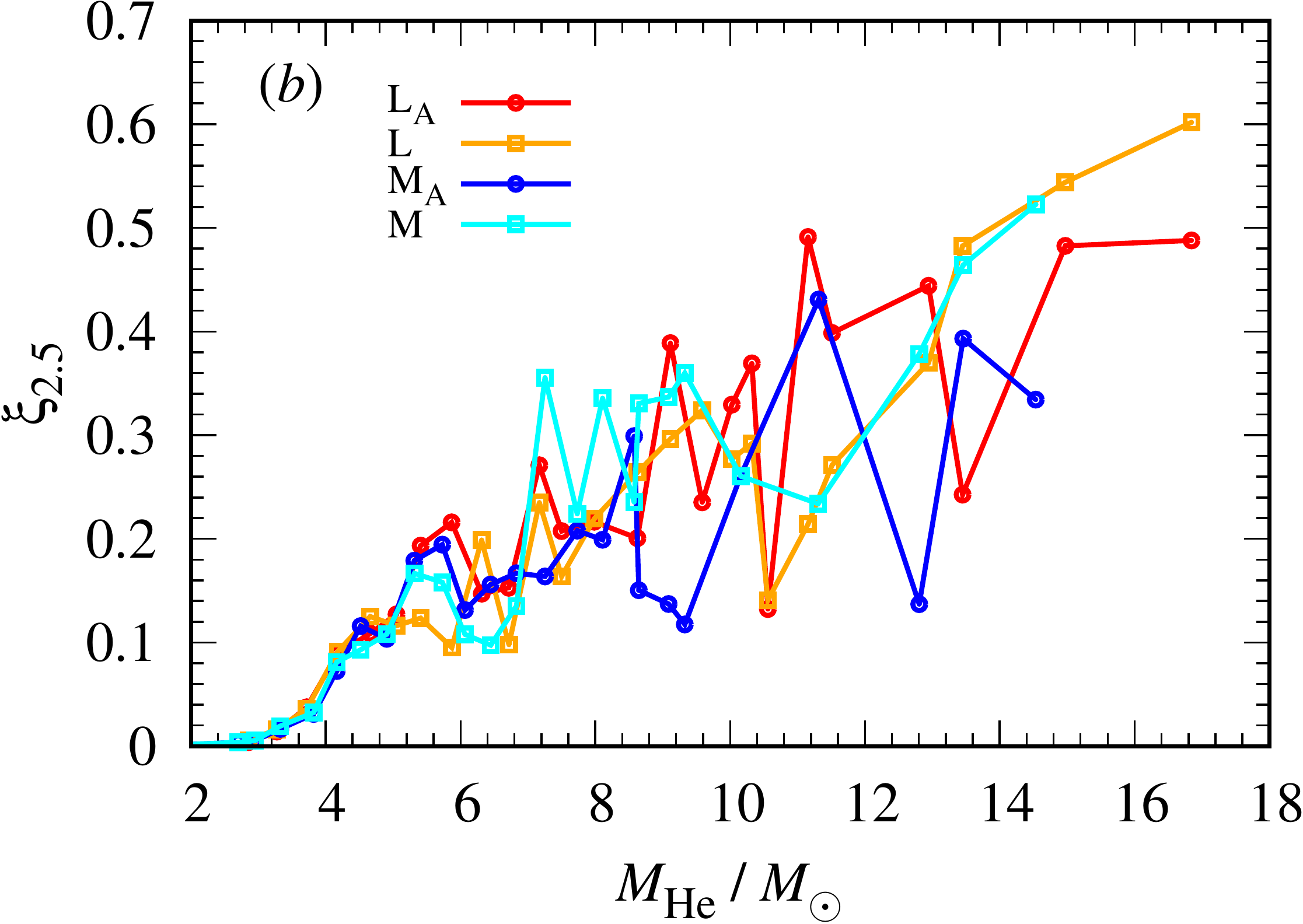}
\caption{Compactness parameter $\xi_{2.5}$ as a function of $M_{\rm ZAMS}$ (panel ({\it a})) and of $M_{\rm He}$ (panel ({\it b})).
\label{fig:f16}}
\end{figure}

The compactness parameter, $\xi_{2.5}$, is considered as a quantity that correlates to the dynamics during the gravitational collapse \citep{Oconnor11}.
It is defined as an inverse of the radius, inside which the mass of 2.5 $M_\odot$ is enclosed,
\begin{equation}
    \xi_{2.5} = \frac{2.5}{R(M_r=2.5 M_\odot)/1000{\rm km}}
\end{equation}
Several works have reported that a small compactness parameter is favorable for supernova explosions \citep{Nakamura15,Pejha15,Sukhbold16,Horiuchi14}, although the criterion has not yet converged \citep{Ugliano12,Suwa16,Ertl16,Burrows18}.
This parameter is also important to predict the neutrino emission \citep{Horiuchi17}.
Tight correlation of $\xi_{2.5}$ to the CO core mass has been shown \citep{Limongi18}, and detailed study on the compactness is performed by \citet{Sukhbold14,Sukhbold18}.
Effects of convective boundary mixing in the advanced evolutionary stages regarding $\xi_{2.5}$ is studied by \citet{Davis19}.

Figure \ref{fig:f16} shows the compactness parameter $\xi_{2.5}$ of our models.
Although the results show a large scatter, $\xi_{2.5}$ roughly increases with ZAMS mass.
The scatter is reduced when the x-axis is changed from the ZAMS mass to the He-core mass (panel ($b$)) as shown in \citet{Sukhbold18}.
\KT{As for the correlation between $\xi_{2.5}$ and the He core mass,
the scatter is small for the models with $M_{\rm He} \la 5 M_\odot$.
Besides, we do not see clear dependence of $\xi_{2.5}$} on $f_{\rm A}$ in this mass range.
For more massive models, the scatter is larger, and the different overshoot parameters also give an influence on $\xi_{2.5}$ in a complicated manner.

Because of the limited number of our models, it is difficult to make a detailed comparison between our results on the compactness parameter (Figure \ref{fig:f16}) with \citet{Sukhbold18} (see their Figure 8).  
We do see a diversity of $\xi_{2.5}$ in $M_{\rm He} \ga 10 M_\odot$, however, the range of $\xi_{2.5}$ for $M_{\rm He} \la 6 M_\odot$ is almost within the same range as in \citet{Sukhbold18}. Although a more systematic study needs to be done as in \citet{Sukhbold16,Sukhbold18}, the overall feature (e.g., the increasing trend of the compactness parameter with the ZAMS masses) is roughly in accordance with \citet{Sukhbold18}.
 
Finally, Figure \ref{fig:f18} shows the mass fraction distributions as a function of the radius at the last step of nine models, of which 2D hydrodynamics simulations were performed which are not shown in Figure \ref{fig:mfrad}.

\begin{figure}[htbp]
\includegraphics[width=0.5\linewidth]{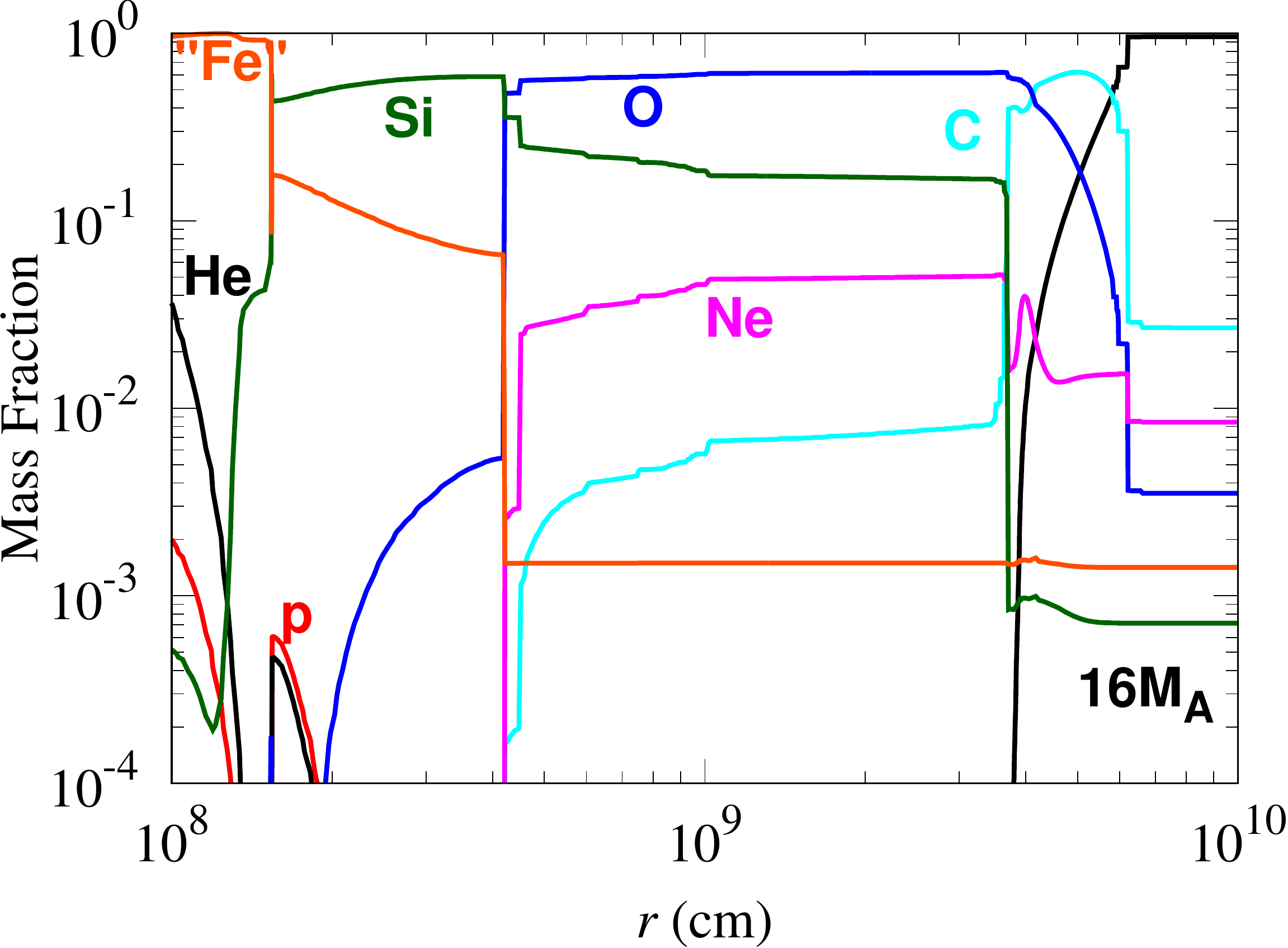}
\includegraphics[width=0.5\linewidth]{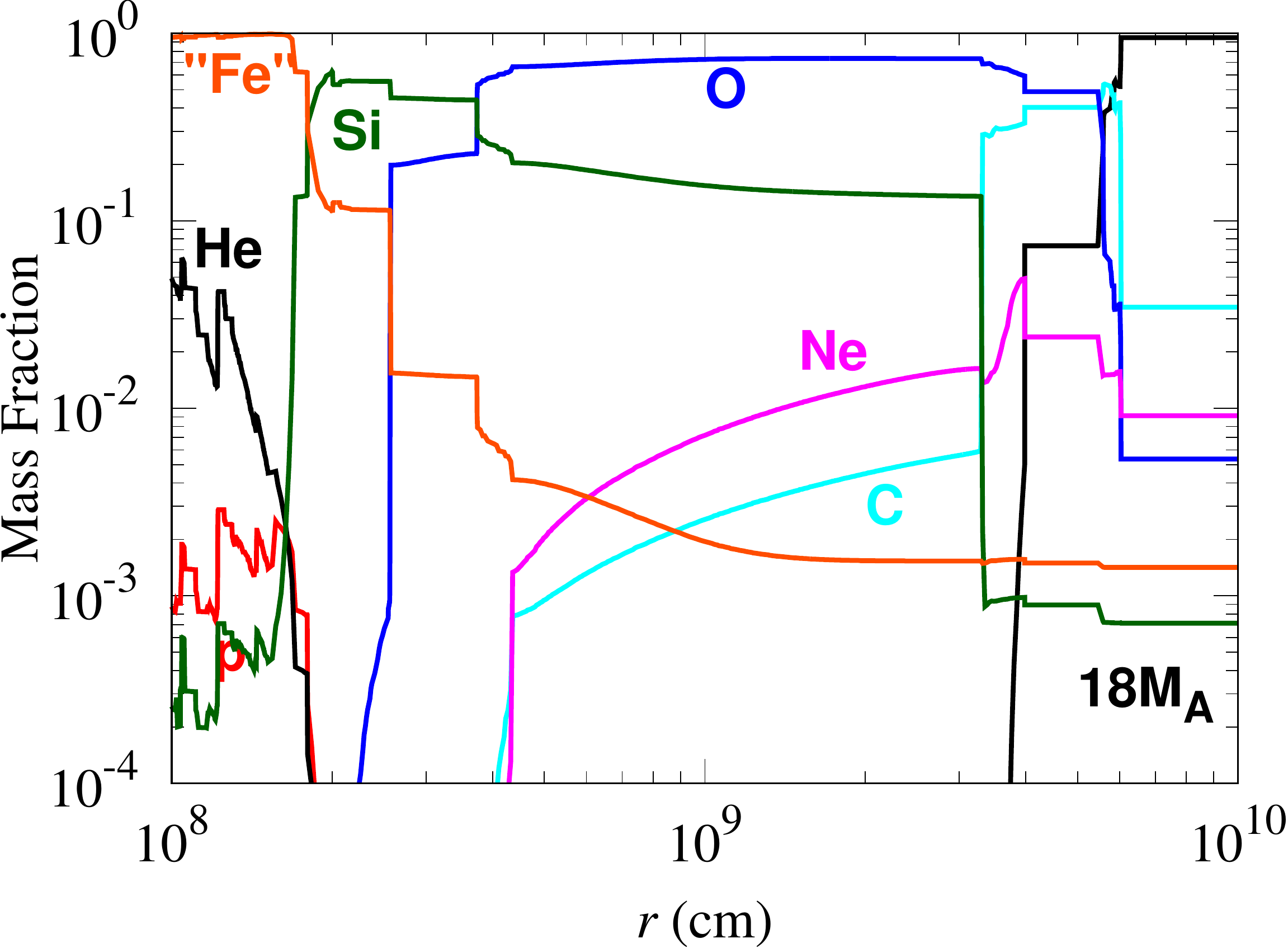}
\includegraphics[width=0.5\linewidth]{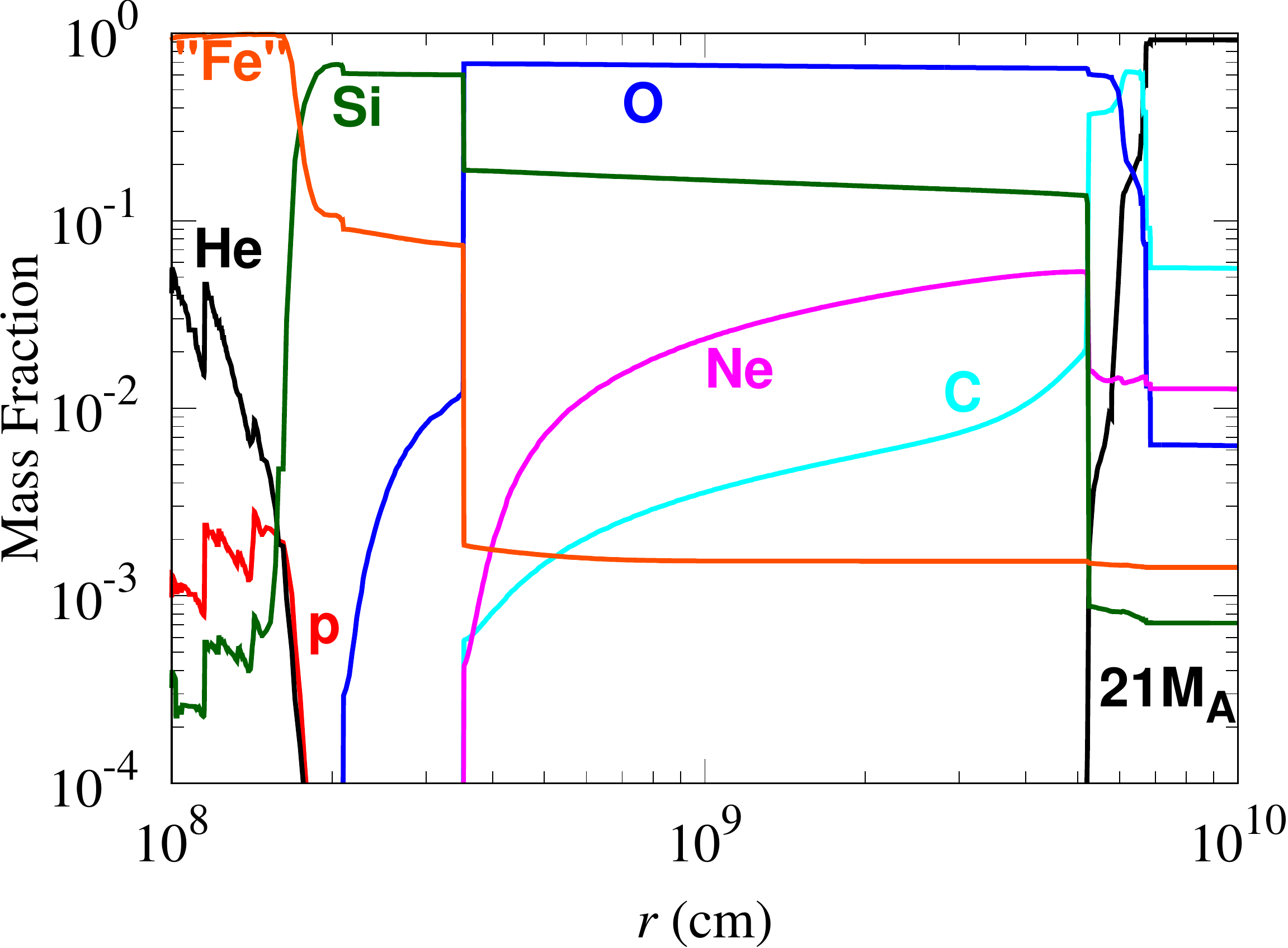}
\includegraphics[width=0.5\linewidth]{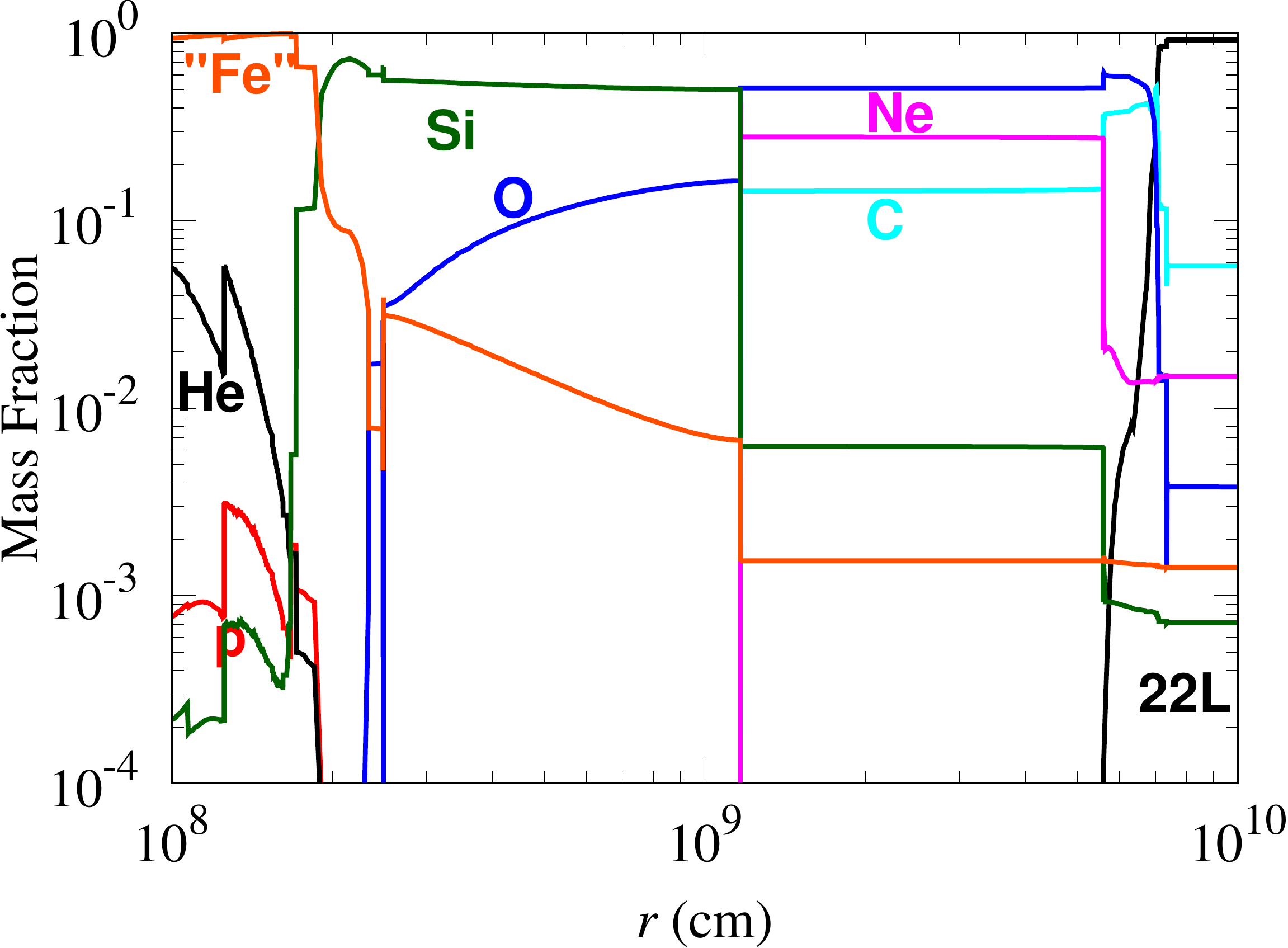}
\includegraphics[width=0.5\linewidth]{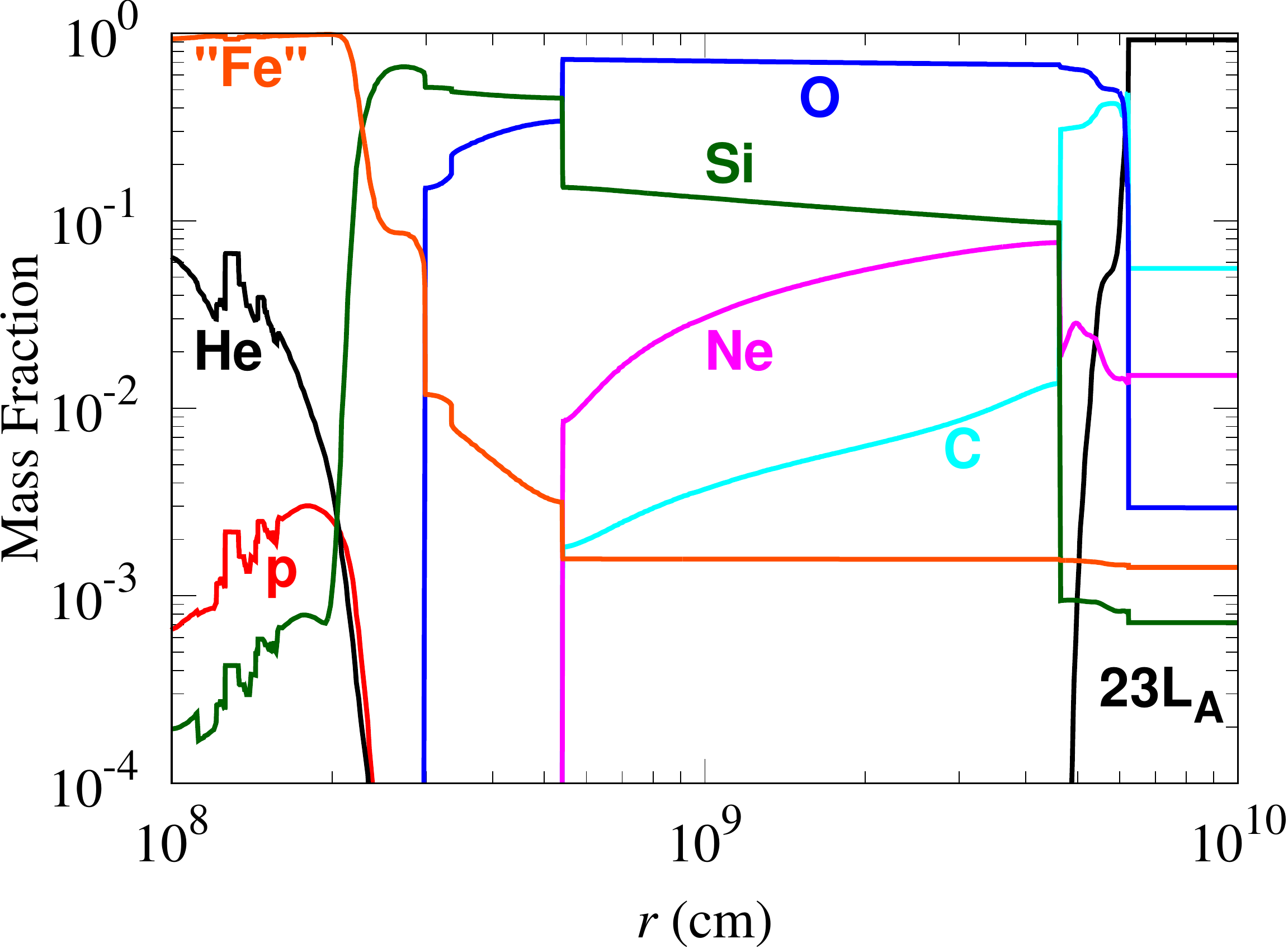}
\includegraphics[width=0.5\linewidth]{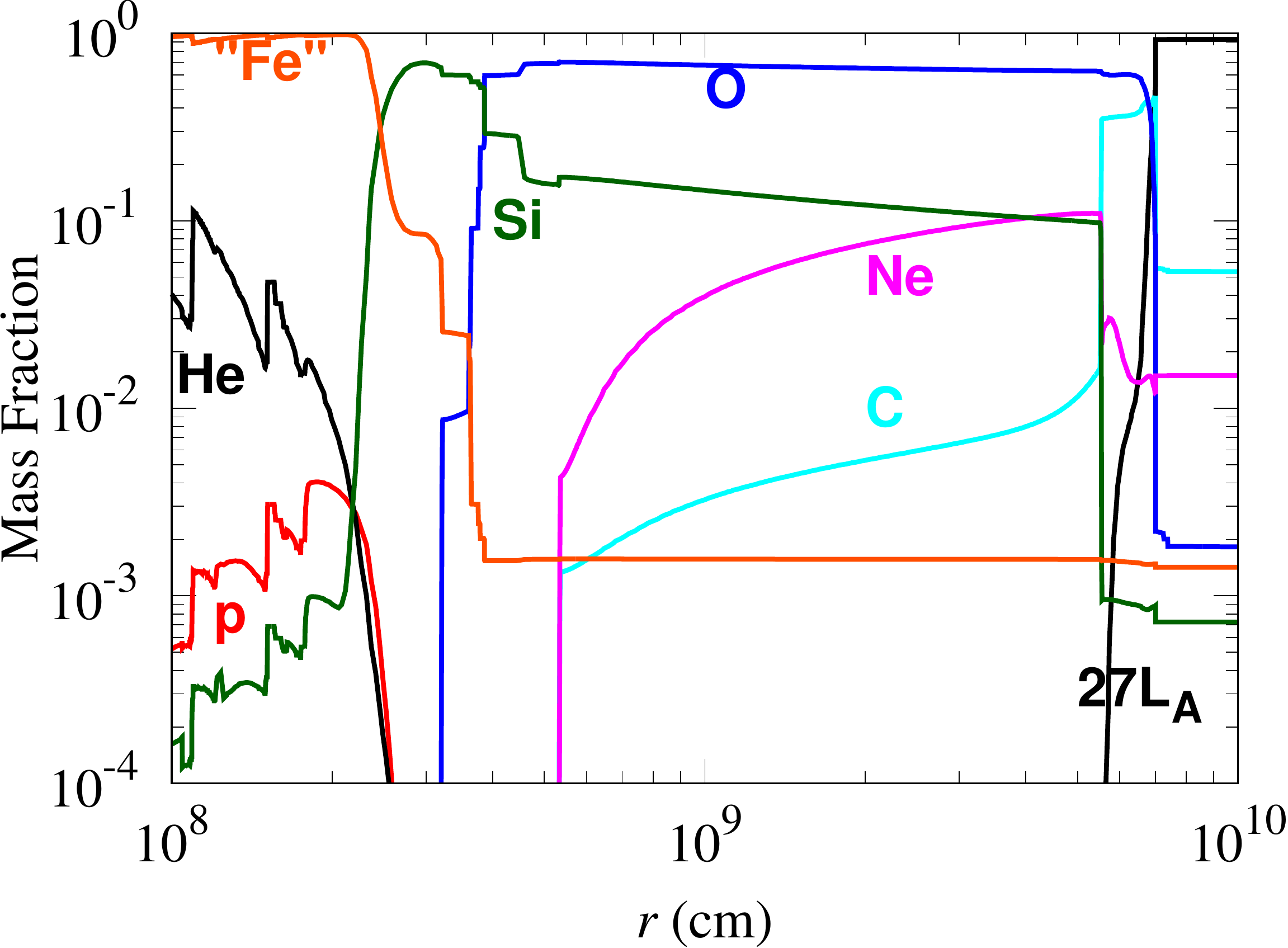}
\caption{Same as Figure \ref{fig:mfrad} but for models 16M$_{\rm A}$ (top left), 18M$_{\rm A}$ (top right), 21M$_{\rm A}$ (middle left), 22L (middle right),
23L$_{\rm A}$ (bottom left), and 27L$_{\rm A}$ (bottom right).
\label{fig:f18}}
\end{figure}

\addtocounter{figure}{-1}
\begin{figure}[htbp]
\includegraphics[width=0.5\linewidth]{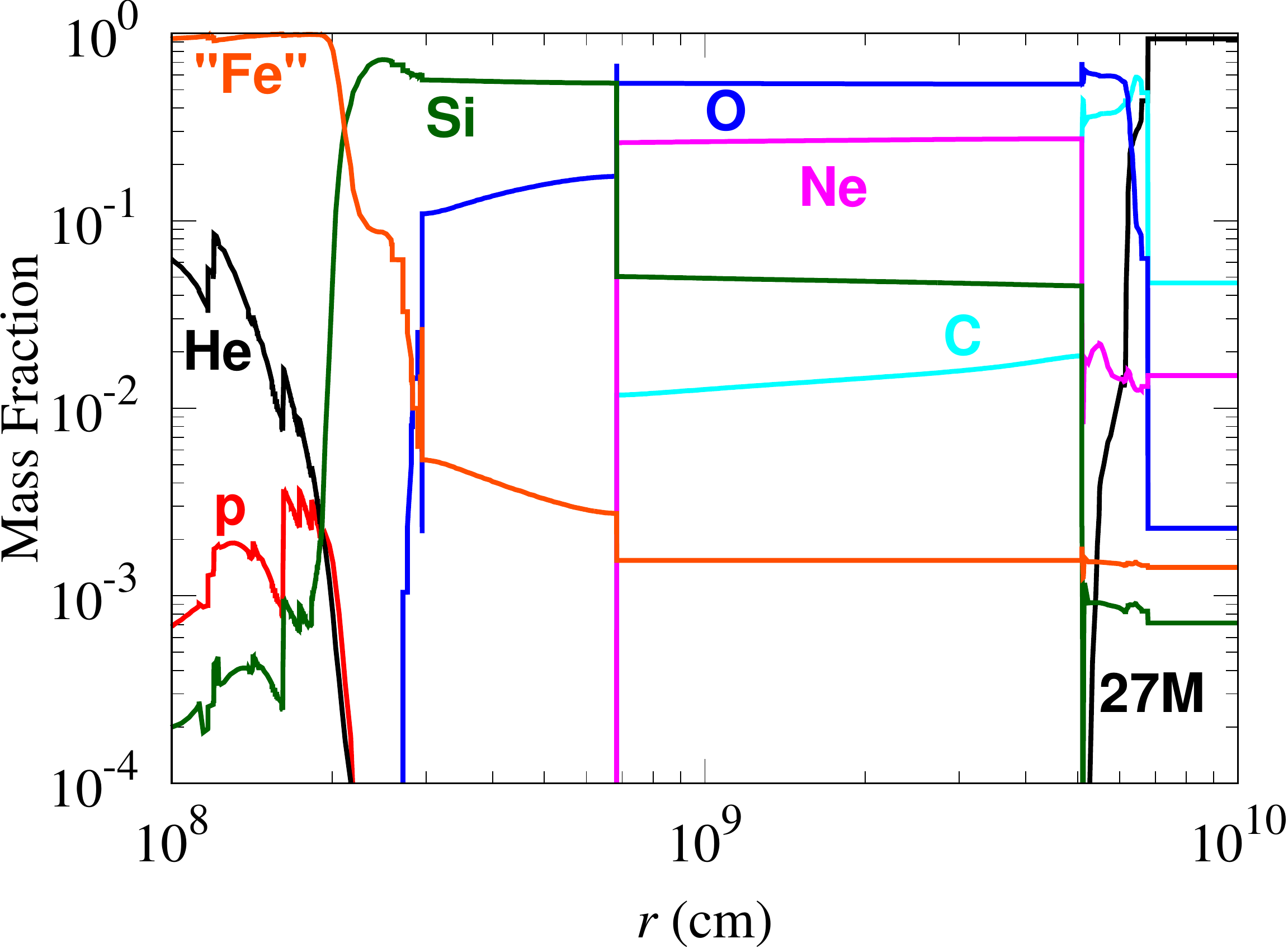}
\includegraphics[width=0.5\linewidth]{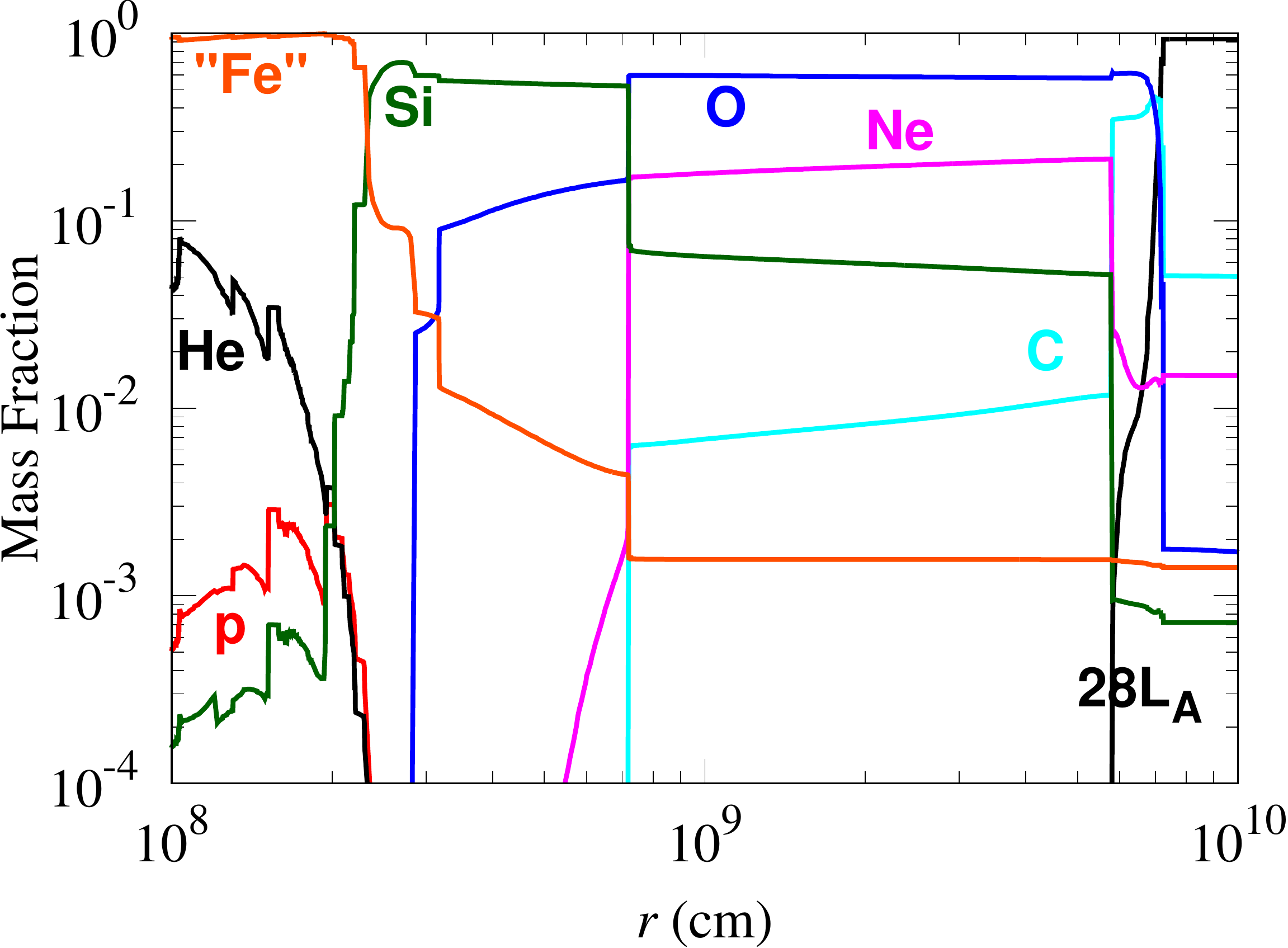}
\includegraphics[width=0.5\linewidth]{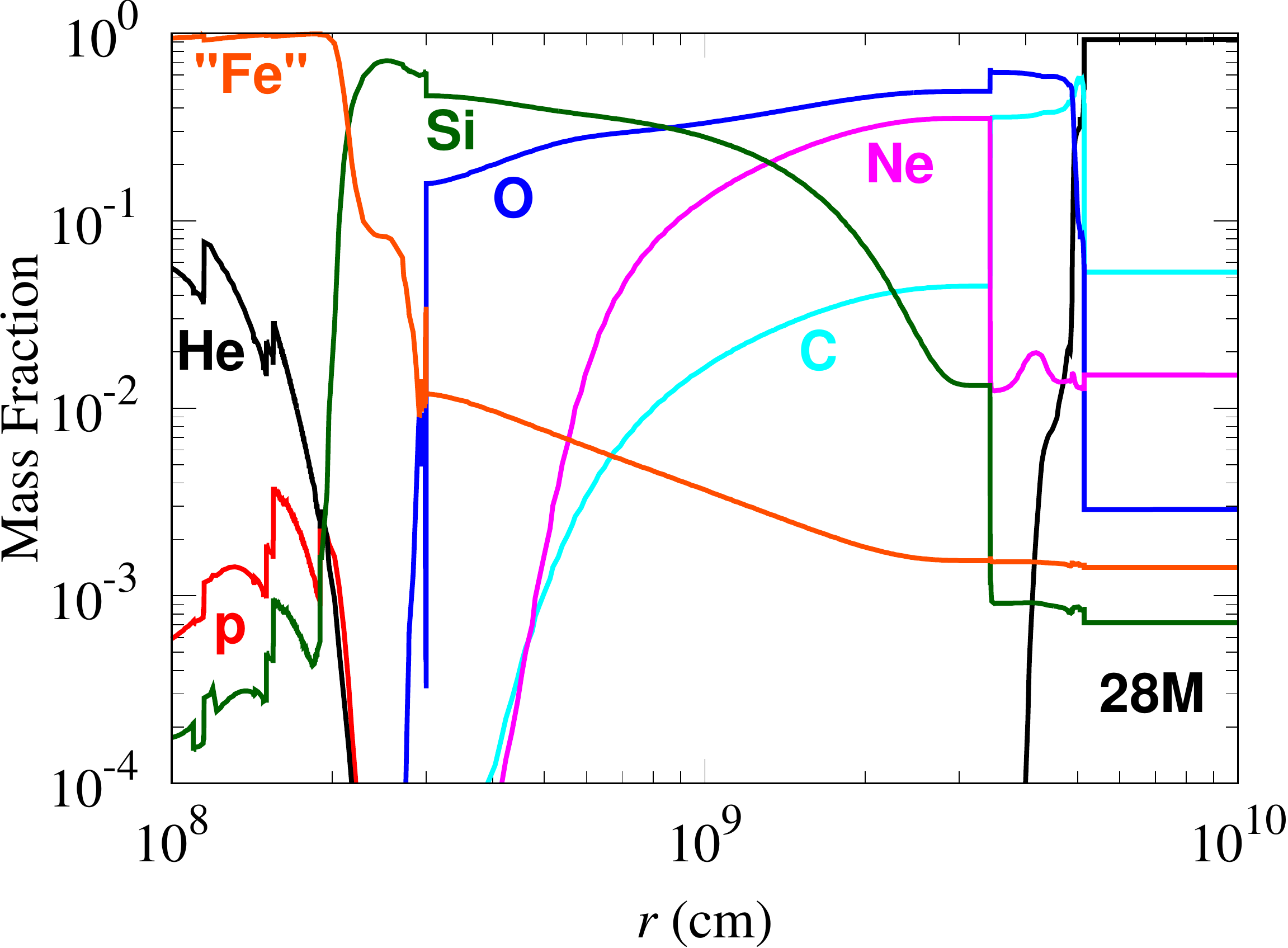}
\caption{({\it Continued.})
Same as Figure \ref{fig:mfrad} but for 27M (top left), 28L$_{\rm A}$ (top right), and 28M (bottom).
}
\end{figure}




\bibliographystyle{aasjournal} 
\bibliography{sample62.bbl}

\begin{thebibliography}{}
\expandafter\ifx\csname natexlab\endcsname\relax\def\natexlab#1{#1}\fi
\providecommand{\url}[1]{\href{#1}{#1}}

\bibitem[{{Abdikamalov} {et~al.}(2016){Abdikamalov}, {Zhaksylykov}, {Radice},
  \& {Berdibek}}]{ernazar16}
{Abdikamalov}, E., {Zhaksylykov}, A., {Radice}, D., \& {Berdibek}, S. 2016,
  \mnras, 461, 3864

\bibitem[{{Arnett}(1994)}]{arnett94}
{Arnett}, D. 1994, \apj, 427, 932

\bibitem[{{Arnett} \& {Meakin}(2011)}]{arnett11}
{Arnett}, W.~D., \& {Meakin}, C. 2011, \apj, 733, 78

\bibitem[{{Arnett} \& {Meakin}(2016)}]{arnett16}
---. 2016, Reports on Progress in Physics, 79, 102901

\bibitem[{{Asida} \& {Arnett}(2000)}]{asida00}
{Asida}, S.~M., \& {Arnett}, D. 2000, \apj, 545, 435

\bibitem[{{Bazan} \& {Arnett}(1994)}]{bazan94}
{Bazan}, G., \& {Arnett}, D. 1994, \apjl, 433, L41

\bibitem[{{Baz{\'a}n} \& {Arnett}(1998)}]{bazan98}
{Baz{\'a}n}, G., \& {Arnett}, D. 1998, \apj, 496, 316

\bibitem[{{Beacom} \& {Vagins}(2004)}]{Beacom04}
{Beacom}, J.~F., \& {Vagins}, M.~R. 2004, Physical Review Letters, 93, 171101

\bibitem[{{Bethe} \& {Wilson}(1985)}]{bethe85}
{Bethe}, H.~A., \& {Wilson}, J.~R. 1985, \apj, 295, 14

\bibitem[{{Blondin} {et~al.}(2003){Blondin}, {Mezzacappa}, \&
  {DeMarino}}]{Blondin03}
{Blondin}, J.~M., {Mezzacappa}, A., \& {DeMarino}, C. 2003, \apj, 584, 971

\bibitem[{{Bollig} {et~al.}(2017){Bollig}, {Janka}, {Lohs},
  {Mart{\'{\i}}nez-Pinedo}, {Horowitz}, \& {Melson}}]{bollig17}
{Bollig}, R., {Janka}, H.-T., {Lohs}, A., {et~al.} 2017, Physical Review
  Letters, 119, 242702

\bibitem[{{Brott} {et~al.}(2011){Brott}, {de Mink}, {Cantiello}, {Langer}, {de
  Koter}, {Evans}, {Hunter}, {Trundle}, \& {Vink}}]{Brott11}
{Brott}, I., {de Mink}, S.~E., {Cantiello}, M., {et~al.} 2011, \aap, 530, A115

\bibitem[{{Burrows} {et~al.}(2018){Burrows}, {Vartanyan}, {Dolence}, {Skinner},
  \& {Radice}}]{Burrows18}
{Burrows}, A., {Vartanyan}, D., {Dolence}, J.~C., {Skinner}, M.~A., \&
  {Radice}, D. 2018, \ssr, 214, 33

\bibitem[{{Campbell} {et~al.}(2016){Campbell}, {Constantino}, {D'Orazi},
  {Meakin}, {Stello}, {Christensen-Dalsgaard}, {Kuehn}, {De Silva}, {Arnett},
  {Lattanzio}, \& {MacLean}}]{simon15}
{Campbell}, S.~W., {Constantino}, T.~N., {D'Orazi}, V., {et~al.} 2016,
  Astronomische Nachrichten, 337, 788

\bibitem[{{Chatzopoulos} {et~al.}(2016){Chatzopoulos}, {Couch}, {Arnett}, \&
  {Timmes}}]{chatz16}
{Chatzopoulos}, E., {Couch}, S.~M., {Arnett}, W.~D., \& {Timmes}, F.~X. 2016,
  \apj, 822, 61

\bibitem[{{Chatzopoulos} {et~al.}(2014){Chatzopoulos}, {Graziani}, \&
  {Couch}}]{chatz14}
{Chatzopoulos}, E., {Graziani}, C., \& {Couch}, S.~M. 2014, \apj, 795, 92

\bibitem[{{Collins} {et~al.}(2018){Collins}, {M{\"u}ller}, \&
  {Heger}}]{Collins18}
{Collins}, C., {M{\"u}ller}, B., \& {Heger}, A. 2018, \mnras, 473, 1695

\bibitem[{{Couch}(2017)}]{couch_book}
{Couch}, S.~M. 2017, {Influence of Non-spherical Initial Stellar Structure on
  the Core-Collapse Supernova Mechanism}, ed. A.~W. {Alsabti} \& P.~{Murdin},
  1791

\bibitem[{{Couch} {et~al.}(2015){Couch}, {Chatzopoulos}, {Arnett}, \&
  {Timmes}}]{Couch15}
{Couch}, S.~M., {Chatzopoulos}, E., {Arnett}, W.~D., \& {Timmes}, F.~X. 2015,
  \apjl, 808, L21

\bibitem[{{Couch} \& {Ott}(2013)}]{couch_ott13}
{Couch}, S.~M., \& {Ott}, C.~D. 2013, \apjl, 778, L7

\bibitem[{{Couch} \& {Ott}(2015)}]{couch_ott15}
---. 2015, \apj, 799, 5

\bibitem[{{Cristini} {et~al.}(2019){Cristini}, {Hirschi}, {Meakin}, {Arnett},
  {Georgy}, \& {Walkington}}]{cristini19}
{Cristini}, A., {Hirschi}, R., {Meakin}, C., {et~al.} 2019, \mnras, 484, 4645

\bibitem[{{Cristini} {et~al.}(2017){Cristini}, {Meakin}, {Hirschi}, {Arnett},
  {Georgy}, {Viallet}, \& {Walkington}}]{cristini17}
{Cristini}, A., {Meakin}, C., {Hirschi}, R., {et~al.} 2017, \mnras, 471, 279

\bibitem[{{Crowther}(2007)}]{Crowther07}
{Crowther}, P.~A. 2007, \araa, 45, 177

\bibitem[{{Cyburt} {et~al.}(2010){Cyburt}, {Amthor}, {Ferguson}, {Meisel}, \&
  {others}}]{Cyburt10}
{Cyburt}, R.~H., {Amthor}, A.~M., {Ferguson}, R., {Meisel}, Z., \& {others}.
  2010, \apjs, 189, 240

\bibitem[{{Davis} {et~al.}(2019){Davis}, {Jones}, \& {Herwig}}]{Davis19}
{Davis}, A., {Jones}, S., \& {Herwig}, F. 2019, \mnras, 484, 3921

\bibitem[{{de Jager} {et~al.}(1988){de Jager}, {Nieuwenhuijzen}, \& {van der
  Hucht}}]{deJager88}
{de Jager}, C., {Nieuwenhuijzen}, S., \& {van der Hucht}, K.~A. 1988, \aaps,
  72, 259

\bibitem[{{Decressin} {et~al.}(2009){Decressin}, {Mathis}, {Palacios}, {Siess},
  {Talon}, {Charbonnel}, \& {Zahn}}]{Decressin09}
{Decressin}, T., {Mathis}, S., {Palacios}, A., {et~al.} 2009, \aap, 495, 271

\bibitem[{{Ebinger} {et~al.}(2019){Ebinger}, {Curtis}, {Fr{\"o}hlich},
  {Hempel}, {Perego}, {Liebend{\"o}rfer}, \& {Thielemann}}]{Ebinger18}
{Ebinger}, K., {Curtis}, S., {Fr{\"o}hlich}, C., {et~al.} 2019, \apj, 870, 1

\bibitem[{{Ekstr{\"o}m} {et~al.}(2012){Ekstr{\"o}m}, {Georgy}, {Eggenberger},
  {Meynet}, {Mowlavi}, {Wyttenbach}, {Granada}, {Decressin}, {Hirschi},
  {Frischknecht}, {Charbonnel}, \& {Maeder}}]{Ekstroem12}
{Ekstr{\"o}m}, S., {Georgy}, C., {Eggenberger}, P., {et~al.} 2012, \aap, 537,
  A146

\bibitem[{{Ertl} {et~al.}(2016){Ertl}, {Janka}, {Woosley}, {Sukhbold}, \&
  {Ugliano}}]{Ertl16}
{Ertl}, T., {Janka}, H.-T., {Woosley}, S.~E., {Sukhbold}, T., \& {Ugliano}, M.
  2016, \apj, 818, 124

\bibitem[{{Fern{\'a}ndez} {et~al.}(2014){Fern{\'a}ndez}, {M{\"u}ller},
  {Foglizzo}, \& {Janka}}]{rodorigo14}
{Fern{\'a}ndez}, R., {M{\"u}ller}, B., {Foglizzo}, T., \& {Janka}, H.-T. 2014,
  \mnras, 440, 2763

\bibitem[{{Foglizzo} {et~al.}(2015){Foglizzo}, {Kazeroni}, {Guilet}, {Masset},
  {Gonz{\'a}lez}, {Krueger}, {Novak}, {Oertel}, {Margueron}, {Faure}, {Martin},
  {Blottiau}, {Peres}, \& {Durand}}]{foglizzo_15}
{Foglizzo}, T., {Kazeroni}, R., {Guilet}, J., {et~al.} 2015, \pasa, 32, e009

\bibitem[{{Fuller} {et~al.}(2015){Fuller}, {Cantiello}, {Lecoanet}, \&
  {Quataert}}]{fuller15}
{Fuller}, J., {Cantiello}, M., {Lecoanet}, D., \& {Quataert}, E. 2015, \apj,
  810, 101

\bibitem[{{Gando} {et~al.}(2013){Gando}, {Gando}, {Hanakago}, {Ikeda}, {Inoue},
  {Ishidoshiro}, {Ishikawa}, {Koga}, {Matsuda}, {Matsuda}, {Mitsui}, {Motoki},
  {Nakamura}, {Obata}, {Oki}, {Oki}, {Otani}, {Shimizu}, {Shirai}, {Suzuki},
  {Takemoto}, {Tamae}, {Ueshima}, {Watanabe}, {Xu}, {Yamada}, {Yamauchi},
  {Yoshida}, {Kozlov}, {Yoshida}, {Piepke}, {Banks}, {Fujikawa}, {Han},
  {O'Donnell}, {Berger}, {Learned}, {Matsuno}, {Sakai}, {Efremenko},
  {Karwowski}, {Markoff}, {Tornow}, {Detwiler}, {Enomoto}, \&
  {Decowski}}]{Gando13}
{Gando}, A., {Gando}, Y., {Hanakago}, H., {et~al.} 2013, \prd, 88, 033001

\bibitem[{{Guilet} \& {M{\"u}ller}(2015)}]{jerome15}
{Guilet}, J., \& {M{\"u}ller}, E. 2015, \mnras, 450, 2153

\bibitem[{{Hanke} {et~al.}(2013){Hanke}, {M{\"u}ller}, {Wongwathanarat},
  {Marek}, \& {Janka}}]{Hanke13}
{Hanke}, F., {M{\"u}ller}, B., {Wongwathanarat}, A., {Marek}, A., \& {Janka},
  H.-T. 2013, \apj, 770, 66

\bibitem[{{Harada} {et~al.}(2018){Harada}, {Nagakura}, {Iwakami}, {Okawa},
  {Furusawa}, {Matsufuru}, {Sumiyoshi}, \& {Yamada}}]{harada18}
{Harada}, A., {Nagakura}, H., {Iwakami}, W., {et~al.} 2018, arXiv e-prints,
  arXiv:1810.12316

\bibitem[{{Horiuchi} {et~al.}(2017){Horiuchi}, {Nakamura}, {Takiwaki}, \&
  {Kotake}}]{Horiuchi17}
{Horiuchi}, S., {Nakamura}, K., {Takiwaki}, T., \& {Kotake}, K. 2017, Journal
  of Physics G Nuclear Physics, 44, 114001

\bibitem[{{Horiuchi} {et~al.}(2014){Horiuchi}, {Nakamura}, {Takiwaki},
  {Kotake}, \& {Tanaka}}]{Horiuchi14}
{Horiuchi}, S., {Nakamura}, K., {Takiwaki}, T., {Kotake}, K., \& {Tanaka}, M.
  2014, \mnras, 445, L99

\bibitem[{{Hyper-Kamiokande Proto-Collaboration}
  {et~al.}(2018){Hyper-Kamiokande Proto-Collaboration}, {:}, {Abe}, {Abe},
  {Aihara}, {Aimi}, {Akutsu}, {Andreopoulos}, {Anghel}, {Anthony}, \&
  et~al.}]{HKPC18}
{Hyper-Kamiokande Proto-Collaboration}, {:}, {Abe}, K., {et~al.} 2018, arXiv
  e-prints, arXiv:1805.04163

\bibitem[{{Itoh} {et~al.}(1996){Itoh}, {Hayashi}, {Nishikawa}, \&
  {Kohyama}}]{Itoh96}
{Itoh}, N., {Hayashi}, H., {Nishikawa}, A., \& {Kohyama}, Y. 1996, \apjs, 102,
  411

\bibitem[{{Janka} {et~al.}(2016){Janka}, {Melson}, \& {Summa}}]{janka16_rev}
{Janka}, H.-T., {Melson}, T., \& {Summa}, A. 2016, Annual Review of Nuclear and
  Particle Science, 66, 341

\bibitem[{{Jones} {et~al.}(2017){Jones}, {Andrassy}, {Sandalski}, {Davis},
  {Woodward}, \& {Herwig}}]{jones17}
{Jones}, S., {Andrassy}, R., {Sandalski}, S., {et~al.} 2017, \mnras, 465, 2991

\bibitem[{{Just} {et~al.}(2018){Just}, {Bollig}, {Janka}, {Obergaulinger},
  {Glas}, \& {Nagataki}}]{just18}
{Just}, O., {Bollig}, R., {Janka}, H.-T., {et~al.} 2018, \mnras, 481, 4786

\bibitem[{{Kato} {et~al.}(2017){Kato}, {Nagakura}, {Furusawa}, {Takahashi},
  {Umeda}, {Yoshida}, {Ishidoshiro}, \& {Yamada}}]{kato17}
{Kato}, C., {Nagakura}, H., {Furusawa}, S., {et~al.} 2017, \apj, 848, 48

\bibitem[{{Kippenhahn} {et~al.}(2012){Kippenhahn}, {Weigert}, \&
  {Weiss}}]{kippenhahn12}
{Kippenhahn}, R., {Weigert}, A., \& {Weiss}, A. 2012, {Stellar Structure and
  Evolution}, doi:10.1007/978-3-642-30304-3

\bibitem[{{Kotake} {et~al.}(2018){Kotake}, {Takiwaki}, {Fischer}, {Nakamura},
  \& {Mart{\'{\i}}nez-Pinedo}}]{kotake18}
{Kotake}, K., {Takiwaki}, T., {Fischer}, T., {Nakamura}, K., \&
  {Mart{\'{\i}}nez-Pinedo}, G. 2018, \apj, 853, 170

\bibitem[{{Kuhlen} {et~al.}(2003){Kuhlen}, {Woosley}, \&
  {Glatzmaier}}]{kuhlen03}
{Kuhlen}, M., {Woosley}, W.~E., \& {Glatzmaier}, G.~A. 2003, in Astronomical
  Society of the Pacific Conference Series, Vol. 293, 3D Stellar Evolution, ed.
  S.~{Turcotte}, S.~C. {Keller}, \& R.~M. {Cavallo}, 147

\bibitem[{{Kuroda} {et~al.}(2012){Kuroda}, {Kotake}, \& {Takiwaki}}]{KurodaT12}
{Kuroda}, T., {Kotake}, K., \& {Takiwaki}, T. 2012, \apj, 755, 11

\bibitem[{{Kuroda} {et~al.}(2016){Kuroda}, {Takiwaki}, \& {Kotake}}]{KurodaT16}
{Kuroda}, T., {Takiwaki}, T., \& {Kotake}, K. 2016, \apjs, 222, 20

\bibitem[{{Lentz} {et~al.}(2015){Lentz}, {Bruenn}, {Hix}, {Mezzacappa},
  {Messer}, {Endeve}, {Blondin}, {Harris}, {Marronetti}, \&
  {Yakunin}}]{Lentz15}
{Lentz}, E.~J., {Bruenn}, S.~W., {Hix}, W.~R., {et~al.} 2015, \apjl, 807, L31

\bibitem[{{Limongi} \& {Chieffi}(2018)}]{Limongi18}
{Limongi}, M., \& {Chieffi}, A. 2018, \apjs, 237, 13

\bibitem[{{Maeder} \& {Meynet}(1989)}]{Maeder89}
{Maeder}, A., \& {Meynet}, G. 1989, \aap, 210, 155

\bibitem[{{Marek} \& {Janka}(2009)}]{Marek09}
{Marek}, A., \& {Janka}, H.-T. 2009, \apj, 694, 664

\bibitem[{{Martins} \& {Palacios}(2013)}]{Martins13}
{Martins}, F., \& {Palacios}, A. 2013, \aap, 560, A16

\bibitem[{{Masada} {et~al.}(2015){Masada}, {Takiwaki}, \& {Kotake}}]{masada15}
{Masada}, Y., {Takiwaki}, T., \& {Kotake}, K. 2015, \apjl, 798, L22

\bibitem[{{Meakin} \& {Arnett}(2006)}]{meakin06}
{Meakin}, C.~A., \& {Arnett}, D. 2006, \apjl, 637, L53

\bibitem[{{Meakin} \& {Arnett}(2007)}]{meakin07}
---. 2007, \apj, 667, 448

\bibitem[{{Melson} {et~al.}(2015{\natexlab{a}}){Melson}, {Janka}, {Bollig},
  {Hanke}, {Marek}, \& {M{\"u}ller}}]{melson15b}
{Melson}, T., {Janka}, H.-T., {Bollig}, R., {et~al.} 2015{\natexlab{a}}, \apjl,
  808, L42

\bibitem[{{Melson} {et~al.}(2015{\natexlab{b}}){Melson}, {Janka}, \&
  {Marek}}]{melson15a}
{Melson}, T., {Janka}, H.-T., \& {Marek}, A. 2015{\natexlab{b}}, \apjl, 801,
  L24

\bibitem[{{Mignone}(2014)}]{Mignone14}
{Mignone}, A. 2014, Journal of Computational Physics, 270, 784

\bibitem[{{Mirizzi} {et~al.}(2016){Mirizzi}, {Tamborra}, {Janka}, {Saviano},
  {Scholberg}, {Bollig}, {H{\"u}depohl}, \& {Chakraborty}}]{Mirizzi16}
{Mirizzi}, A., {Tamborra}, I., {Janka}, H.-T., {et~al.} 2016, Nuovo Cimento
  Rivista Serie, 39, 1

\bibitem[{{M{\"o}sta} {et~al.}(2014){M{\"o}sta}, {Richers}, {Ott}, {Haas},
  {Piro}, {Boydstun}, {Abdikamalov}, {Reisswig}, \& {Schnetter}}]{moesta14}
{M{\"o}sta}, P., {Richers}, S., {Ott}, C.~D., {et~al.} 2014, \apjl, 785, L29

\bibitem[{{M{\"u}ller} \& {Janka}(2015)}]{bernhard15}
{M{\"u}ller}, B., \& {Janka}, H.-T. 2015, \mnras, 448, 2141

\bibitem[{{M{\"u}ller} {et~al.}(2012){M{\"u}ller}, {Janka}, \&
  {Marek}}]{BMuller12a}
{M{\"u}ller}, B., {Janka}, H.-T., \& {Marek}, A. 2012, \apj, 756, 84

\bibitem[{{M{\"u}ller} {et~al.}(2017){M{\"u}ller}, {Melson}, {Heger}, \&
  {Janka}}]{bernhard17}
{M{\"u}ller}, B., {Melson}, T., {Heger}, A., \& {Janka}, H.-T. 2017, \mnras,
  472, 491

\bibitem[{{M{\"u}ller} {et~al.}(2016){M{\"u}ller}, {Viallet}, {Heger}, \&
  {Janka}}]{bernhard16_prog}
{M{\"u}ller}, B., {Viallet}, M., {Heger}, A., \& {Janka}, H.-T. 2016, \apj,
  833, 124

\bibitem[{{M{\"u}ller} {et~al.}(2018){M{\"u}ller}, {Tauris}, {Heger},
  {Banerjee}, {Qian}, {Powell}, {Chan}, {Gay}, \& {Langer}}]{bernhard18_prog}
{M{\"u}ller}, B., {Tauris}, T.~M., {Heger}, A., {et~al.} 2018, arXiv e-prints,
  arXiv:1811.05483

\bibitem[{{Nagakura} {et~al.}(2019){Nagakura}, {Takahashi}, \&
  {Yamamoto}}]{nagakura19}
{Nagakura}, H., {Takahashi}, K., \& {Yamamoto}, Y. 2019, \mnras, 483, 208

\bibitem[{{Nagakura} {et~al.}(2013){Nagakura}, {Yamamoto}, \&
  {Yamada}}]{nagakura13}
{Nagakura}, H., {Yamamoto}, Y., \& {Yamada}, S. 2013, \apj, 765, 123

\bibitem[{{Nagakura} {et~al.}(2018){Nagakura}, {Iwakami}, {Furusawa}, {Okawa},
  {Harada}, {Sumiyoshi}, {Yamada}, {Matsufuru}, \& {Imakura}}]{nagakura18}
{Nagakura}, H., {Iwakami}, W., {Furusawa}, S., {et~al.} 2018, \apj, 854, 136

\bibitem[{{Nakamura} {et~al.}(2016){Nakamura}, {Horiuchi}, {Tanaka}, {Hayama},
  {Takiwaki}, \& {Kotake}}]{Nakamura16}
{Nakamura}, K., {Horiuchi}, S., {Tanaka}, M., {et~al.} 2016, \mnras, 461, 3296

\bibitem[{{Nakamura} {et~al.}(2015){Nakamura}, {Takiwaki}, {Kuroda}, \&
  {Kotake}}]{Nakamura15}
{Nakamura}, K., {Takiwaki}, T., {Kuroda}, T., \& {Kotake}, K. 2015, \pasj, 67,
  107

\bibitem[{{Nugis} \& {Lamers}(2000)}]{Nugis00}
{Nugis}, T., \& {Lamers}, H.~J.~G.~L.~M. 2000, \aap, 360, 227

\bibitem[{{Obergaulinger} \& {Aloy}(2017)}]{martin17}
{Obergaulinger}, M., \& {Aloy}, M.~{\'A}. 2017, \mnras, 469, L43

\bibitem[{{Obergaulinger} {et~al.}(2006){Obergaulinger}, {Aloy}, {Dimmelmeier},
  \& {M{\"u}ller}}]{martin06}
{Obergaulinger}, M., {Aloy}, M.~A., {Dimmelmeier}, H., \& {M{\"u}ller}, E.
  2006, Astron. Astrophys., 457, 209

\bibitem[{{O'Connor} \& {Ott}(2011)}]{Oconnor11}
{O'Connor}, E., \& {Ott}, C.~D. 2011, \apj, 730, 70

\bibitem[{{O'Connor} \& {Couch}(2018)}]{oconnor18b}
{O'Connor}, E.~P., \& {Couch}, S.~M. 2018, \apj, 865, 81

\bibitem[{{Ott} {et~al.}(2013){Ott}, {Abdikamalov}, {M{\"o}sta}, {Haas},
  {Drasco}, {O'Connor}, {Reisswig}, {Meakin}, \& {Schnetter}}]{Ott13}
{Ott}, C.~D., {Abdikamalov}, E., {M{\"o}sta}, P., {et~al.} 2013, \apj, 768, 115

\bibitem[{{Patat}(2017)}]{patat_book}
{Patat}, F. 2017, {Introduction to Supernova Polarimetry}, ed. A.~W. {Alsabti}
  \& P.~{Murdin}, 1017

\bibitem[{{Paxton} {et~al.}(2011){Paxton}, {Bildsten}, {Dotter}, {Herwig},
  {Lesaffre}, \& {Timmes}}]{Paxton11}
{Paxton}, B., {Bildsten}, L., {Dotter}, A., {et~al.} 2011, \apjs, 192, 3

\bibitem[{{Pejcha} \& {Thompson}(2015)}]{Pejha15}
{Pejcha}, O., \& {Thompson}, T.~A. 2015, \apj, 801, 90

\bibitem[{{Plewa} \& {M{\"u}ller}(1999)}]{Plewa99}
{Plewa}, T., \& {M{\"u}ller}, E. 1999, \aap, 342, 179

\bibitem[{{Richers} {et~al.}(2017){Richers}, {Nagakura}, {Ott}, {Dolence},
  {Sumiyoshi}, \& {Yamada}}]{richers17}
{Richers}, S., {Nagakura}, H., {Ott}, C.~D., {et~al.} 2017, \apj, 847, 133

\bibitem[{{Roberts} {et~al.}(2016){Roberts}, {Ott}, {Haas}, {O'Connor},
  {Diener}, \& {Schnetter}}]{Roberts16}
{Roberts}, L.~F., {Ott}, C.~D., {Haas}, R., {et~al.} 2016, \apj, 831, 98

\bibitem[{{Sasaki} {et~al.}(2017){Sasaki}, {Kajino}, {Takiwaki}, {Hayakawa},
  {Balantekin}, \& {Pehlivan}}]{Sasaki17}
{Sasaki}, H., {Kajino}, T., {Takiwaki}, T., {et~al.} 2017, \prd, 96, 043013

\bibitem[{{Sekiya}(2013)}]{Sekiya13}
{Sekiya}, H. 2013, in Journal of Physics Conference Series, Vol. 460, Journal
  of Physics Conference Series, 012017

\bibitem[{{Sukhbold} {et~al.}(2016){Sukhbold}, {Ertl}, {Woosley}, {Brown}, \&
  {Janka}}]{Sukhbold16}
{Sukhbold}, T., {Ertl}, T., {Woosley}, S.~E., {Brown}, J.~M., \& {Janka}, H.-T.
  2016, \apj, 821, 38

\bibitem[{{Sukhbold} \& {Woosley}(2014)}]{Sukhbold14}
{Sukhbold}, T., \& {Woosley}, S.~E. 2014, \apj, 783, 10

\bibitem[{{Sukhbold} {et~al.}(2018){Sukhbold}, {Woosley}, \&
  {Heger}}]{Sukhbold18}
{Sukhbold}, T., {Woosley}, S.~E., \& {Heger}, A. 2018, \apj, 860, 93

\bibitem[{{Sumiyoshi} \& {Yamada}(2012)}]{sumi12}
{Sumiyoshi}, K., \& {Yamada}, S. 2012, \apjs, 199, 17

\bibitem[{{Summa} {et~al.}(2016){Summa}, {Hanke}, {Janka}, {Melson}, {Marek},
  \& {M{\"u}ller}}]{Summa16}
{Summa}, A., {Hanke}, F., {Janka}, H.-T., {et~al.} 2016, \apj, 825, 6

\bibitem[{{Summa} {et~al.}(2018){Summa}, {Janka}, {Melson}, \&
  {Marek}}]{Summa18}
{Summa}, A., {Janka}, H.-T., {Melson}, T., \& {Marek}, A. 2018, \apj, 852, 28

\bibitem[{{Suwa} {et~al.}(2010){Suwa}, {Kotake}, {Takiwaki}, {Whitehouse},
  {Liebend{\"o}rfer}, \& {Sato}}]{Suwa10}
{Suwa}, Y., {Kotake}, K., {Takiwaki}, T., {et~al.} 2010, \pasj, 62, L49

\bibitem[{{Suwa} {et~al.}(2016){Suwa}, {Yamada}, {Takiwaki}, \&
  {Kotake}}]{Suwa16}
{Suwa}, Y., {Yamada}, S., {Takiwaki}, T., \& {Kotake}, K. 2016, \apj, 816, 43

\bibitem[{{Takahashi} {et~al.}(2001){Takahashi}, {Watanabe}, {Sato}, \&
  {Totani}}]{Takahashi01}
{Takahashi}, K., {Watanabe}, M., {Sato}, K., \& {Totani}, T. 2001, \prd, 64,
  093004

\bibitem[{{Takahashi} \& {Yamada}(2014)}]{takahashi14}
{Takahashi}, K., \& {Yamada}, S. 2014, \apj, 794, 162

\bibitem[{{Takahashi} {et~al.}(2018){Takahashi}, {Yoshida}, \&
  {Umeda}}]{Takahashi18}
{Takahashi}, K., {Yoshida}, T., \& {Umeda}, H. 2018, \apj, 857, 22

\bibitem[{{Takahashi} {et~al.}(2016){Takahashi}, {Yoshida}, {Umeda},
  {Sumiyoshi}, \& {Yamada}}]{Takahashi16}
{Takahashi}, K., {Yoshida}, T., {Umeda}, H., {Sumiyoshi}, K., \& {Yamada}, S.
  2016, \mnras, 456, 1320

\bibitem[{{Takiwaki} {et~al.}(2014){Takiwaki}, {Kotake}, \&
  {Suwa}}]{Takiwaki14}
{Takiwaki}, T., {Kotake}, K., \& {Suwa}, Y. 2014, \apj, 786, 83

\bibitem[{{Takiwaki} {et~al.}(2016){Takiwaki}, {Kotake}, \&
  {Suwa}}]{Takiwaki16}
---. 2016, \mnras, 461, L112

\bibitem[{{Tanaka} {et~al.}(2009){Tanaka}, {Kawabata}, {Maeda}, {Iye},
  {Hattori}, {Pian}, {Nomoto}, {Mazzali}, \& {Tominaga}}]{Tanaka09}
{Tanaka}, M., {Kawabata}, K.~S., {Maeda}, K., {et~al.} 2009, \apj, 699, 1119

\bibitem[{{Timmes} \& {Swesty}(2000)}]{Timmes00}
{Timmes}, F.~X., \& {Swesty}, F.~D. 2000, \apjs, 126, 501

\bibitem[{{Toro} {et~al.}(1994){Toro}, {Spruce}, \& {Speares}}]{Toro94}
{Toro}, E.~F., {Spruce}, M., \& {Speares}, W. 1994, Shock Waves, 4, 25

\bibitem[{{Ugliano} {et~al.}(2012){Ugliano}, {Janka}, {Marek}, \&
  {Arcones}}]{Ugliano12}
{Ugliano}, M., {Janka}, H.-T., {Marek}, A., \& {Arcones}, A. 2012, \apj, 757,
  69

\bibitem[{{Vartanyan} {et~al.}(2019){Vartanyan}, {Burrows}, {Radice},
  {Skinner}, \& {Dolence}}]{vartanyan19}
{Vartanyan}, D., {Burrows}, A., {Radice}, D., {Skinner}, M.~A., \& {Dolence},
  J. 2019, \mnras, 482, 351

\bibitem[{{Viallet} {et~al.}(2013){Viallet}, {Meakin}, {Arnett}, \&
  {Moc{\'a}k}}]{viallet13}
{Viallet}, M., {Meakin}, C., {Arnett}, D., \& {Moc{\'a}k}, M. 2013, \apj, 769,
  1

\bibitem[{{Vink} {et~al.}(2000){Vink}, {de Koter}, \& {Lamers}}]{Vink00}
{Vink}, J.~S., {de Koter}, A., \& {Lamers}, H.~J.~G.~L.~M. 2000, \aap, 362, 295

\bibitem[{{Vink} {et~al.}(2001){Vink}, {de Koter}, \& {Lamers}}]{Vink01}
---. 2001, \aap, 369, 574

\bibitem[{{Woosley} \& {Heger}(2007)}]{Woosley07}
{Woosley}, S.~E., \& {Heger}, A. 2007, \physrep, 442, 269

\bibitem[{{Woosley} {et~al.}(2002){Woosley}, {Heger}, \& {Weaver}}]{woosley02}
{Woosley}, S.~E., {Heger}, A., \& {Weaver}, T.~A. 2002, Reviews of Modern
  Physics, 74, 1015

\bibitem[{{Yoon}(2017)}]{Yoon2017}
{Yoon}, S.-C. 2017, \mnras, 470, 3970

\bibitem[{{Yoshida} {et~al.}(2016){Yoshida}, {Takahashi}, {Umeda}, \&
  {Ishidoshiro}}]{Yoshida16}
{Yoshida}, T., {Takahashi}, K., {Umeda}, H., \& {Ishidoshiro}, K. 2016, \prd,
  93, 20

\end{thebibliography}



\end{document}